\documentclass[]{pasj01}
\usepackage{caption}

\newcommand{\radns}{R_{6}}

\begin{document} 
\Received{}
\Accepted{}

\title{
Correlation between the luminosity and spin-period changes
during outbursts of 12 Be binary pulsars 
observed by the MAXI/GSC and the Fermi/GBM
}

\author{
  Mutsumi \textsc{Sugizaki},\altaffilmark{1}
  Tatehiro \textsc{Mihara},\altaffilmark{1}
  Motoki \textsc{Nakajima},\altaffilmark{2}
  Kazuo \textsc{Makishima},\altaffilmark{1}
}
\altaffiltext{1}{MAXI team, RIKEN, 2-1 Hirosawa, Wako, Saitama 351-0198}
\altaffiltext{2}{School of Dentistry at Matsudo, Nihon University, 
  2-870-1 Sakaecho-nishi, Matsudo, Chiba 101-8308}
\email{sugizaki@riken.jp}



\maketitle

\begin{abstract}

To observationally study spin-period changes of accreting pulsars
caused by the accretion torque,
the present work analyzes
X-ray light curves of 12 Be binary
pulsars obtained by the MAXI/GSC all-sky survey and their pulse periods
measured by the Fermi/GBM pulsar project, both covering more than 6 years
from 2009 August to 2016 March. 
The 12 objects were selected because they are accompanied by clear
optical identification, and accurate measurements of surface magnetic
fields.
The luminosity $L$ and the spin-frequency derivatives $\dot{\nu}$,
measured during large outbursts with $L\gtsim 1\times 10^{37}$ erg s$^{-1}$,
were found to
approximately follow the theoretical relations
in the accretion torque models,
represented by  $\dot{\nu} \propto L^{\alpha}$ ($\alpha\simeq 1$), 
and the coefficient of proportionality between $\dot{\nu}$ and $L^{\alpha}$,
agrees, within a factor of $\sim 3$,
with that proposed by Ghosh \& Lamb (1979).
In the course of the present study, the orbital elements 
of several sources were refined.

\end{abstract}

\section{Introduction}
\label{sec:intro}

X-ray binary pulsars (XBPs) are systems consisting of magnetized
neutron stars and mass-donating stellar companions.
In the vicinity of the neutron star, matter flows from the companion
are guided by the magnetic fields, and are finally funneled onto the
magnetic poles of the neutron star.
Because the accreting matter meanwhile transfers its angular momentum to
the neutron star, the pulsar's period-change rate should correlate with the
mass accretion rate, i.e. the X-ray luminosity.
The relation is thought to reflect the mode of accretion flows,
whether thin-disk or nearly spherical, and also the fundamental
neutron-star parameters including the mass, radius, and magnetic fields.
Consequently, this important issue
has been studied from both theoretical 
and observational points of view.

From theoretical viewpoints, Ghosh \& Lamb (1979a, 1979b, hereafter
GL79) developed a comprehensive theoretical model, which extends those
proposed by \citet{1973ApJ...184..271L} and
\citet{1977Natur.266..683R}.  The GL79 model assumes that magnetic
field lines from the neutron star thread the disk in a broad
transition zone.
Then,
Wang (1987, 1995), 
Lovelace et al. (1995, hereafter LRB95), 
Klu{\'z}niak \& Rappaport (2007, hereafter KR07), and other authors
proposed their revised models, which assume different physical
conditions (see \cite{2009A&A...493..809B},
\cite{2015ApJ...813...91S}, and references therein).

Although a number of observations have so far been performed to
examine how the period changes of XBPs depend on their luminosities
(e.g. \cite{1996ApJ...459..288F, 1996A&A...312..872R,
  1997ApJS..113..367B}),
the results are still inconclusive to
answer whether the phenomenon can be adequately explained 
by any of the proposed theoretical models, 
and if so, whether they can be differentiated.
This is mainly because these
observations have been limited in the sample size, used different
energy bands, or employed different assumptions.  To obtain a clearer
result, we need to carry out unified observations of a reasonable
number of objects that satisfy the following four requirements.  First, the
sample objects must show relatively large changes in their X-ray
fluxes, so that the effects of accretion torque are clearly manifested
in their spin period changes.  Second, the objects must have well
established orbital elements (to remove the orbital Doppler effects in
their period changes), and reasonably accurate distances (to convert
the flux to the mass accretion rate).  Third, we need to have
preliminary knowledge of the objects' magnetic-field strength,
because this is a key quantity that determines the efficiency of the
angular-momentum transfer from the accreting matter to the neutron
star.  Finally, we need to measure the X-ray intensity, spin period,
and the period-change rate of the sample objects for a sufficiently
long time in a unified manner.

The first requirement for our study, 
i.e., large intensity changes,
is accomplished by focusing on Be XBPs.
Being one of the major XBP subclasses, 
they form binaries with Be companion stars, which host
a circumstellar disk along their equator (e.g. \cite{Reig2011}).
These XBPs often exhibit large outbursts lasting for a few weeks to a
few months, mostly at a limited orbital phase near the 
pulsar periastron passage.
These outbursts, together with spin-up episodes which are 
often associated with them (e.g. \cite{1997ApJS..113..367B}),
are naturally explained by an increase in the
accretion rate as the pulsar gets through the stellar
disk, and the associated increase in transfer of the angular 
momentum to the neutron star.
If the luminosity and the spin-period changes
are monitored throughout these outbursts, 
the obtained data will become of great value.
As detailed later in section \ref{sec:target_select},
a fair fraction of the currently known Be XBPs have known 
orbital parameters and estimated distances.
As a result, our second requirement is satisfied automatically.

The third requirement, i.e. the surface magnetic field of neutron
stars in XBPs, is best measured with the cyclotron-resonance
scattering feature (CRSF) in X-ray spectra
(e.g. \cite{Makishima1999}).
Thanks to the recent high-sensitivity instruments
covering the hard X-ray band onboard the INTEGRAL, Suzaku, and NuSTAR
satellites,
the number of XBPs with confirmed CRSFs 
increased in these years
(e.g. 
\cite{2011PASJ...63S.751Y}, 
\cite{2012A&A...542L..28K}, 
\cite{2014PASJ...66...59Y}, 
\cite{2014ApJ...795..154T}, 
\cite{2015ApJ...815...44M}, 
\cite{2016MNRAS.457..258T}). 
So far, the CRSF has been detected from about 25 XBPs
altogether, of which 15 are Be XBPs
(e.g. \cite{2015SSRv..191..293R, 2015A&ARv..23....2W}).
Therefore, focusing on Be XBPs will also
satisfy the third requirement.

Let us consider the final requirement.  Since 2009, the MAXI (Monitor
of All-sky X-ray Image; \cite{Matsuoka_pasj2009}) mission on the
International Space Statin has been scanning the whole X-ray sky every
92-minute orbital cycle with the GSC (Gas Slit Camera;
\cite{Mihara_pasj2011}).
Meanwhile, the GBM (Gamma-ray Burst Monitor; \cite{Meegan2009})
onboard the Fermi Gamma-Ray Space Telescope has been monitoring the
whole sky in the X-ray to gamma-ray band since 2008.  The timing
analysis of the GBM data provides us information on the pulsed
emission from bright XBPs in our Galaxy \citep{2009arXiv0912.3847F,
  2010ApJ...708.1500C}.
Data taken by these two missions for over 6 years satisfy the
requirement.
In fact, we analyzed the data of two XBPs, GX 304$-$1
\citep{2015PASJ...67...73S} and 4U 1626$-$67
\citep{2016PASJ...68S..13T}, and found that the results on both these
sources agree reasonably with the disk-accretion model proposed by
GL79.

In the present paper, we investigate the correlation between the
luminosity and pulse-period changes of 12 Be XBPs using the long-term
($> 6$ years) X-ray data, which were obtained by the MAXI GSC survey
and the Fermi GBM pulsar project.
The observations and target selection are described in section
\ref{sec:obs_data}, and the analysis in section \ref{sec:analysis}.
We discuss the obtained results in section \ref{sec:discussion}.

\section{Observations}
\label{sec:obs_data}

\subsection{MAXI GSC}

Since the MAXI in-orbit operation started in 2009 August, the GSC
light curves of $\sim 300$ pre-registered sources have been processed,
typically every day, in the 2--4 keV, 4--10 keV, and 10--20 keV bands, 
and the results are uploaded on the archive web
site\footnote{http://maxi.riken.jp/}.
The data provide 3-energy-band photon fluxes
for each scan transit of 30--50 s duration,
every 92 minutes synchronized with the ISS orbital cycle,
as well as those averaged for every MJD (Modified Julian Date) time bin.
In the standard data processing, the time-dependent effective area
for each target is calculated by assuming that it has
a nominal Crab-like spectrum, in the 2--20 keV band,
represented by a power-law with a photon index $\Gamma =2.1$.
Among these sources, some 52 are XBPs;
their long-term ($\gtsim 6$ years) intensity histories 
can be constructed from the GSC data.

We can also analyze X-ray energy spectra of bright sources
using all available event data
from the MAXI GSC.
Together with the distance, this information is necessary to
quantitatively estimate the bolometric source luminosity, 
and hence the accretion rate.
The GSC response functions are calculated with the standard
tools \citep{sugizaki_pasj2011, 2012PASJ...64...13N}, and 
the model fits to the GSC spectra are carried out 
on the XSPEC software version 12.8 
\citep{1996ASPC..101...17A}
released as a part of the HEASOFT software package, version 6.19.

Since the MAXI GSC scans over each target on the sky only for 30--50 s
every 92 minutes, it is not suited for pulse-period
measurements, unless the period is $\lesssim 30$ s or longer than $\gtsim 92$ min
(e.g. \cite{2016PASJ...68S..13T}).  Therefore, we rely on the Fermi GBM data
as described below.

\subsection{Fermi GBM pulsar data}
The Fermi GBM pulsar project \citep{2009arXiv0912.3847F,2010ApJ...708.1500C}
provides, on their web
site\footnote{http://gammaray.nsstc.nasa.gov/gbm/science/pulsars/},
results of the timing analysis
for pulsating X-ray sources in the energy band above 8 keV.
The data consist of pulse frequencies and pulsed fluxes of 
positively detected $\sim 50$ XBPs in our Galaxy and the Large Magellanic Cloud (LMC),
obtained since the in-orbit operation started in 2008 July. 
When the binary orbital elements of the objects are accurately known,
the released pulse periods are already corrected for the 
expected orbital Doppler shifts.
We utilize the pulse frequency data of the Be XBPs to be studied,
but not their pulsed fluxes.

\subsection{Target selection}
\label{sec:target_select}

The high-mass X-ray binary catalogues,
given by \citet{2006A&A...455.1165L} and \citet{2015A&ARv..23....2W}, 
list 60 Be X-ray binaries and its candidates
in our Galaxy.
Out of them, 
29 objects have securely been established as Be XBPs 
bases on the detection of periodic X-ray pulsations 
and the optical identification with Be-star companions.
These objects hence constitute a starting point of our sample, 
because Be XBPs are considered to provide an ideal opportunity
for our purpose (section \ref{sec:intro}).

Among the 29 Be XBPs, the GBM have detected significant pulsed
emission from 14 sources, each at least on one occasion, since the
MAXI in-orbit operation started in 2009.  Table \ref{tab:bexbpobs}
lists their source names, pulse periods, orbital periods and
eccentricities, spectral types of their optical companions, and the
source distances estimated from the optical data.
The table also gives the time period over which each source was 
positively detected by both the MAXI/GSC and the Fermi/GBM.

Among these 14 objects, the binary orbital elements (as represented by
the eccentricity in table \ref{tab:bexbpobs}) are still unavailable
for two sources, Cep X-4 and LS V $+$44 17.  Therefore, we cannot
remove the orbital Doppler effects from their pulse-period data.
Furthermore, useful period-change measurements require the source to
be detected over a sufficiently long period, typically 10 days.
As seen in table \ref{tab:bexbpobs}, the data of two other sources,
MXB 0656-072 and SAX J2103.5+4545, do not satisfy the condition.  We
thus excluded these four sources, and chose the remaining 11 Be XBPs
as the primary analysis targets.

In addition to these, we included, into our final sample, one more
object, RX J0520.5$-$6547, which is not in our Galaxy but in the LMC.
It showed a large outburst activity in 2013--2014, which was observed
by both the GSC and the GBM, and also allowed the CRSF detection
\citep{2014ApJ...795..154T}.  Table \ref{tab:bexbpobs} hence includes
data of the object.

Among the 12 objects in our final sample, 
the CRSF has been detected from 9 sources. 
The surface magnetic field $B_{12}$ in units of $10^{12}$ G
is estimated from the fundamental CRSF energy $E_\mathrm{a}$  as 
\begin{equation}
B_{12} = \frac{1}{\sqrt{1-x^{-1}}}\left(\frac{E_\mathrm{a}}{\mathrm{11.6\, keV}}\right)
\label{equ:bcyc}
\end{equation}
where
\begin{equation}
x = \frac{Rc^2}{2GM}
\label{equ:rsch}
\end{equation}
is the surface redshift parameter,
namely, the ratio of the neutron-star radius $R$ 
to the Schwarzschild radius,
with the neutron-star mass $M$,
the gravitational constant $G$, and the velocity of light $c$.
In several XBPs, the observed $E_\mathrm{a}$ values are known to depend
to some extent on the source luminosity 
\citep{2004ApJ...610..390M, 2006ApJ...646.1125N, 2010ApJ...710.1755N, 2011PASJ...63S.751Y, 2012A&A...542L..28K}.
This behavior is understood by considering that
the scattering region responsible for the CRSF formation
changes its height along the field lines, 
depending on the balance between the radiation and accretion pressures 
\citep{1998AdSpR..22..987M, 2007A&A...465L..25S}.
To best estimate the field strength on the neutron-star surface,
we employed the highest $E_\mathrm{a}$ value that has ever been recorded 
in each source.
In table \ref{tab:bexbpobs}, the selected $E_\mathrm{a}$ values from
the past literature are listed.

\begin{table*}
\caption{
%
%
Properties of Be XBPs detected by Fermi GBM and MAXI GSC
since 2009 August 15 to 2015 December 31.
}
\label{tab:bexbpobs}
\footnotesize
\begin{center}
\begin{tabular}{rlcccccccccc}
\hline
\hline
$^*$No. & Source name    & $^\dagger$$P_{\rm s}$ & $^\dagger$$P_{\rm orb}$ & $^\dagger$$e$ & Active epoch & $^\dagger$$T_{\rm out}$ & Spec.Type & $^\dagger$$D$ & $^\dagger$$E_{\rm a}$ \\
        &                 & ( s )      & ( d )        &     &  ( MJD )     & ( d )              &         &(kpc)&(keV)\\
\hline
1 &         4U 0115$+$63 & ~~3.6 & ~24.3  & 0.34$^{*1}$ & 55701 -- 57343 &  64.9 & B0.2 Ve$^{*2}$  &  $7.0\pm 0.3$$^{*44}$  & $16$$^{*3,4}$ \\
2 &          X 0331$+$53 & ~~4.4 & ~33.9  & 0.37$^{*5}$ & 57193 -- 57290 &  56.0 & O8.5 Ve$^{*6}$  &  $6.0\pm 1.5$$^{*45}$  & $31$$^{*7}$   \\
3 &  RX J0520.5$-$6932   & ~~8.0 & ~23.93 & 0.03$^{*8}$ & 56644 -- 56725 &  77.9 & O8 Ve$^{*9}$   &  $50\pm 2$$^{*9}$    & $31.5$$^{*10}$   \\
4 &         H 1553$-$542 & ~~9.3 & 31.34  & 0.04$^{*12}$ & 57046 -- 57144 &  94.0 & B1-2 V$^{*11}$   &  $20\pm 4$$^{*11}$   & $27.3$$^{*12}$  \\
5 &        GS 0834$-$430 & ~12.3 & 105.8  & 0.12$^{*14}$ & 56106 -- 56146 &  34.0 & B0-2 III-Ve$^{*13}$ &  $5^{+1}_{-2}$$^{*13}$   &  ---   \\
6 &      XTE J1946$+$274 & ~15.8 & 172.0  & 0.33$^{*15}$ & 55352 -- 55682 & 119.9 & B0–1 IV–Ve$^{*16}$  &  $8.7\pm 1.2$$^{*44}$  & $35$$^{*17,15}$ \\
7 &        2S 1417$-$624 & ~17.5 & ~42.2  & 0.45$^{*18}$ & 55124 -- 55218 &  94.0 & B1 Ve$^{*19}$    &  $11^{+1}_{-9}$$^{*19}$   &  ---   \\
8 &        KS 1947$+$300 & ~18.8 & ~40.4  & 0.02$^{*20}$ & 56567 -- 57089 & 135.9 & B0 Ve$^{*21}$    &  $10.4\pm 0.9$$^{*44}$  & $12.2$$^{*22}$   \\
9 &       EXO 2030$+$375 & ~41.3 & ~46.0  & 0.41$^{*23}$ & 55057 -- 57279 & 368.4 & B0e$^{*24}$      &  $6.5\pm 2.5$$^{*45}$   &  ---  \\     
$\cdot$ &    Cep X-4$^*$ & ~66.3 &  ---   & ---  & 56813 -- 56843 &   6.0 & B1-B2 Ve$^{*25}$        &  $5.9\pm 0.9$$^{*44}$   & $30.4$$^{*26,27}$ \\ 
10&       GRO J1008$-$57 & ~93.7 & 249.5  & 0.68$^{*28}$ & 55157 -- 57170 & 365.4 & B0e$^{*29}$      &  $5.8\pm 0.5$$^{*44}$   & $76$$^{*30}$  & \\
11&          A 0535$+$262 & 103.5 & 111.1 & 0.47$^{*31}$ & 55050 -- 57070 & 245.6 & O9.7 IIIe$^{*32}$ &  $2.1\pm 0.5$$^{*32}$  & $46.8$$^{*33,34}$  \\
$\cdot$ &  MXB 0656$-$072$^*$ & 160.7 & 101.2 & ---  & 55288 -- 55292 &   3.9 & O9.7 Ve$^{*35}$    &  $3.9\pm 0.1$$^{*35}$  & $32.8$$^{*35}$    \\
$\cdot$ &   LS V $+$44 17$^*$ & 205.2 &  ---  & ---  & 55284 -- 55716 &  33.9 & B0.2 Ve$^{*36}$    &  $2.2\pm 0.5$$^{*36}$  & $31.9$$^{*37}$    \\
12&            GX 304$-$1 & 275.5 & 132.2 & 0.52$^{*38}$ & 55286 -- 57145 & 209.1 & B0.7 Ve$^{*39}$  &  $2.4\pm 0.5$$^{*40}$ & $53.7$$^{*41}$   \\
$\cdot$ & SAX J2103.5$+$4545$^*$ & 358.6 & ~12.7 & 0.4$^{*42}$  & 55483 -- 56965 &  23.9 & B0 Ve$^{*43}$ &  $6.5\pm 0.9$$^{*43}$   & ---      \\
\hline                                                                         
\end{tabular}
\end{center}
$^*$ Objects with numbers (1--12) constitute our final sample.
$^\dagger$$P_{\rm s}$,   
$^\dagger$$P_{\rm orb}$, 
$^\dagger$$e$,    
$^\dagger$$T_{\rm out}$, 
$^\dagger$$D$, and
$^\dagger$$E_{\rm a}$
are 
the pulse period,
the orbital period,
the orbital eccentricity,
the total period for which both the MAXI GSC and the Fermi GBM detected the source,
the source distance estimated from the optical companion, 
and the fundamental cyclotron-resonance energy, respectively.
%
%
%
\\
References: 
%
%
*1. \citet{1997ApJS..113..367B},  
*2. \citet{2001A&A...369..108N},  
*3. \citet{2004ApJ...610..390M},  
*4. \citet{2006ApJ...646.1125N},  
%
%
%
*5. \citet{2016A&A...589A..72D}, 
*6. \citet{1999MNRAS.307..695N}, 
*7. \citet{2010ApJ...710.1755N}, 
%
%
%
*8. \citet{2014ATel.5856....1K},  
*9. \citet{2001MNRAS.324..623C},  
*10. \citet{2014ApJ...795..154T}, 
%
%
%
*11. \citet{2016MNRAS.462.3823L}   
*12. \citet{2016MNRAS.457..258T},  
%
%
*13. \citet{2000MNRAS.314...87I}, 
*14. \citet{1997ApJ...479..388W}, 
%
%
*15. \citet{2015ApJ...815...44M}, 
*16. \citet{2002A&A...393..983V}, 
*17. \citet{2001ApJ...563L..35H}, 
%
%
*18. \citet{2004MNRAS.349..173I}, 
*19. \citet{1984ApJ...276..621G}, 
%
%
*20. \citet{2004ApJ...613.1164G}, 
*21. \citet{2003A&A...397..739N}, 
*22. \citet{2014ApJ...784L..40F}, 
%
%
*23. \citet{2008ApJ...678.1263W}, 
*24. \citet{1988MNRAS.232..865C}, 
%
%
%
*25. \citet{1998A&A...332L...9B} 
*26. \citet{1991ApJ...379L..61M} 
*27. \citet{2015ApJ...806L..24F} 
%
%
%
*28. \citet{2013A&A...555A..95K}, 
*29. \citet{1994MNRAS.270L..57C}, 
*30. \citet{2014PASJ...66...59Y}, 
%
*31. \citet{1996ApJ...459..288F}, 
*32. \citet{1998MNRAS.297L...5S}, 
*33. \citet{2006ApJ...648L.139T}, 
*34. \citet{2007A&A...465L..21C}, 
%
*35. \citet{2006A&A...451..267M} 
%
%
*36. \citet{2005A&A...440.1079R} 
*37. \citet{2012MNRAS.421.2407T} 
%
%
*38. \citet{2015PASJ...67...73S},  
*39. \citet{1978MNRAS.184P..45M},  
*40. \citet{1980MNRAS.190..537P},  
*41. \citet{2011PASJ...63S.751Y},  
%
%
%
*42. \citet{2000ApJ...544L.129B} 
*43. \citet{2004A&A...421..673R} 
%
%
%
*44. \citet{2012A&A...539A.114R}, 
%
*45. \citet{2015A&A...574A..33R}, 
\end{table*}

\section{Analysis}
\label{sec:analysis}

\subsection{X-ray light curves and pulse-frequency changes}

Figure \ref{fig:lcper} 
show the 2--20 keV light curves
of the selected 12 Be XBPs, measured by the MAXI GSC, 
from 2009 August to 2016 March,
and the pulse-frequency $\nu_\mathrm{s}$ measured with the Fermi GBM
during outbursts in the same period. 
All the pulse frequencies 
are first converted to their barycentric values, and then
corrected for the orbital Doppler effects,
to so-called ``spin frequencies'', 
employing the orbital elements
summarized in table \ref{tab:beparam}.
These corrections were performed,
prior to the data release, by the Fermi/GBM team,
expect for GS 0834$-$430, GRO J1008$-$57 and XTE J1946$+$274
for which the orbital parameters were unavailable.
Since the orbital parameters of two of them, 
GRO J1008$-$57 and XTE J1946$+$274,
were later published 
(see references in table \ref{tab:beparam}),
we conducted the orbital Doppler corrections by ourselves 
using the reported parameters.

Through this analysis process, we found that the spin
frequencies of 4U 0115$+$63, GS 0834$-$430, KS 1947$+$300,
and GRO J1008$-$57
show, as presented in figures
\ref{fig:pmodfit_4u0115}--\ref{fig:pmodfit_groj1008} in Appendix,
some modulations synchronized with the binary period, even
though the data had already been corrected for
the orbital Doppler shifts.
This means that the orbital effects may not have been
adequately removed.
As described in Appendix,
the present study 
allows us to refine the orbital elements in a
self-consistent way.
Then, as listed in table \ref{tab:beparam}
(in comparison with the previous values),
we successfully improved the orbital
parameters of these sources, and by employing them, the residual
orbital modulations were removed
(figures \ref{fig:pmodfit_4u0115}--\ref{fig:pmodfit_groj1008}).
The refined spin-frequency data 
are used in figure \ref{fig:lcper} and all the analysis hereafter.

\begin{table*}
\caption{
Binary orbital elements of the selected Be X-ray binary pulsars.
}
\label{tab:beparam}
\footnotesize
\begin{center}
\begin{tabular}{rllllll@{~~}ll}
\hline
\hline
No. & Source name     & $P_{\rm orb}$ & $^*$$a_{\rm X} \sin i$ & ~$e$ & ~$^*$$\omega$ & \multicolumn{2}{c}{$^*$$T_{\rm peri}$ (P) or $^*$$T_{\pi/2}$ (T) } & Ref.\& Note \\ 
    &           & ( d ) & ( lt-s ) &  & ( $^\circ$ ) &  \multicolumn{2}{c}{( MJD )} & \\
\hline
1 &     4U 0115$+$63 & ~~24.31704(6)  & 140.13(8)  & 0.3402(2) &  47.66(9) & 49279.268(3)     & P & [1] \\
$\cdot$ &    ---     & ~~2.431689     & 141.37   & 0.3401     &  49.225   & 55601.751         & P & Appendix\\
2 &     X 0331$+$53  & ~~33.850(3)  & 77.8(2)    & 0.371(5)   & 277.4(1)  & 57157.38(5)       & P & [2] \\ 
3 & RX J0520.5$-$6932& ~~23.93(7)   & 107.6(8)   & 0.029(10)  & 233(18)   & 56666.41(3)       & T & [3] \\
4 &     H 1553$-$542 & ~~31.303(27) & 201.3(8)   & 0.0351(22)  & 163.4(35) & 57088.921(19)    & T & [4] \\
5 &    GS 0834$-$430 &  105.8(4)    & 128(40)    & 0.12(+8/-4) & 140(40)   & 48809.6(1.5)     & T & [5] \\ 
$\cdot$ &    ---     &  105.8 : fix & 199        & 0.125      & 165      & 56130.0            & P &  Appendix\\
6 &  XTE J1946$+$274 &  172.7(6)    & 471(+3/-4) & 0.246(9)   & 273(2)   & 55514(1)           & P &  [6] \\
7 &  2S 1417$-$624   & ~~42.175     & 188(2)     & 0.446(2)   & 300.3(6) & 51612.17(5)  & P &  [7,8] \\ 
8 &    KS 1947$+$300 & ~~40.415(7)  & 137.4(1.2) & 0.034(7)   & 33(3)    & 51985.31(7)  & T  & [9] \\
$\cdot$ &    ---     & ~~40.50      & 130.2      & 0.008      & 57       & 56550.54     & T  & Appendix\\
9  &   EXO 2030$+$375 & ~~46.0213(3) & 246(2)   & 0.410(1)    & 211.9(4) & 52756.17(1)  & P  & [10] \\
10 &  GRO J1008$-$57  &  249.480(4)  & 530(6)   & 0.68(2)     & 334(8)   & 55424.7(2)   & P  & [11, 12] \\
$\cdot$ &   ---     &  249.480 : fix & 691      & 0.65        & 299      & 55413        & P  & Appendix\\
11 &    A 0535$+$262  &  111.10      & 267(13)  & 0.47(2)     & 130(5)   & 53613.00     & P  & [13] \\
12 &      GX 304$-$1  &  132.189(2)  & 498(6)   & 0.524(7)    & 122.5(4) & 55425.020(1) & P  & [14] \\ 
\hline
\end{tabular}
\end{center}
$^*$$a_{\rm X} \sin i$ is the semi-major axis projected on the line of sight.
$^*$$\omega$ is the argument of periastron.
$^*$$T_{\rm peri}$ (P) or $T_{\rm \pi/2}$ (T) is 
the epoch of periastron passage or mean logitude of $90^\circ$, respectively.
The other symbols have the same meanings as in table \ref{tab:bexbpobs}.
\\
References: 
%
[1] \citet{1997ApJS..113..367B},
%
[2] \citet{2016A&A...589A..72D},
%
[3] \citet{2014ATel.5856....1K},
%
[4] \citet{2016MNRAS.457..258T},
%
[5] \citet{1997ApJ...479..388W},
%
[6] \citet{2015ApJ...815...44M},
%
[7] \citet{1996A&AS..120C.209F},
[8] \citet{2004MNRAS.349..173I},
%
%
%
[9] \citet{2004ApJ...613.1164G},
%
[10] \citet{2008ApJ...678.1263W},
%
%
[11] \citet{2007MNRAS.378.1427C},
[12] \citet{2013A&A...555A..95K},
%
[13] \citet{1996ApJ...459..288F},
%
[14] \citet{2015PASJ...67...73S},
\end{table*}

\begin{figure*}
\includegraphics[width=85mm]{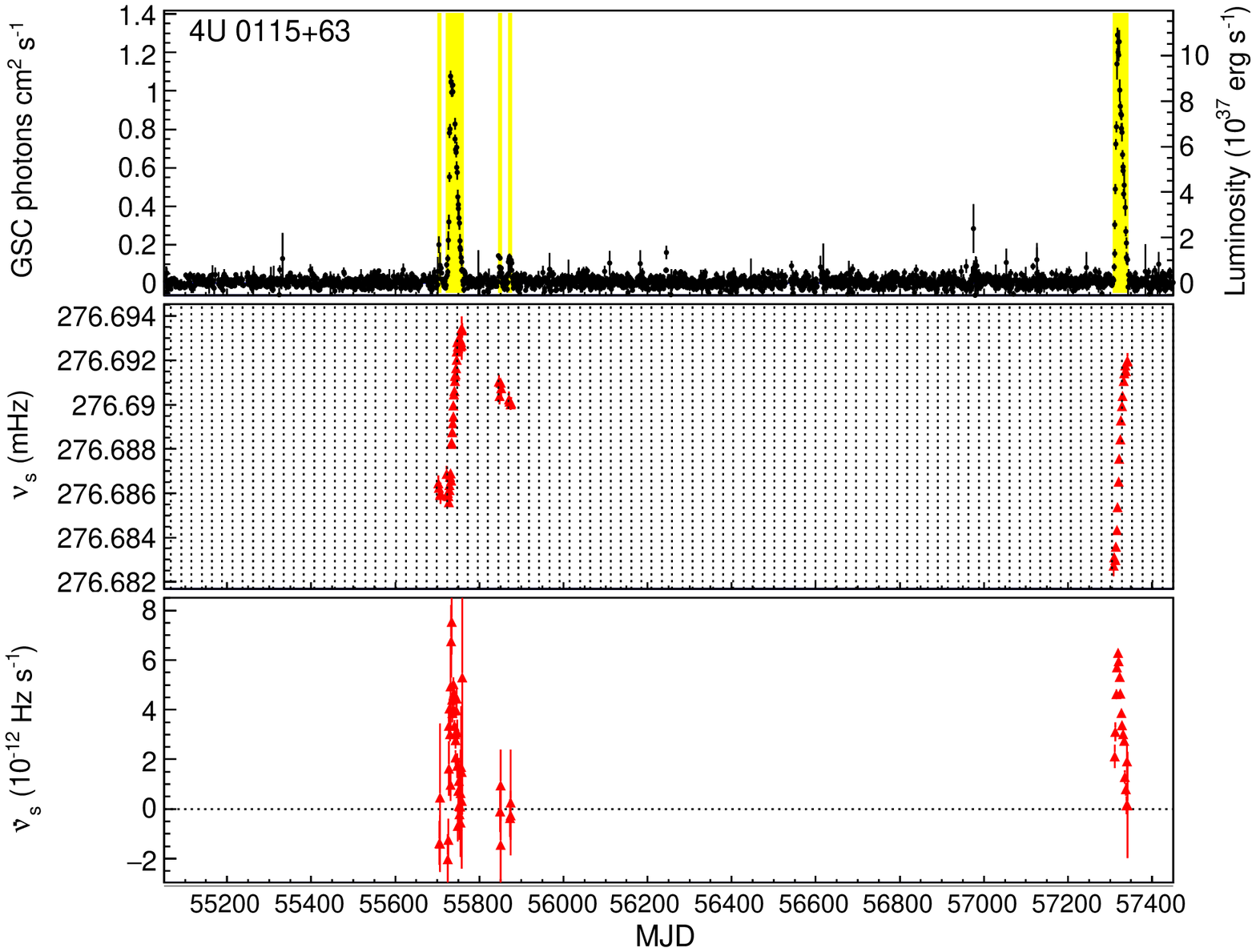}
\hspace{2mm}
\includegraphics[width=85mm]{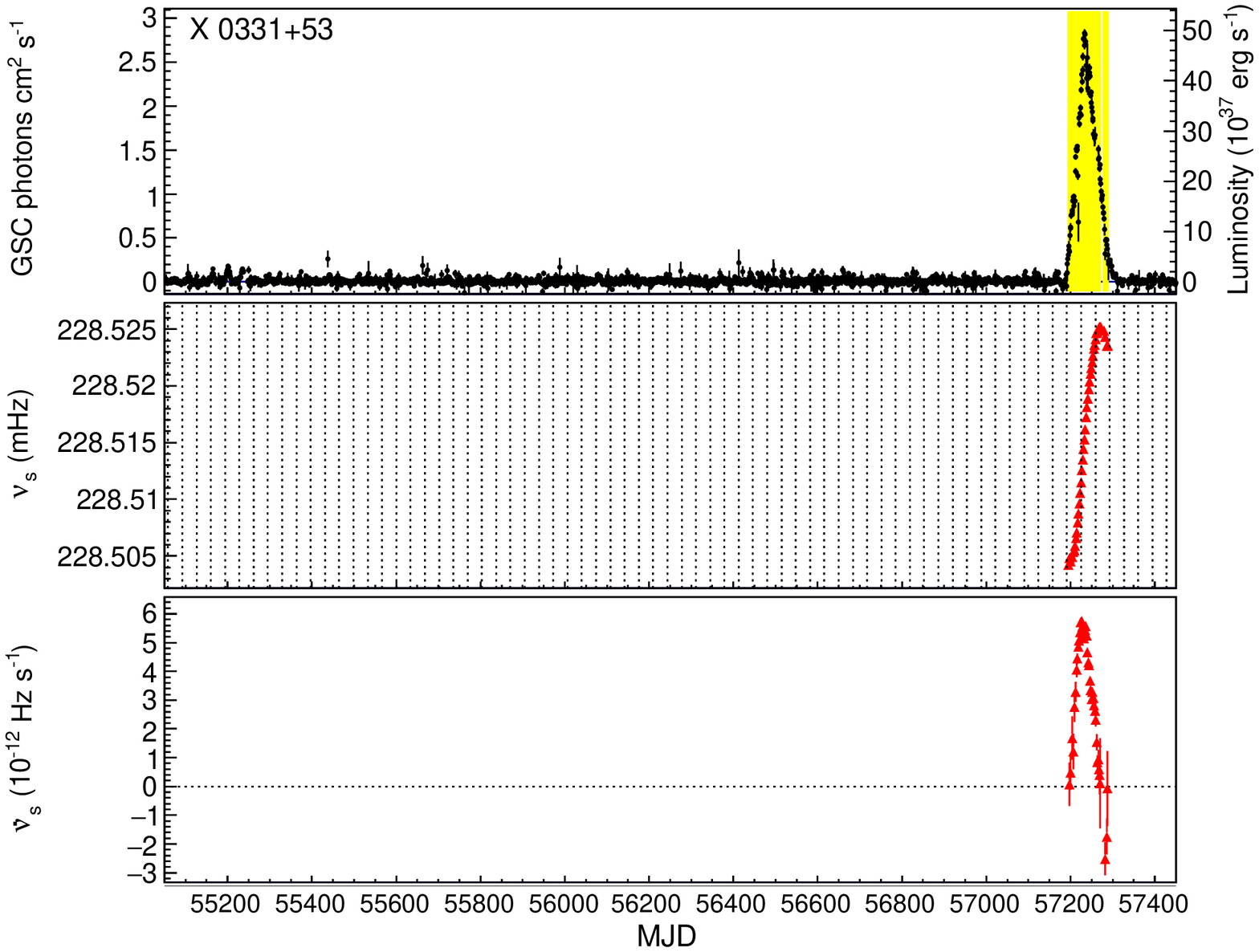}

\vspace{4mm}

\includegraphics[width=85mm]{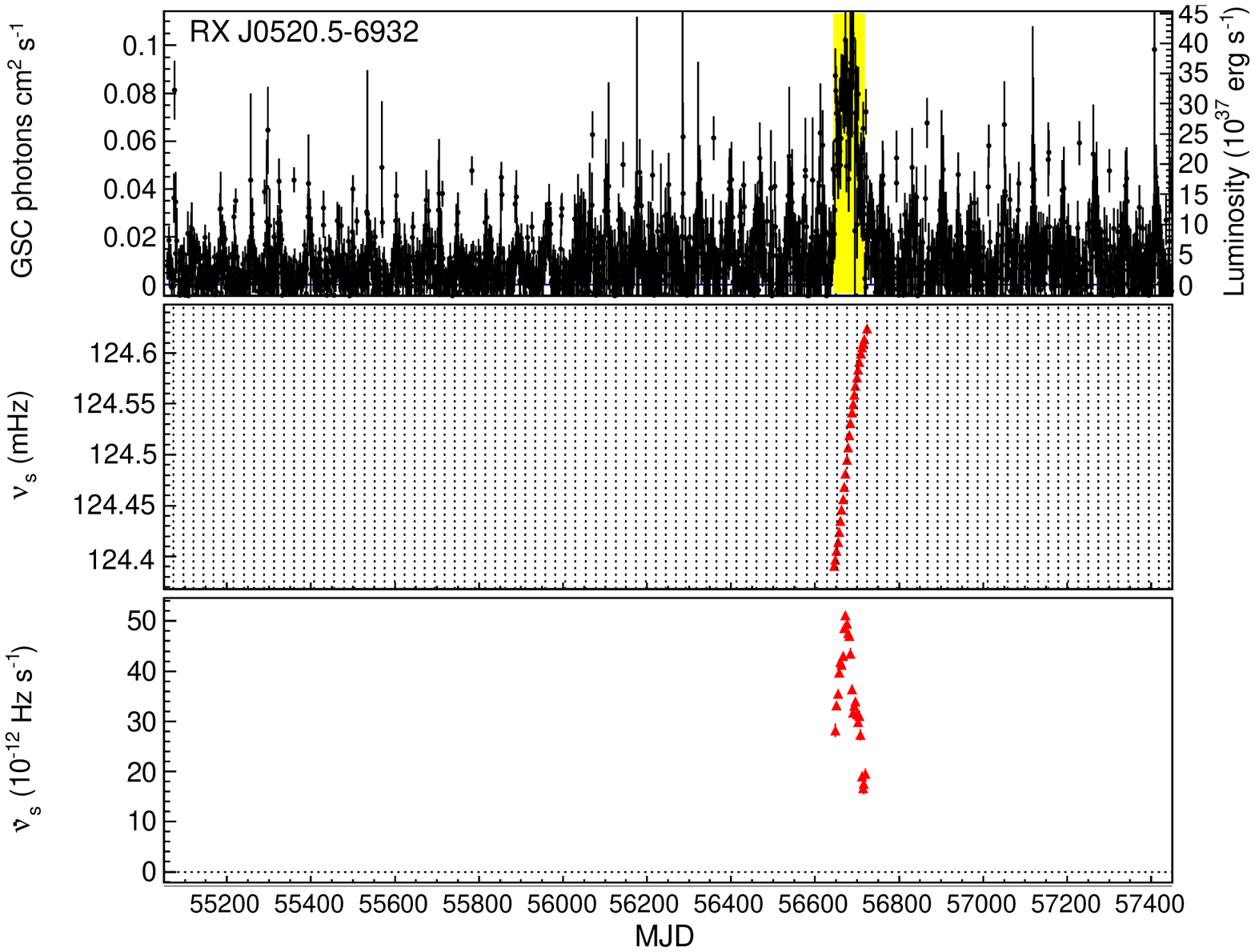}
\hspace{2mm}
\includegraphics[width=85mm]{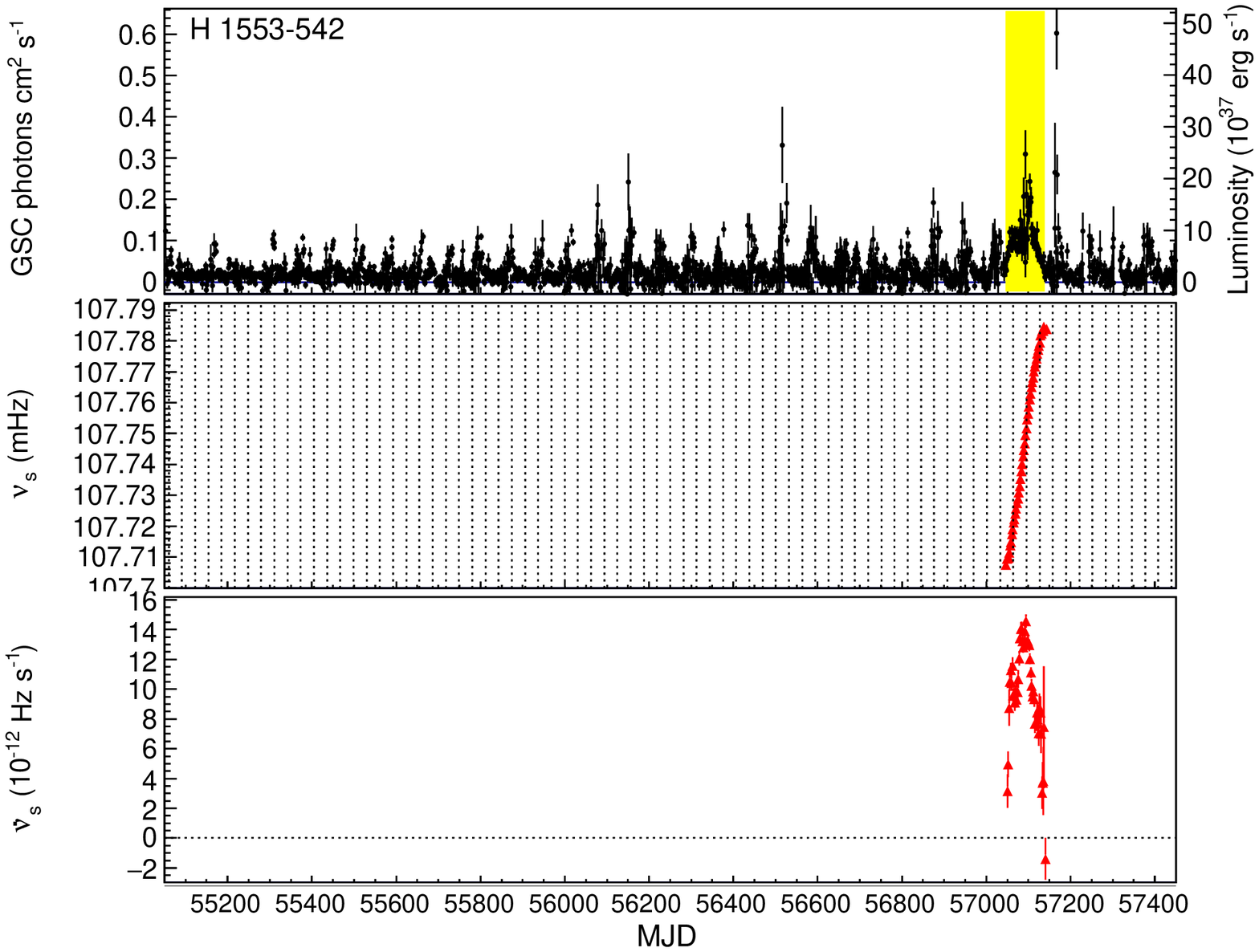}

\vspace{4mm}

\includegraphics[width=85mm]{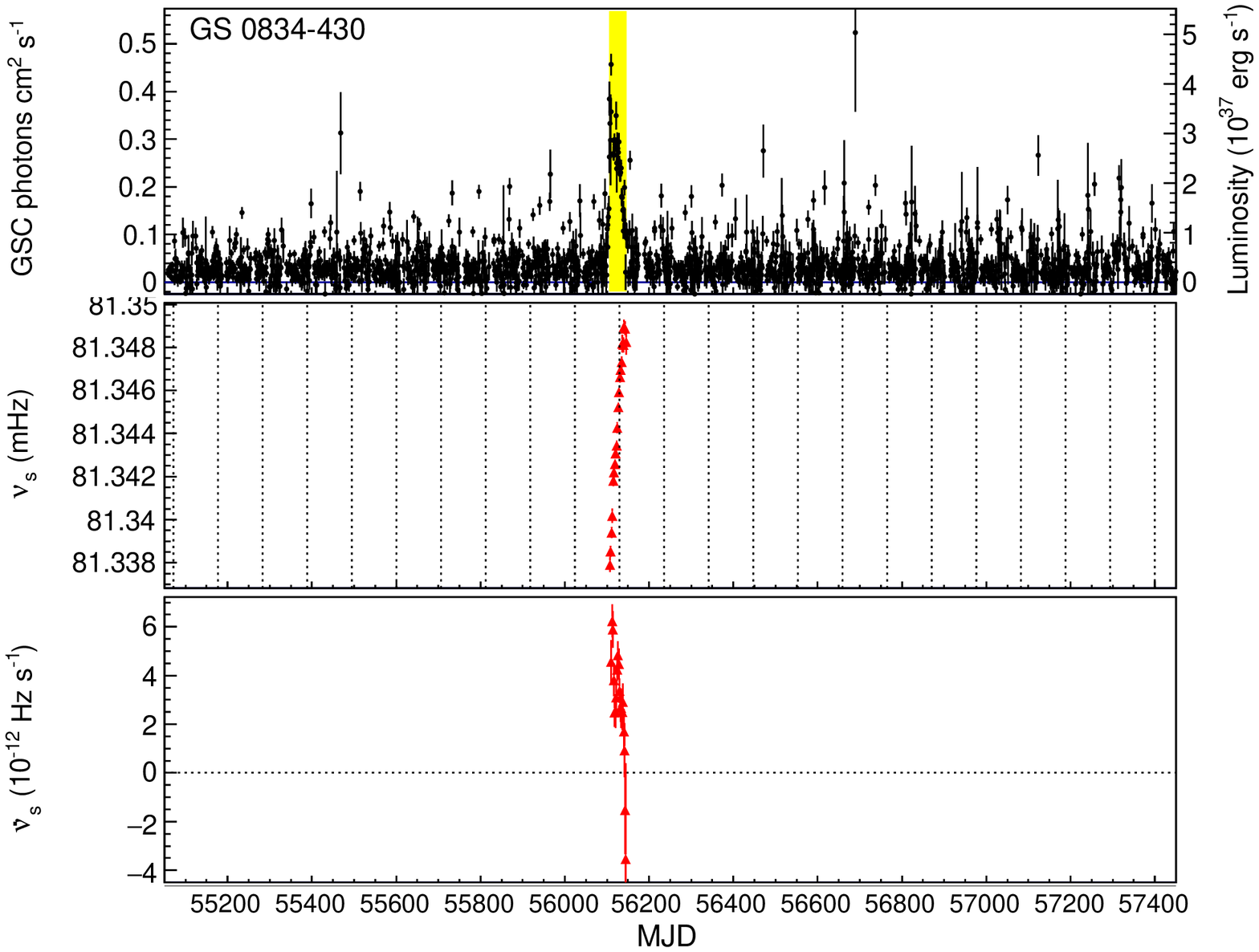}
\hspace{2mm}
\includegraphics[width=85mm]{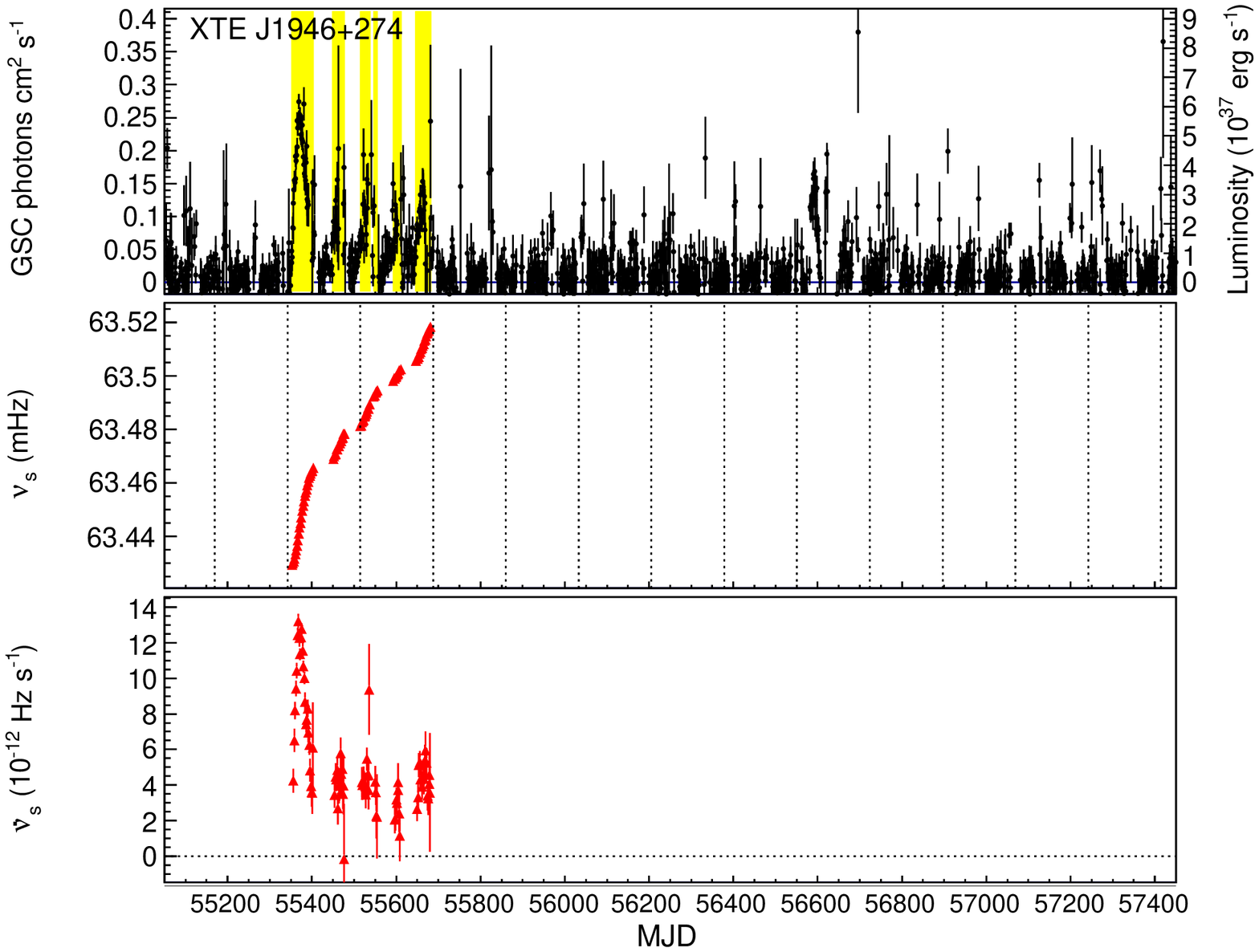}
\caption{
MAXI GSC 2--20 keV light curve in 1-d time bin (top),
Fermi GBM pulse frequency corrected for the orbital Doppler shift 
during outbursts (middle),
and frequency derivative (bottom),
for each of the 12 selected Be XBPs.
All vertical error bars represent 1-$\sigma$ (68 \%) confidence limits of
statistical uncertainty.
The right-side ordinate at the top panel
represents the bolomatric luminosity scale
calculated from the best-fit spectral models. 
Vertical dashed lines in the middle panels 
indicate the epochs of the pulsar periastron passages.
}
\label{fig:lcper}
\end{figure*}

\begin{figure*}
\includegraphics[width=85mm]{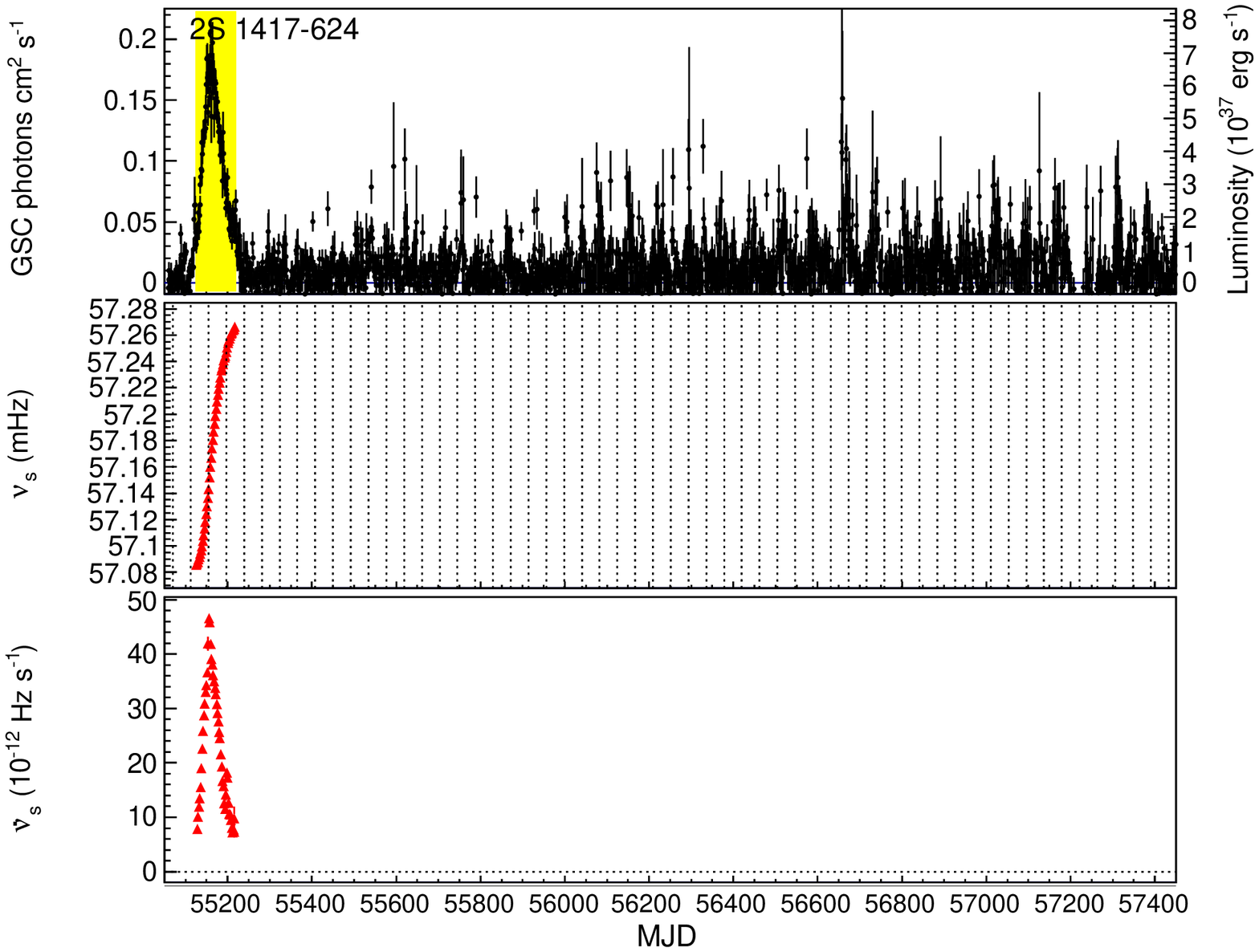}
\hspace{2mm}
\includegraphics[width=85mm]{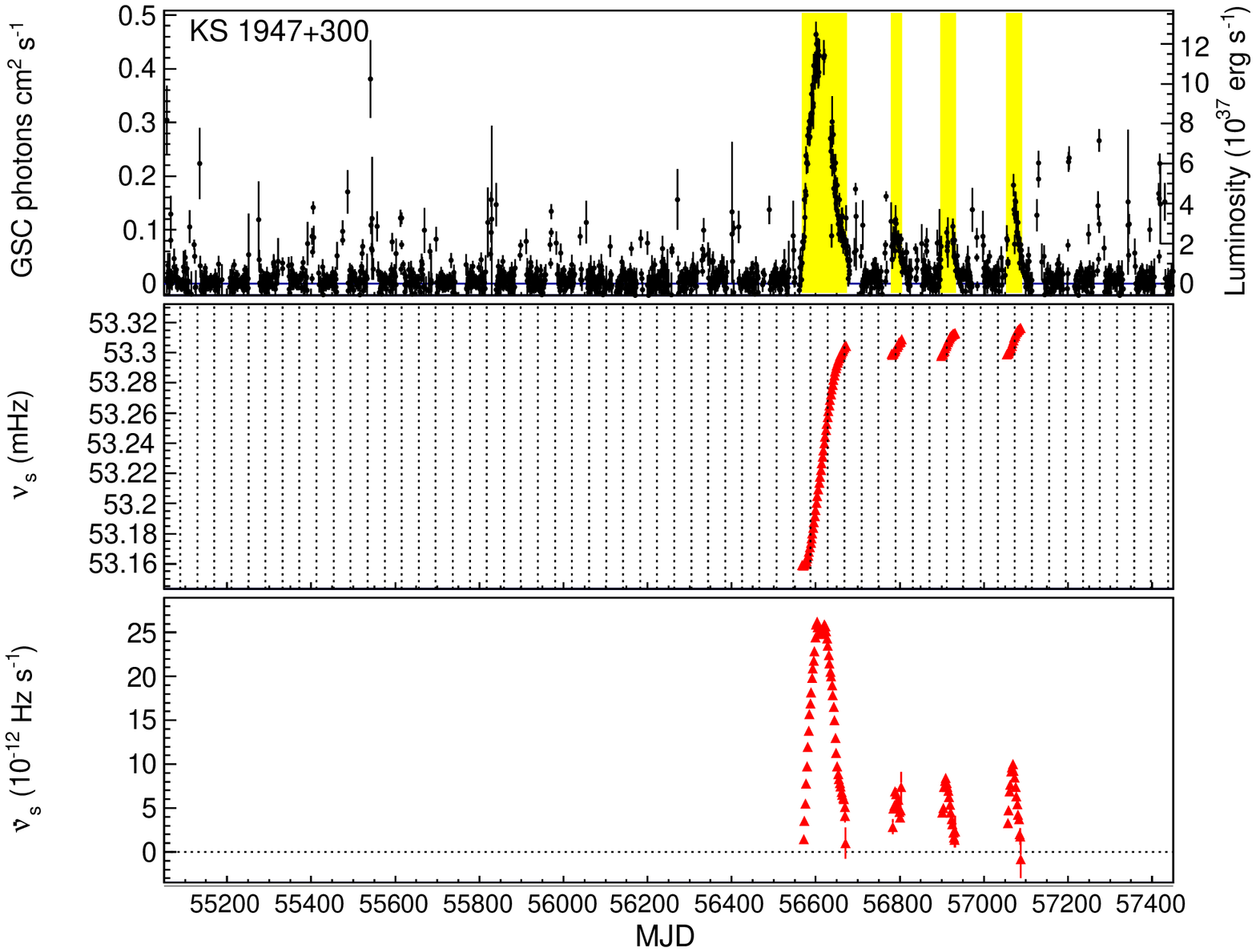}

\vspace{4mm}

\includegraphics[width=85mm]{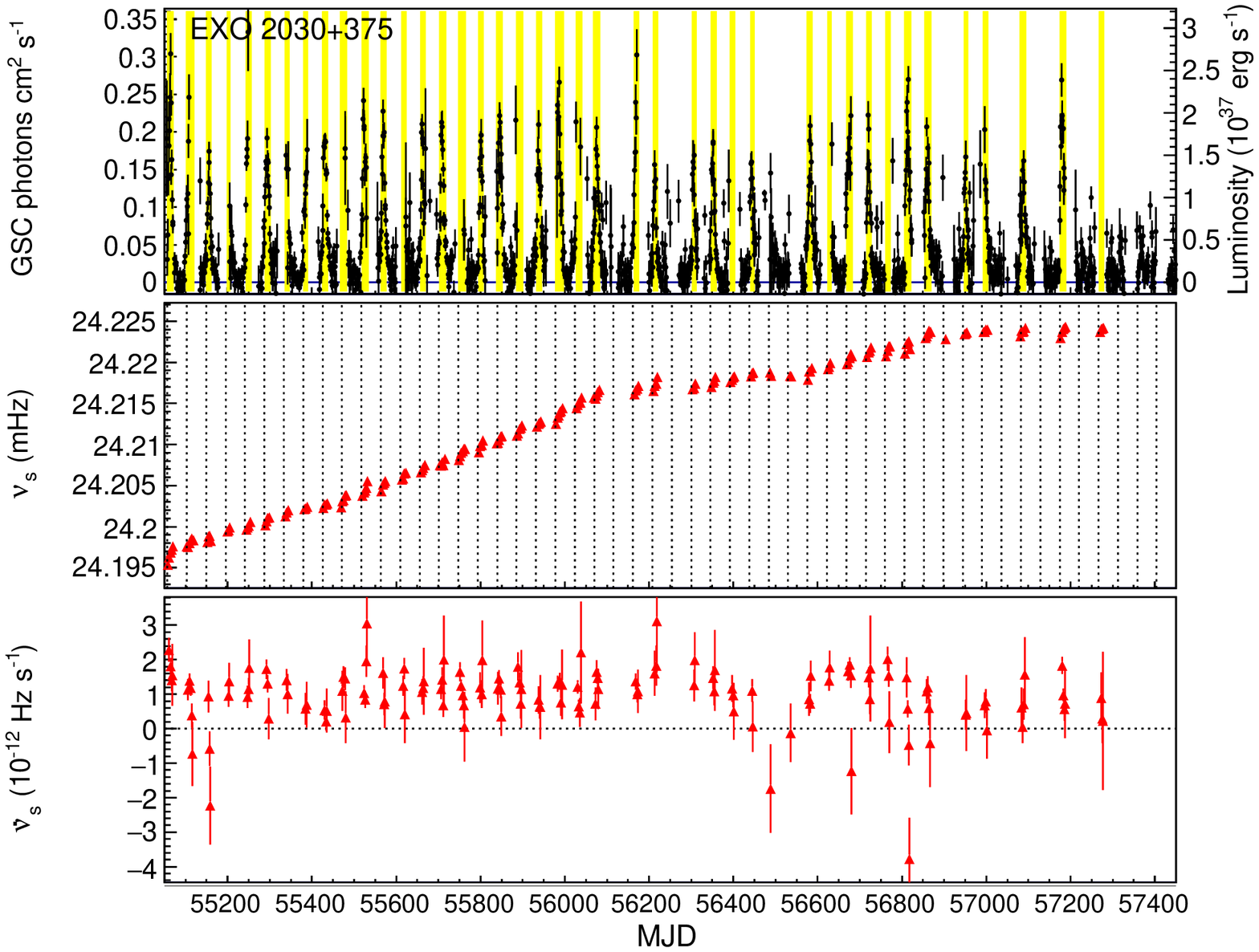}
\hspace{2mm}
\includegraphics[width=85mm]{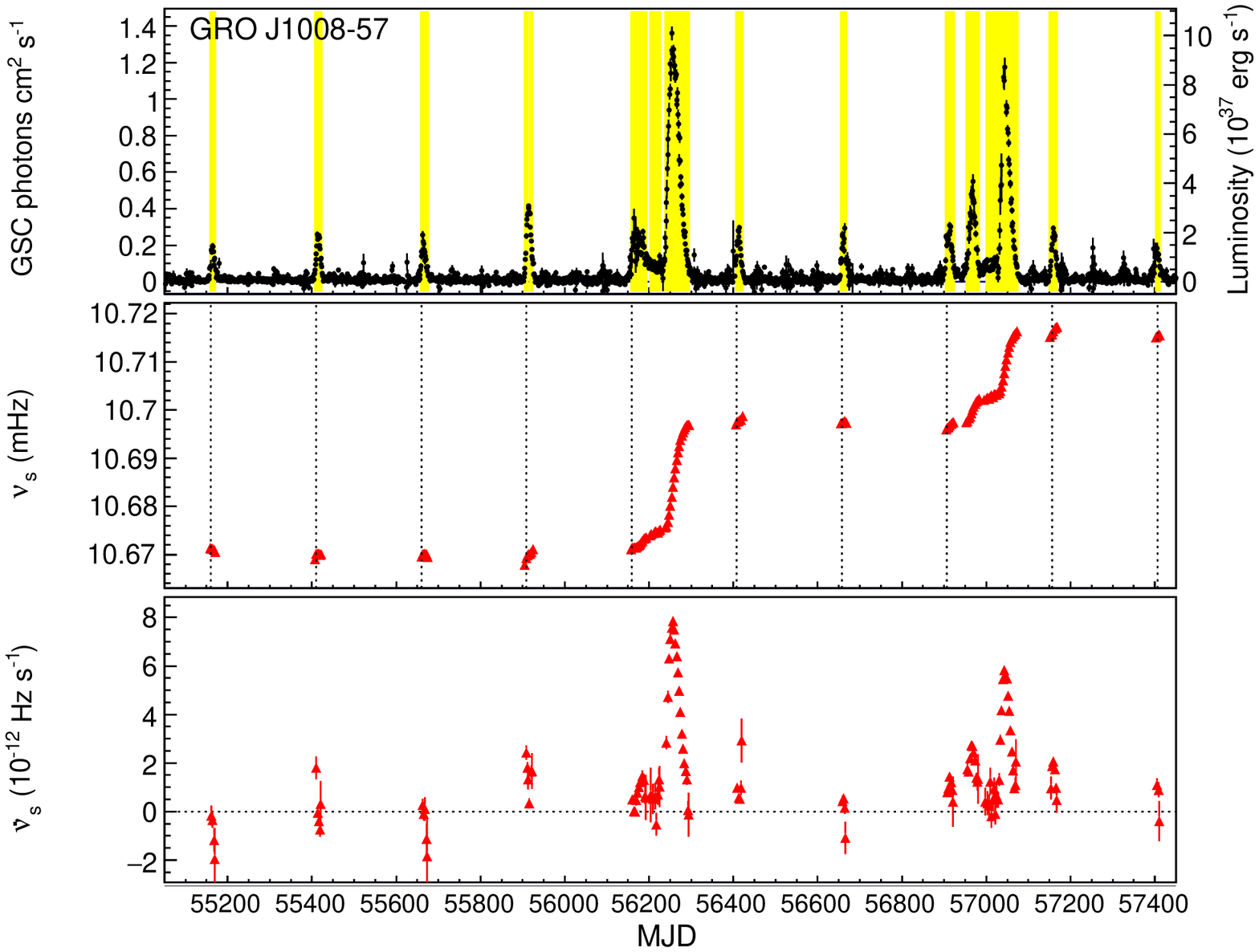}

\vspace{4mm}

\includegraphics[width=85mm]{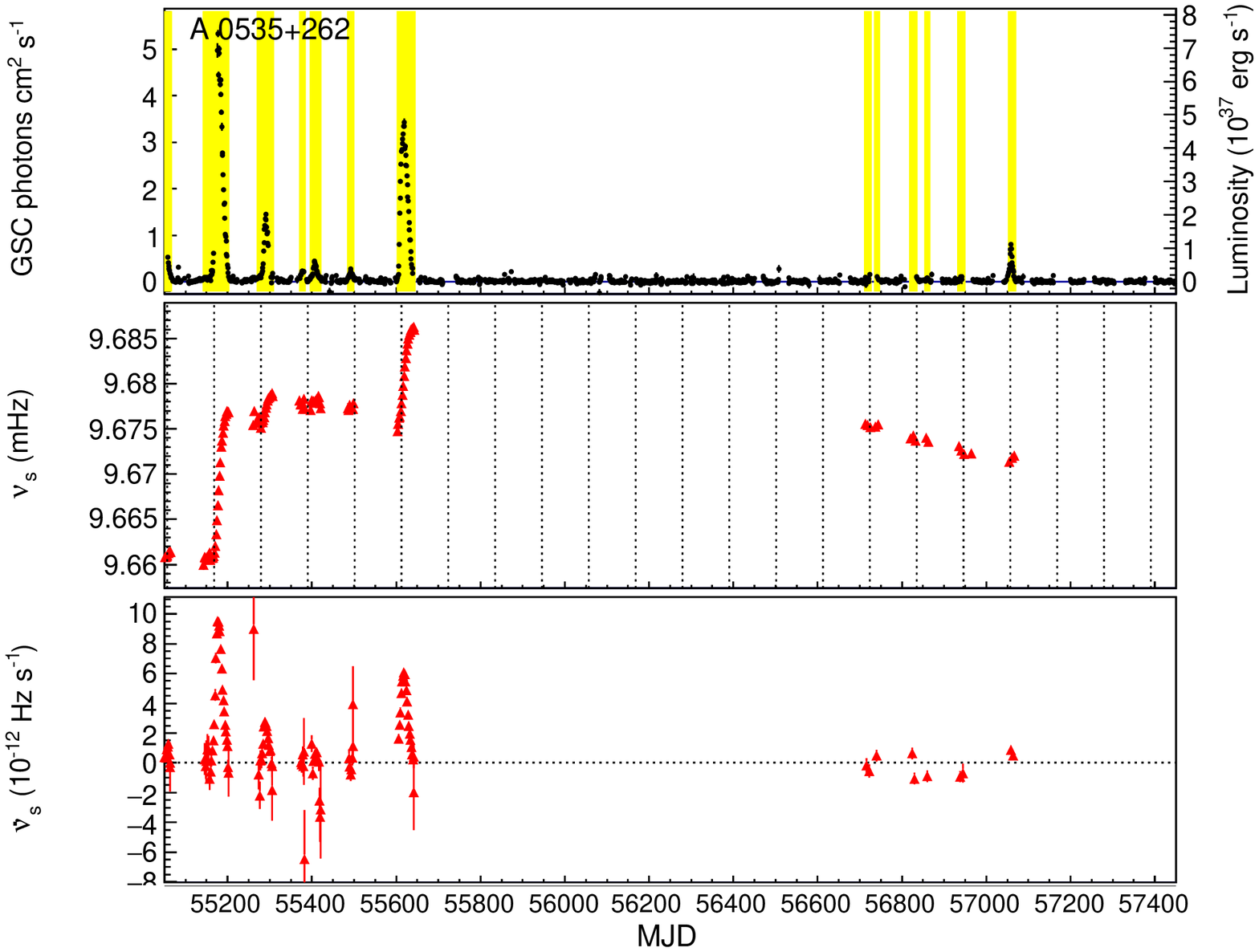}
\hspace{2mm}
\includegraphics[width=85mm]{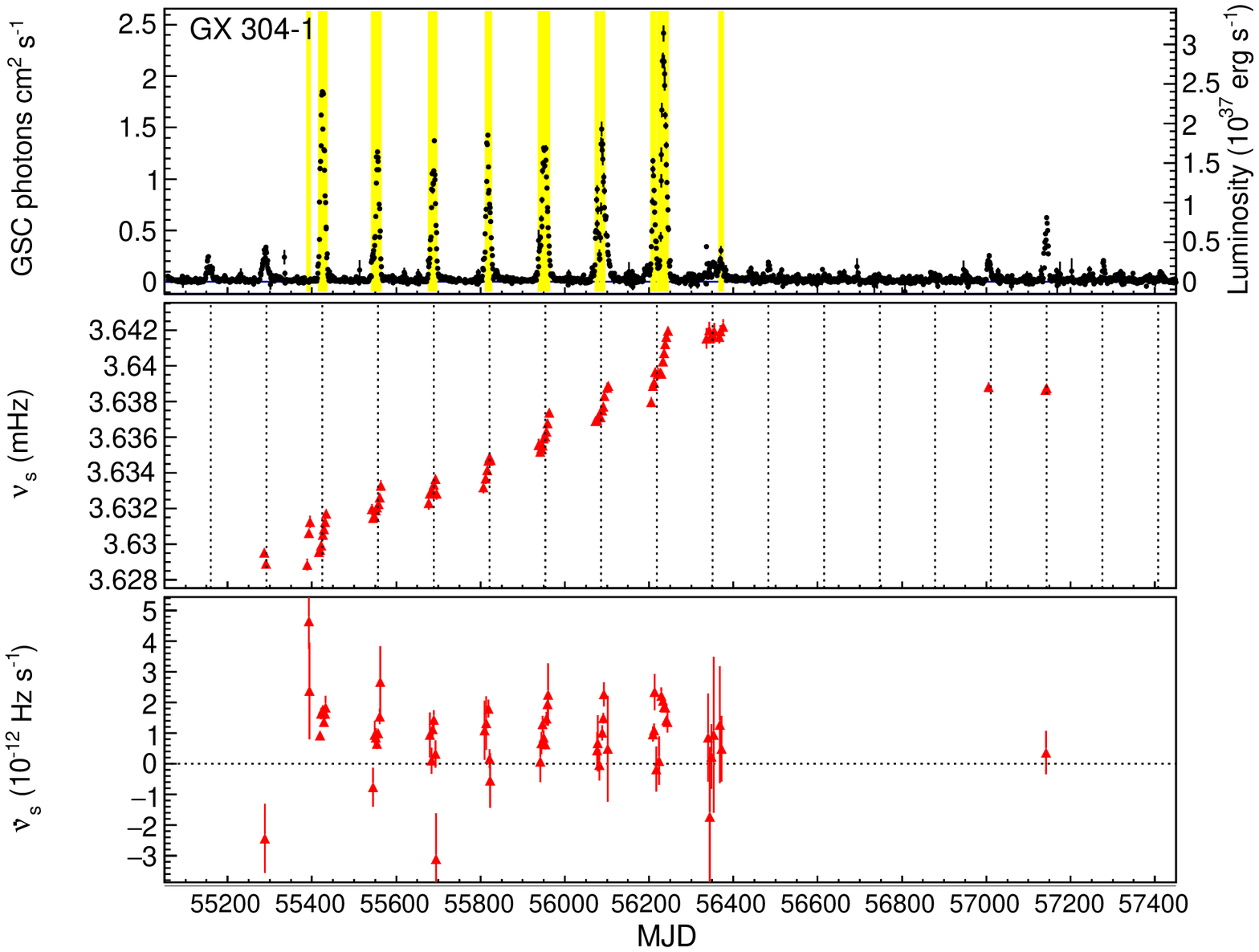}

\addtocounter{figure}{-1}                                         
\caption{(Continued)}
\end{figure*}

\subsection{Pulse-frequency derivative}
\label{sec:nudot}

Figure \ref{fig:lcper} clearly shows a common behavior that the spin
frequency increases, i.e. the pulsar spins up, during each outburst
activity.  We then calculated the pulse-frequency derivative,
$\dot{\nu}_\mathrm{s}$,
in the following way.
In the publicly available GBM pulsar data, 
the pulse periods of various XBPs are determined
typically every 2-d interval.  The obtained
pulse periods are subject to uncertainties which are mainly due
to the limitations in the statistics and the time intervals.
Considering these effects, we determined $\dot{\nu}_\mathrm{s}$
every 6-d interval by fitting several period measurements
in that interval with a linear function,
and then estimate the 1-$\sigma$ statistical error,
$\sigma_{\nu\mathrm{dot}}$, with the $\chi^2$ method.

The obtained values of $\dot{\nu}_\mathrm{s}$ are also plotted 
in figure \ref{fig:lcper} at the bottom panels.
As expected, the time variations of $\dot{\nu}_\mathrm{s}$ clearly
resemble those of the X-ray intensity at the top panels.

\subsection{X-ray spectrum and bolometric luminosity estimate}
\label{sec:anaspec}

For the present study, we need to estimate the instantaneous source
luminosity from the GSC light curve data.  
The factor of conversion from the observed 
count rate to the source luminosity depends on the
emission energy spectrum as well as the instrument response function.
Hence, we analyzed the GSC energy spectra of the 12 sources, 
assuming that
the energy spectrum of each source does not change significantly over
the outburst active periods.
Thus, for each source, 
the spectrum was averaged over all outburst periods, 
which is defined as the periods wherein the GBM data are available.

\begin{figure*}

\includegraphics[width=55mm]{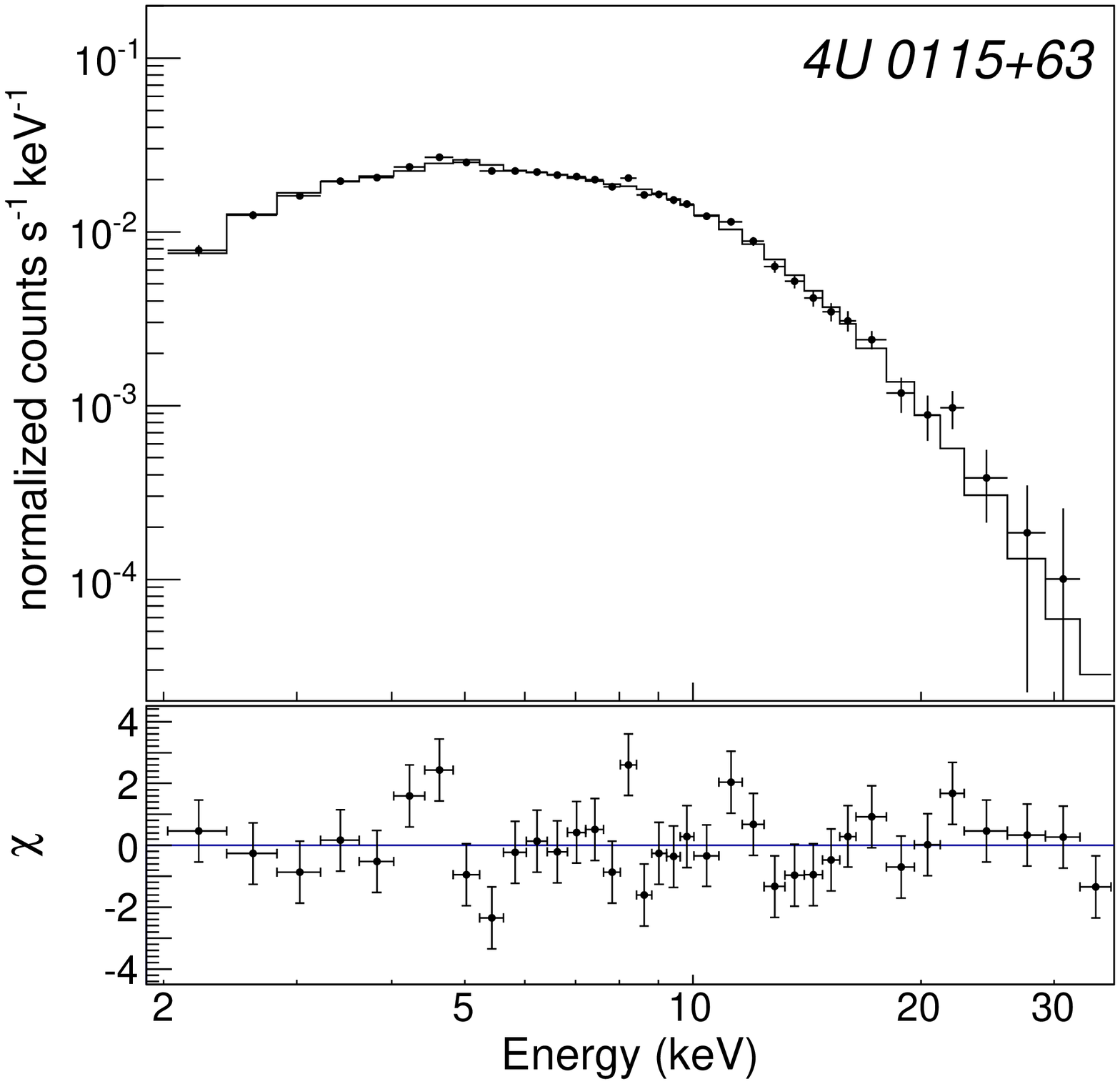}
\includegraphics[width=55mm]{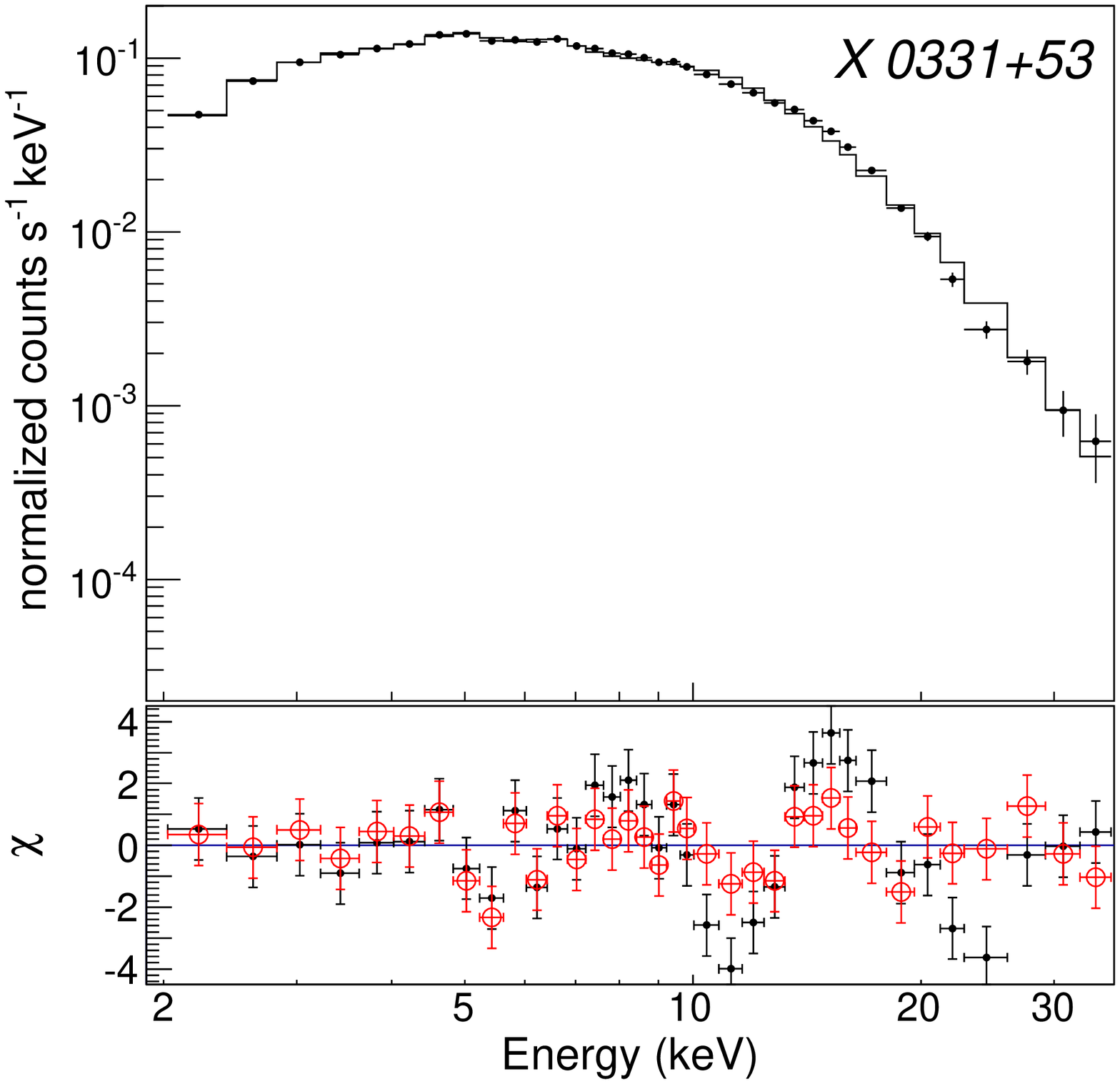}
\includegraphics[width=55mm]{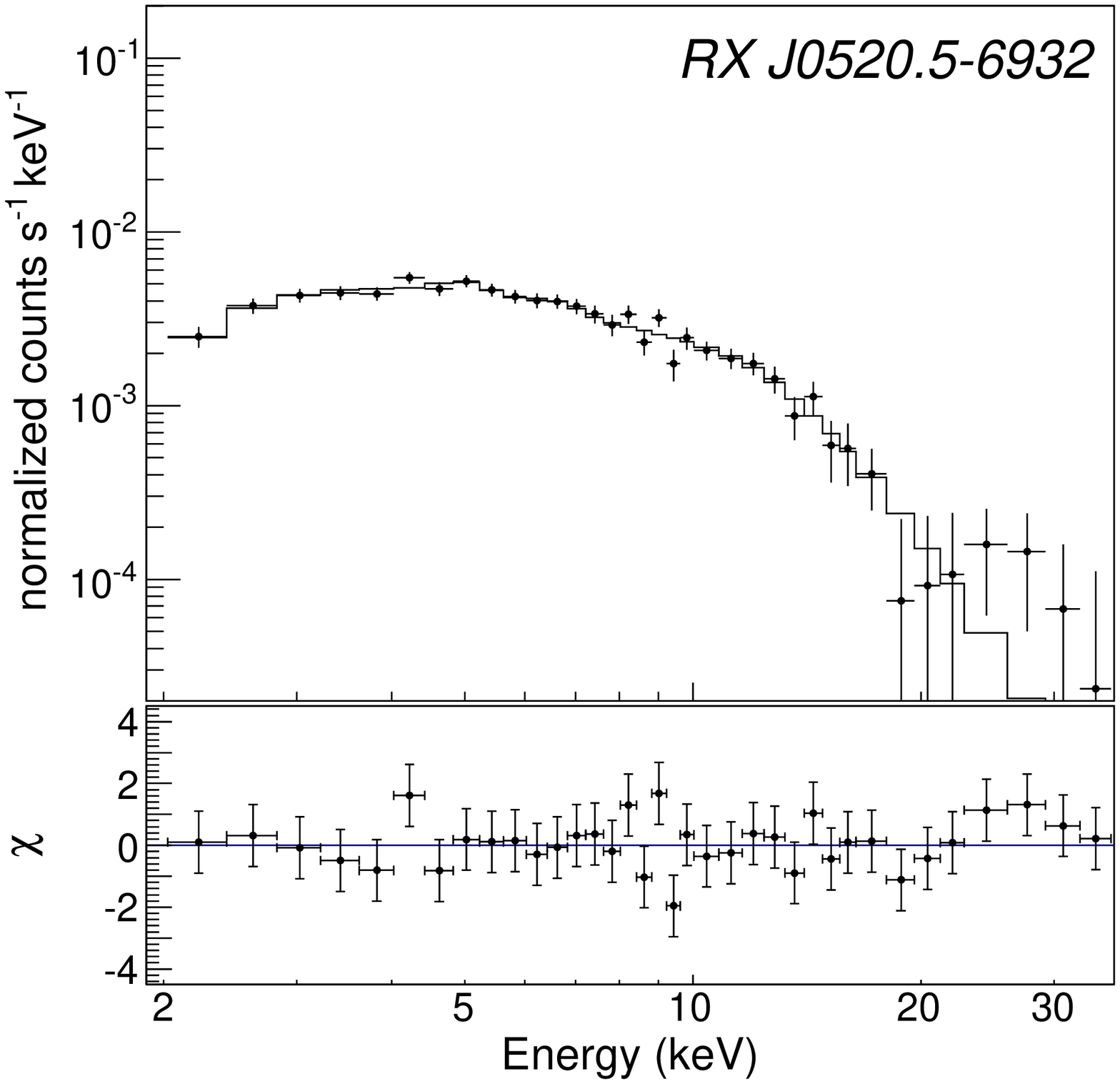}

\vspace{3mm}
\includegraphics[width=55mm]{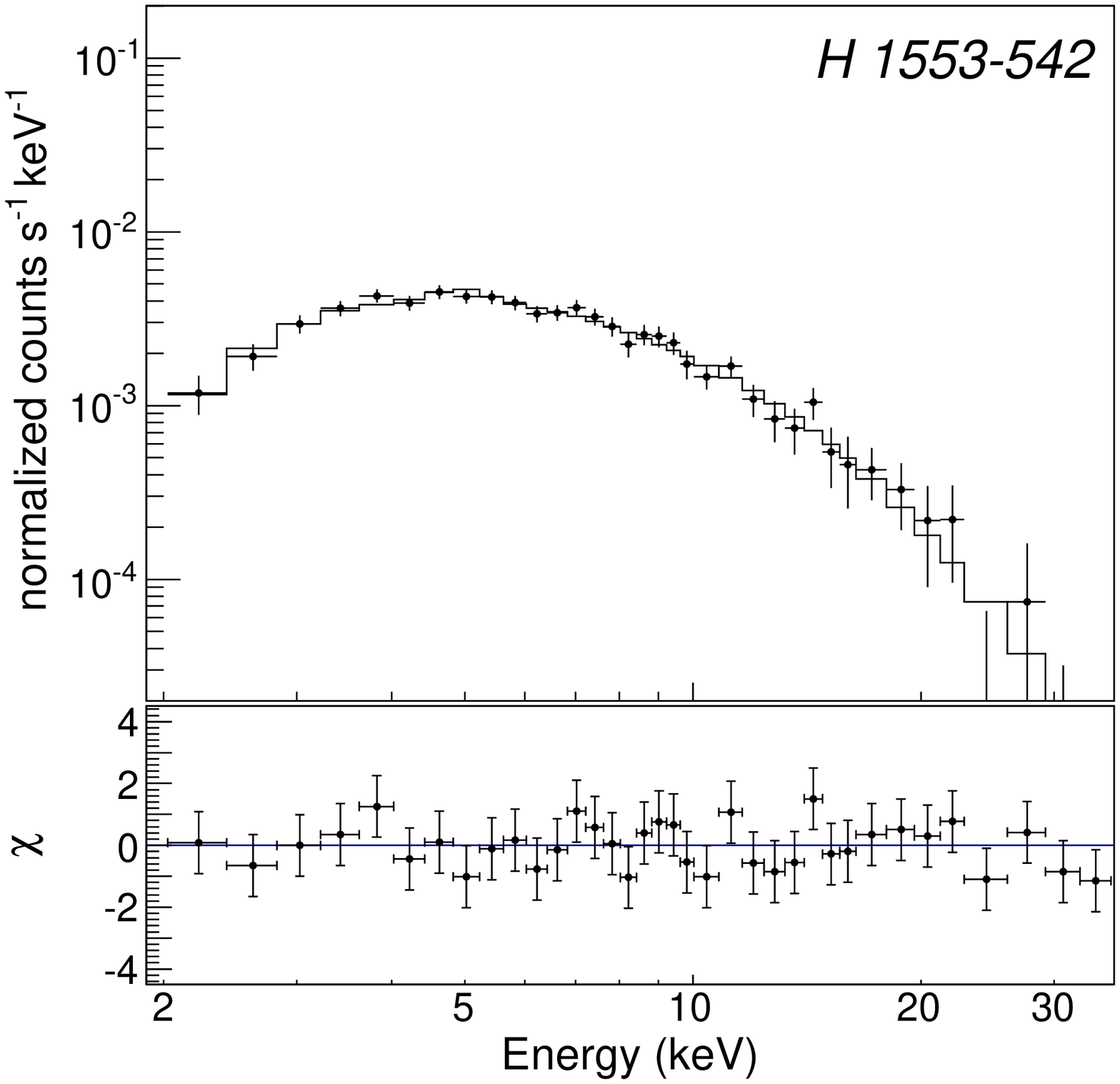}
\includegraphics[width=55mm]{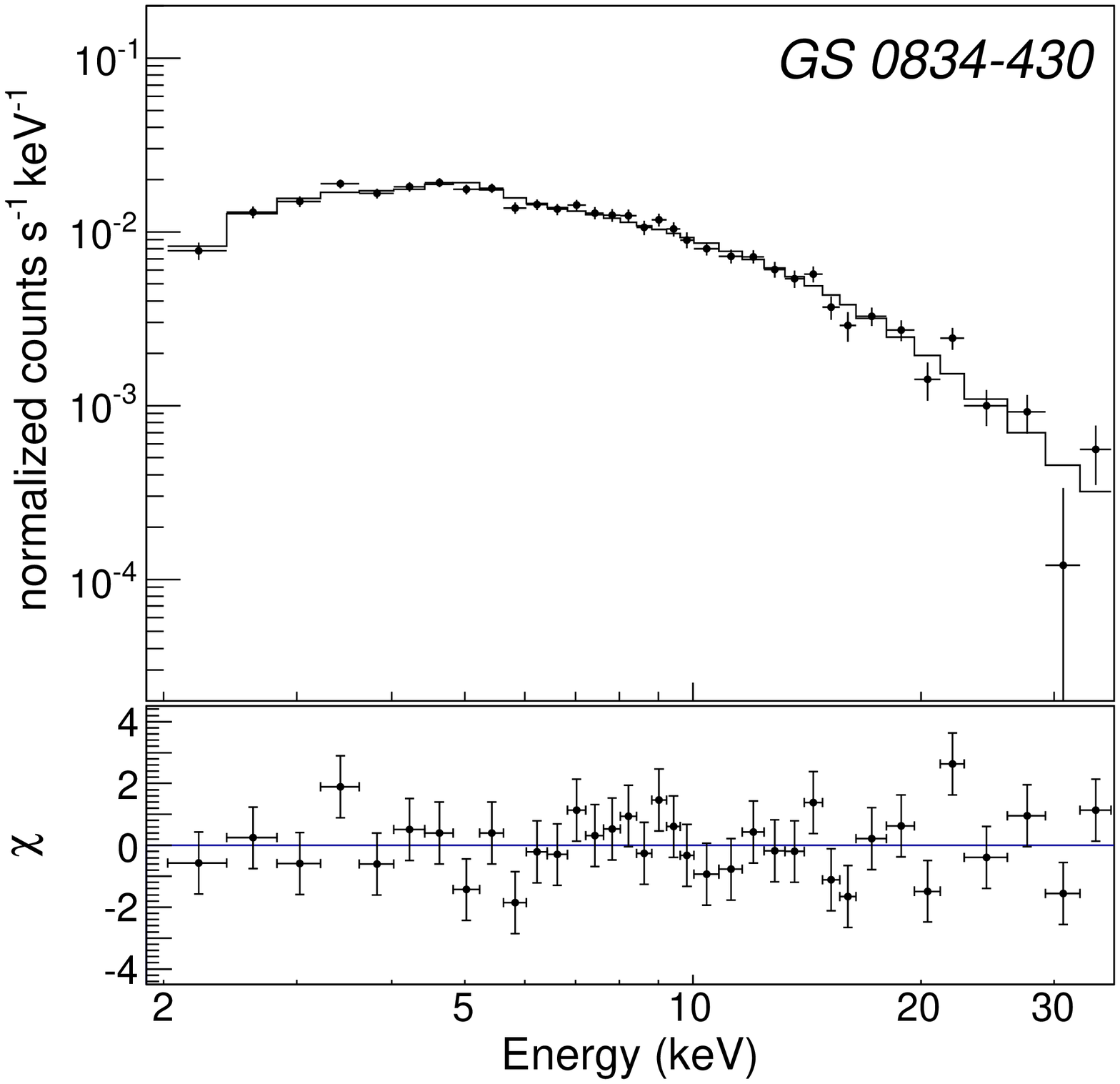}
\includegraphics[width=55mm]{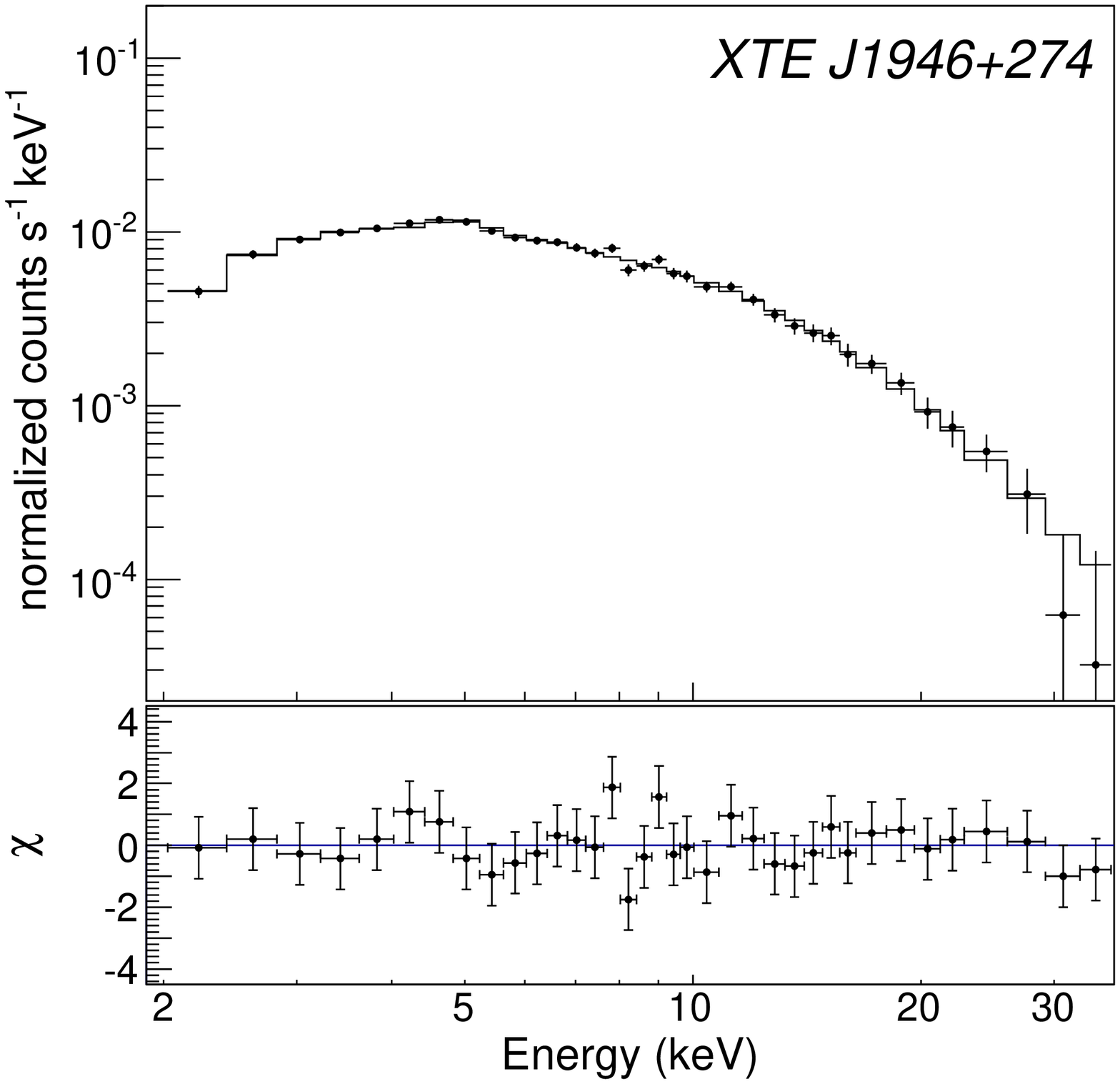}

\vspace{3mm}
\includegraphics[width=55mm]{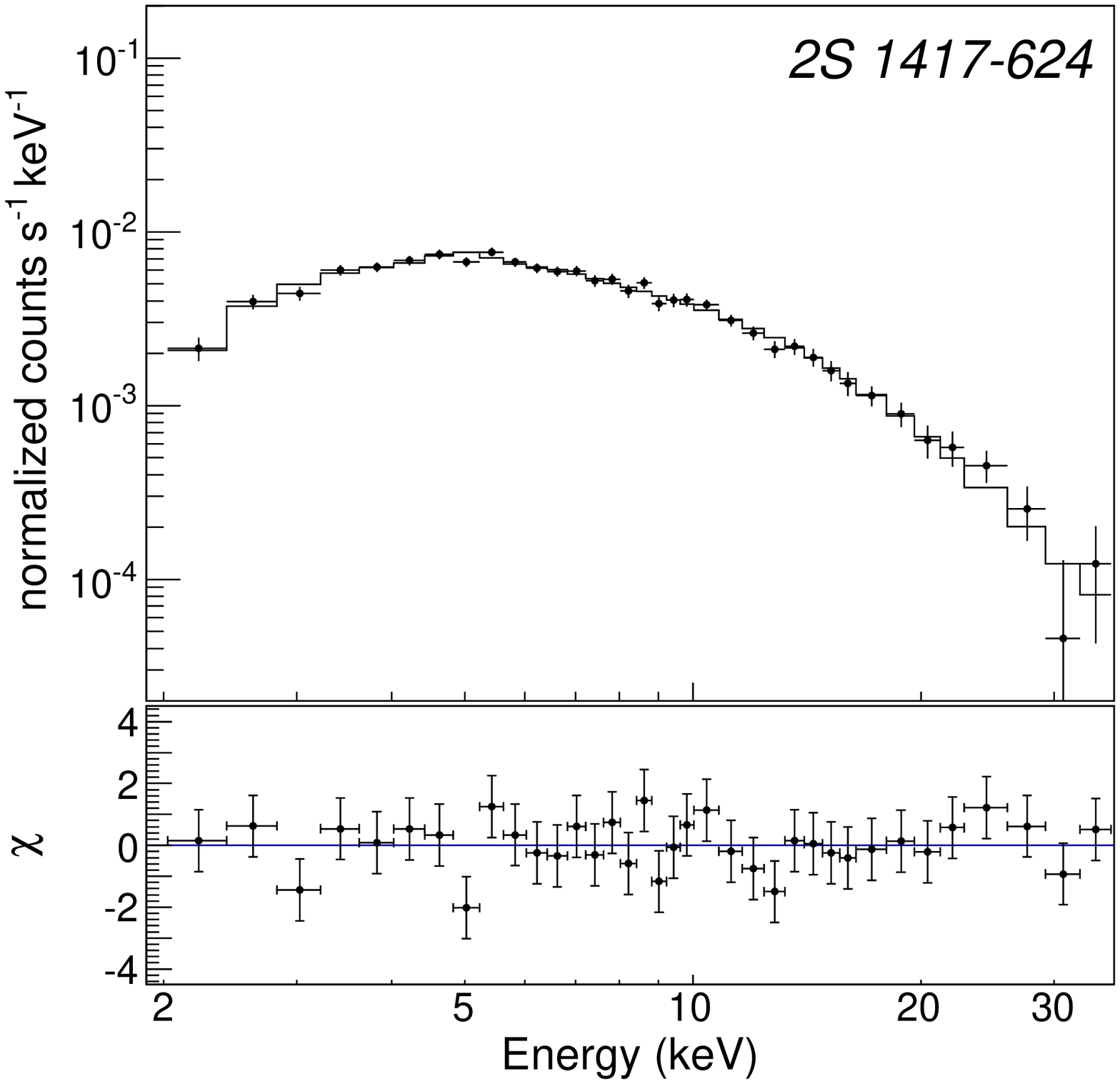}
\includegraphics[width=55mm]{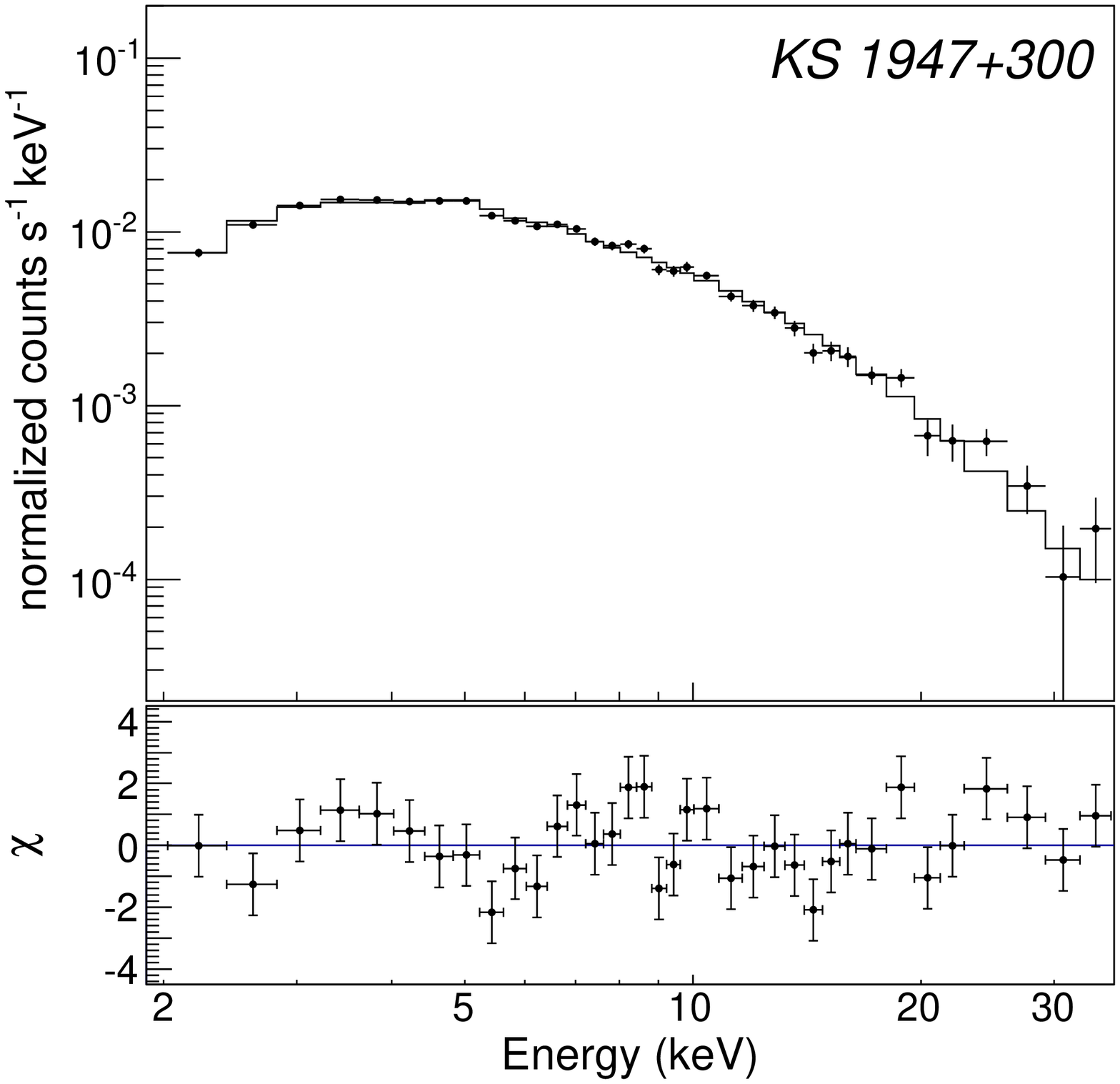}
\includegraphics[width=55mm]{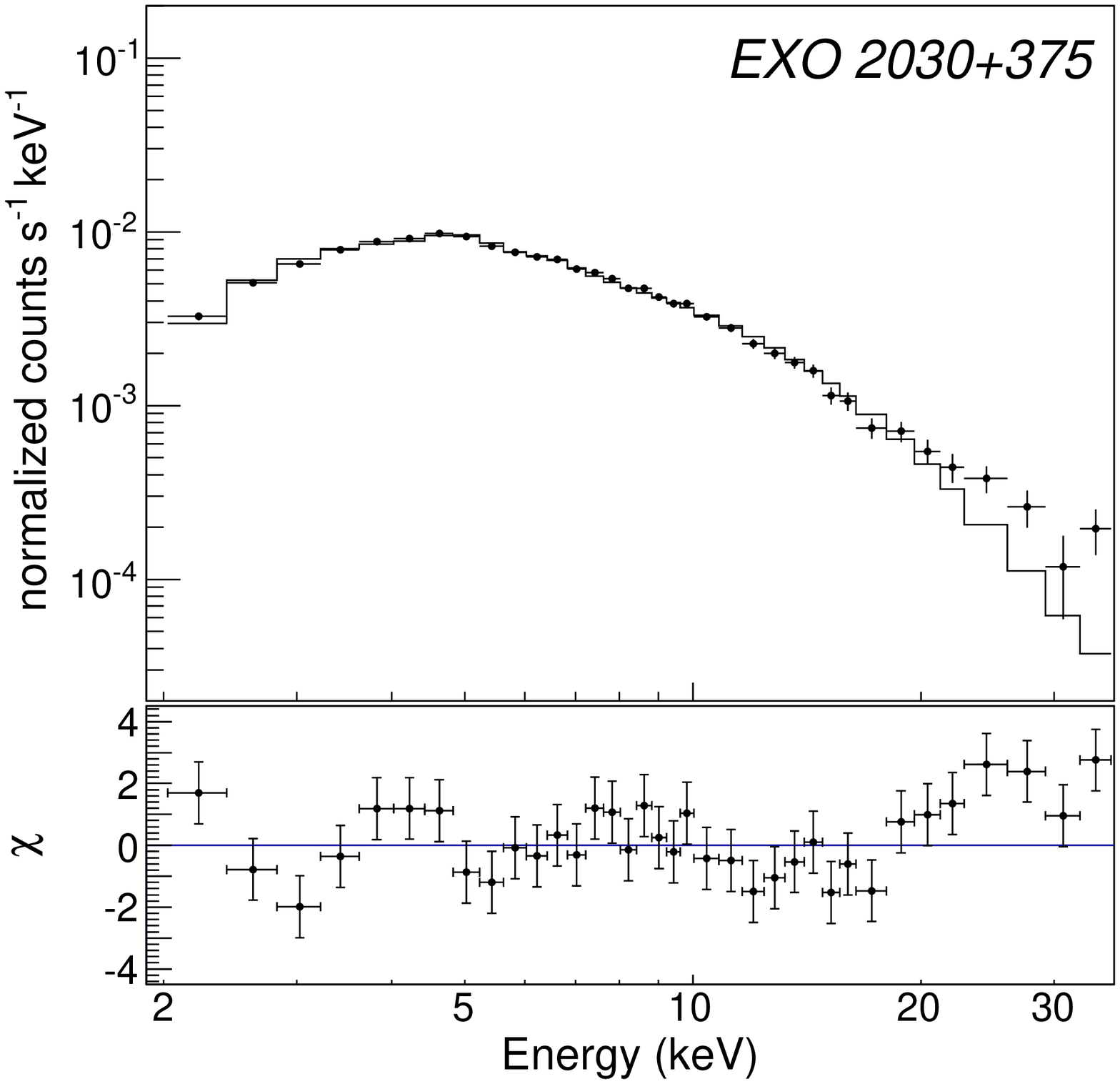}

\vspace{3mm}
\includegraphics[width=55mm]{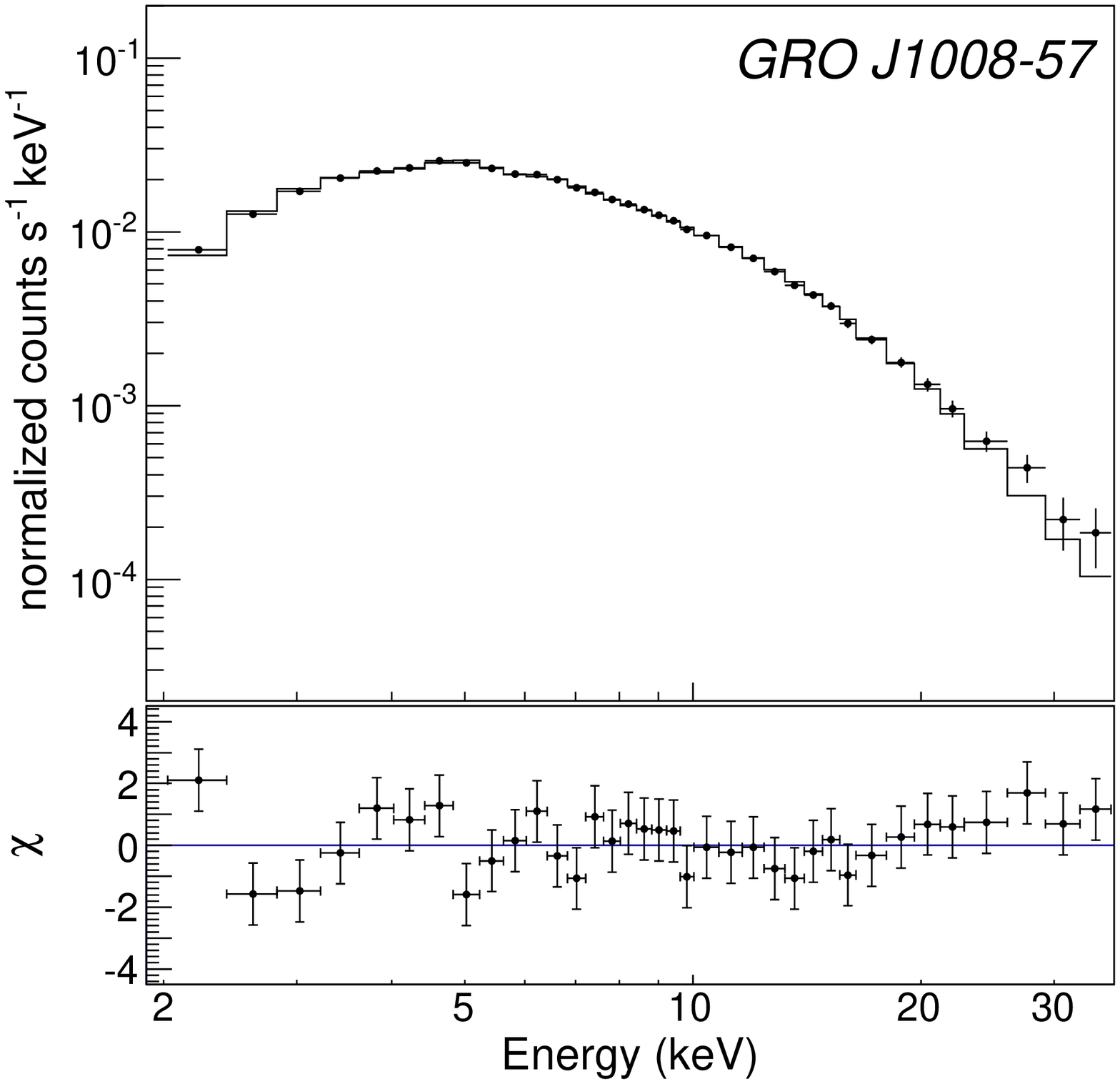}
\includegraphics[width=55mm]{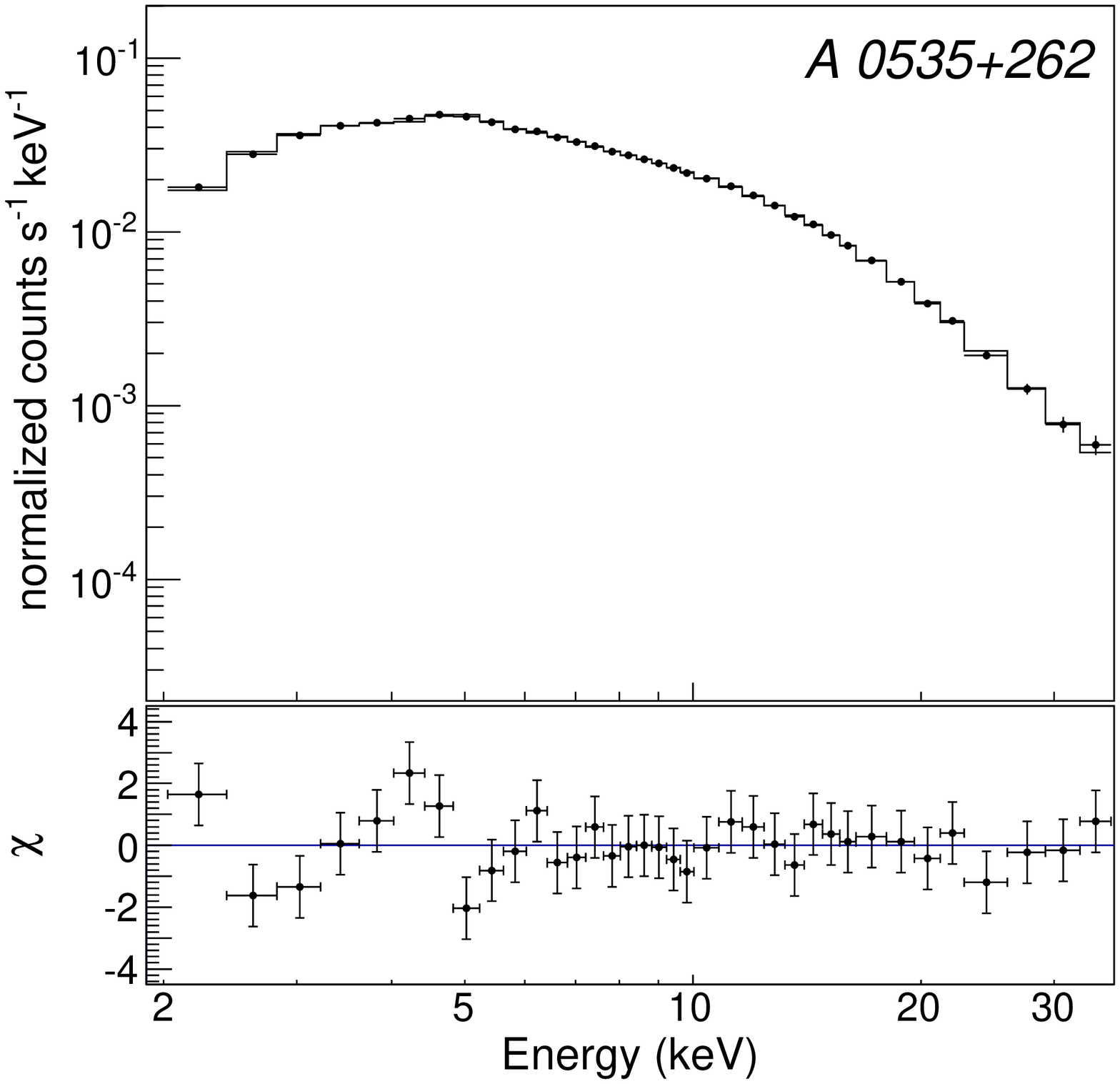}
\includegraphics[width=55mm]{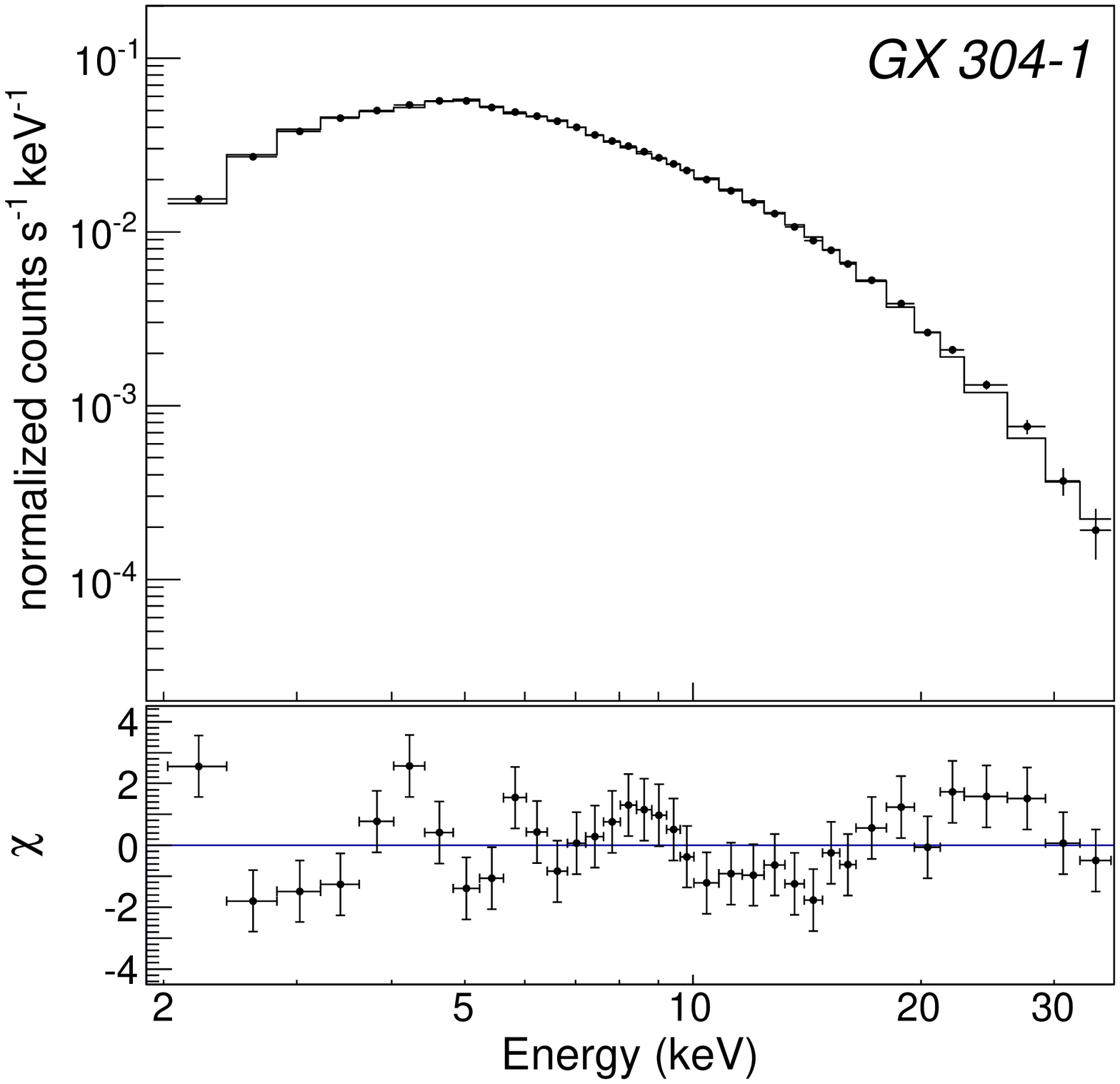}

\vspace{3mm}
\caption{
(Top panels)
GSC 2-30 keV spectra of the selected 12 Be XBPs, 
averaged over the entire outburst period,
compared with the best-fit PLCUT model 
folded with the detector response (solid line).
Crosses represent statistical 1-$\sigma$ errors.
(Bottom panles) 
Date versus model residuals. 
In X 0331$+$53, the red points represent those from 
the PLCUT * CYAB model.
}
\label{fig:spectra}

\end{figure*}

Figure \ref{fig:spectra} shows the 2--30 keV spectra
of the 12 sources, thus obtained with the GSC.
The background has been subtracted, but the instrumental
responses are still inclusive.
We fitted them with a typical model for XBPs, 
consisting of a high-energy-cutoff power-law (PLCUT) continuum,
and a Gaussian for iron-K line emission at 6.4 keV
(e.g. \cite{Makishima1999, 2002ApJ...580..394C}).  
The former is specified by the photon index $\Gamma$, the cutoff energy
$E_\mathrm{cut}$.  the folded energy $E_\mathrm{fold}$, and the normalization
$A$ as
\begin{equation} 
  F_\mathrm{PLCUT}(E) = \left\{
  \begin{array}{ll}
    A E^{-\Gamma} & (E\leq E_\mathrm{cut}) \\
    A E^{-\Gamma}\exp\left(-\frac{E-E_\mathrm{cut}}{E_\mathrm{fold}} \right) & (E_\mathrm{cut}<E).
  \end{array} \right.
\end{equation}
Because of the limited energy resolution of the GSC, 
the centroid and width of the Gaussian
were fixed at their typical values, 6.4 keV and 0.1 keV, respectively.
To account for the interstellar absorption, a photoelectric
absorption factor by a medium with Solar abundances and a free
equivalent-hydrogen column density
$N_\mathrm{H}$ was multiplied.  The overall model is thus expressed as {\tt
  phabs*(gaussian + highecut*powerlaw)} in the XSPEC terminology.

The PLCUT model was accepted,
within 90 \% confidence limits of statistic uncertainty, 
by the GSC spectra of the 11 objects except for X 0331$+$53.
In figure \ref{fig:spectra},
the best-fit model folded with the instrument response 
is shown together with the data, and 
the data versus model residuals are presented at the bottom panels.
Table \ref{tab:specparam} summarizes
the best-fit model parameters.

The residuals of X 0331$+$53
bear an absorption feature at around 25 keV,
which made the fit unacceptable with the reduced chi-squared of $\chi^2_\nu=3.6$ for
31 DOF (degree of freedom).
This must be the
fundamental CRSF detected in past outbursts
(e.g. \cite{1990PASJ...42..295M, 2010ApJ...710.1755N}).  We then multiplied a cyclotron
absorption (CYAB) model ({\tt cyclabs} in XSPEC
terminology; \cite{1990PASJ...42..295M}) to the PLCUT model,
to find that the fit becomes acceptable within the 90\% confidence limit.
In figure \ref{fig:spectra},
the residuals from the fit with the PLCUT$*$CYAB model are shown together.
The best-fit CYAB parameters (table \ref{tab:cyclabparam})
are consistent with those obtained in the past outbursts.

Using the best-fit models and the GSC response functions,
we calculated the factors of conversions $f_\mathrm{bol}$ 
from the 2--20 keV count rates
to the 0.1--100 keV fluxes (considered to approximate the bolometric flux)
corrected for the interstellar absorption.
The obtained values are presented in table \ref{tab:specparam},
together with their 68\% confidence uncertainties
caused by the fitting errors.
Further denoting the beaming factor (the observed flux divided by the spherically averaged
flux) as $f_\mathrm{b}$
and the source distance as $D$,
the 0.1--100 keV luminosity $L_\mathrm{obs}$ is calculated from
the observed 2--20 keV count rate $C_\mathrm{2-20}$ as
\begin{equation}
L=4\pi D^2 f_\mathrm{b} f_\mathrm{bol}  C_\mathrm{2-20} .
\label{equ:lumiobs}
\end{equation}
In figure \ref{fig:lcper} (top),
the ordinate on the right-hand side represents 
the luminosity scale obtained by assuming $D$
from the optical companion (table \ref{tab:beparam}),
$f_\mathrm{b}=1$ (isotropic emission), 
and the value of $f_\mathrm{bol}$ as obtained above.
The validity of these assumptions is evaluated
in section \ref{sec:discussion}.
To make coincident samplings of $L$ and $\nu_\mathrm{s}$,
the $L$ calculation via equation (\ref{equ:lumiobs})
was performed over the same 6-d intervals
as for the determination of $\nu_\mathrm{s}$.

\begin{table*}
\footnotesize
\caption{
Summary of the best-fit spectral parameters$^*$.
}
\label{tab:specparam}
\begin{center}
\begin{tabular}{rlccccccccc}
\hline
\hline
No. & Source ID   & $N_{\rm H}$  & $\Gamma$ & $E_{\rm cut}$ & $E_{\rm fold}$ & $EW_{\rm FeK}$$^\dagger$ & $\chi^2_\nu(\nu)$ & $F_{\rm 2-20}$$^\ddagger$ & $F_{\rm bol}$$^\S$ & $f_{\rm bol}$$^\parallel$\\
    &         & $10^{22}$ cm$^{-2}$ &         & (keV)      & (keV)     & (eV)   &   & \\
\hline
1 &    4U 0115 & $ 0.7^{+0.5}_{-0.5}$ & $ 0.5^{+0.1}_{-0.1}$ & $ 8.9^{+0.5}_{-0.5}$ & $ 5.7^{+0.6}_{-0.6}$ & $<112$ &        1.41(30) & $ 25.7^{+0.4}_{-0.4}$ & $ 36.9^{+1.9}_{-1.8}$ & $ 1.44^{+0.08}_{-0.07}$\\
2 & X 0331     & $  0.2^{+***}_{-***}$ & $  0.4^{+***}_{-***}$ & $ 10.9^{+***}_{-***}$ & $  7.2^{+***}_{-***}$ & $<0$ &    3.73(30) & $  163^{+***}_{-***}$ & $  283^{+***}_{-***}$ & $ 1.74^{+***}_{-***}$\\
  & X 0331$^\#$ & $<0.2$ & $ 0.3^{+0.0}_{-0.0}$ & $ 8.6^{+0.7}_{-0.9}$ & $19.7^{+18.5}_{-6.3}$ & $  134^{+41}_{-43}$ &          1.09(27) & $  161^{+2}_{-2}$ & $  652^{+457}_{-199}$ & $ 4.04^{+2.84}_{-1.23}$\\
3 & RX J0520.5 & $<0.9$ & $ 1.0^{+0.2}_{-0.1}$ & $11.4^{+2.1}_{-3.1}$ & $ 6.0^{+6.5}_{-3.5}$ & $  170^{+163}_{-167}$ &        0.72(30) & $  5.3^{+0.2}_{-0.2}$ & $  7.0^{+1.9}_{-0.9}$ & $ 1.32^{+0.35}_{-0.18}$\\
4 &     H 1553 & $ 1.9^{+1.8}_{-1.7}$ & $<0.4$ & $ 7.5^{+2.5}_{-1.7}$ & $ 9.5^{+7.5}_{-3.0}$ & $<236$ &       0.62(30) & $  4.2^{+0.2}_{-0.2}$ & $  7.0^{+1.8}_{-1.2}$ & $ 1.66^{+0.43}_{-0.28}$\\
5 &    GS 0834 & $<0.7$ & $ 0.8^{+0.1}_{-0.3}$ & $<4.2$ & $  38^{+47}_{-14}$ & $<105$ &       1.27(30) & $ 20.8^{+0.5}_{-0.5}$ & $ 66.8^{+13.1}_{-10.9}$ & $ 3.21^{+0.63}_{-0.52}$\\
6 &  XTE J1946 & $<0.6$ & $ 0.8^{+0.1}_{-0.2}$ & $<2.7$ & $  22^{+11}_{-6}$ & $<137$ &        0.61(30) & $ 12.2^{+0.3}_{-0.3}$ & $ 30.3^{+5.0}_{-4.1}$ & $ 2.48^{+0.41}_{-0.34}$\\
7 &    2S 1417 & $ 1.6^{+1.2}_{-1.4}$ & $ 0.6^{+0.2}_{-0.6}$ & $ 7.5^{+2.3}_{-3.9}$ & $16.5^{+9.6}_{-5.7}$ & $<186$ &        0.74(30) & $  7.8^{+0.2}_{-0.2}$ & $ 20.0^{+3.7}_{-3.1}$ & $ 2.55^{+0.48}_{-0.40}$\\
8 &    KS 1947 & $ 0.7^{+0.4}_{-0.4}$ & $ 1.2^{+0.1}_{-0.1}$ & $ 8.3^{+1.7}_{-1.3}$ & $  27^{+18}_{-8}$ & $  106^{+78}_{-80}$ &       1.42(30) & $ 15.1^{+0.3}_{-0.3}$ & $ 31.3^{+4.3}_{-3.5}$ & $ 2.07^{+0.28}_{-0.23}$\\
9 &   EXO 2030 & $ 0.9^{+0.7}_{-0.7}$ & $ 0.6^{+0.2}_{-0.3}$ & $ 4.1^{+0.6}_{-0.5}$ & $10.7^{+4.4}_{-2.8}$ & $   99^{+55}_{-57}$ &    1.81(30) & $  8.7^{+0.1}_{-0.1}$ & $ 15.3^{+2.2}_{-1.7}$ & $ 1.75^{+0.25}_{-0.20}$\\
10&  GRO J1008 & $ 1.6^{+0.3}_{-0.4}$ & $ 1.0^{+0.1}_{-0.1}$ & $ 7.5^{+0.6}_{-0.8}$ & $13.1^{+1.6}_{-1.5}$ & $  114^{+46}_{-52}$ &    1.01(30) & $ 24.4^{+0.2}_{-0.2}$ & $ 44.9^{+2.0}_{-2.0}$ & $ 1.84^{+0.08}_{-0.08}$\\
11&     A 0535 & $ 1.0^{+0.2}_{-0.3}$ & $ 0.9^{+0.1}_{-0.1}$ & $ 8.2^{+0.9}_{-1.3}$ & $  24^{+3}_{-3}$ & $   53^{+34}_{-45}$ &       0.91(30) & $ 48.9^{+0.3}_{-0.3}$ & $  129^{+4}_{-4}$ & $ 2.63^{+0.09}_{-0.08}$\\
12&     GX 304 & $ 2.2^{+0.3}_{-0.3}$ & $ 1.0^{+0.1}_{-0.1}$ & $ 7.2^{+0.5}_{-0.5}$ & $13.3^{+1.2}_{-1.2}$ & $   96^{+37}_{-36}$ &    1.70(30) & $ 52.2^{+0.3}_{-0.3}$ & $ 98.1^{+3.0}_{-3.1}$ & $ 1.88^{+0.06}_{-0.06}$\\
\hline
\end{tabular}

\end{center}
$^*$All errors represent 90\% confidence limits of statistical uncertainty.\\
$^\dagger$Equivalent width of iron K (6.4 keV) line \\
$^\ddagger$Units in photons cm$^{-1}$ s$^{-1}$ \\
$^\S$Units in $10^{-11}$ erg cm$^{-1}$ s$^{-1}$\\
$^\parallel$Units in $10^{-8}$ erg counts$^{-1}$\\
$^\#$CYAB model is applied. The best-fit parameters of the CYAB model are in table \ref{tab:cyclabparam}.\\
\end{table*}

\begin{table*}
\small
\caption{
CYAB model parameters in X 0331$+$53
}
\label{tab:cyclabparam}
\begin{center}
\begin{tabular}{ccc}
\hline
\hline
$E_{\rm a}$ (keV) & $W$ (keV) & $D$ \\
\hline
$23.4^{+1.0}_{-0.8}$ & $6.2^{+2.5}_{-2.3}$ & $2.0^{+0.8}_{-0.5}$  \\
\hline
\end{tabular}
\end{center}
\end{table*}

\subsection{Relation between the luminosity and the spin-frequency derivative
compared with theoretical models
}
\label{sec:nulx}

We have so far derived $\dot{\nu}_\mathrm{s}$ and $L$ 
of the selected 12 Be XBPs, on almost daily
basis during the outbursts since 2009 August.
Figure \ref{fig:pdotlx} shows their relation,
called $\dot{\nu}_\mathrm{s}$-$L$ diagram, for each of the 12 sources.
All diagrams clearly reveal the expected positive correlations between
$\dot{\nu}_\mathrm{s}$ and $L$, indicating that the pulsars indeed spin
up by the accretion torque.
Furthermore, in a fair fraction of the 12 objects, the correlation is
close to a direct proportionality ($\dot{\nu}_\mathrm{s}\propto L$) in
their luminous phase.
%
%
The behavior largely agrees with the prediction of 
most of the disk-magnetosphere interaction models,
\begin{equation}
 \dot{\nu}_\mathrm{s} \propto L^\alpha,
\label{equ:nudotLpower}
\end{equation}
with $\alpha \simeq 0.85-1$.

\subsubsection{
Brief theoretical reviews
}
\label{sec:modelreview}

Before actually analyzing the $\dot{\nu}_\mathrm{s}$-$L$ relations
in figure \ref{fig:pdotlx}, 
let us briefly revisit the theoretical models.
When a rotating neutron star is spun up by 
mass accretion via a Keplerian disk,
$\dot{\nu}_\mathrm{s}$ is 
expressed as a function of the mass accretion rate $\dot{M}$ as
\begin{equation}
\dot{\nu}_\mathrm{s} = n \dot{M}\sqrt{GMr_0} \left( 2\pi I \right)^{-1} ,
\label{equ:ntorque}
\end{equation}
where $I$ is the moment of inertia, 
$r_0$ is the radius at which the disk terminates
due to the magnetic barrier,
and $n$ is a dimensionless parameter representing
the effect of torque integration over a disk region
that is 
threaded by the
pulsar's magnetic fields.
Although the two parameters, $r_0$ and $n$, depend on
the disk-magnetosphere interaction models,
most of them assume $r_{0}$ to be of the order of the 
Alfven radius $r_\mathrm{a}$.
By introducing a dimensionless parameter $\zeta\sim 1$,
it is hence written as
\begin{equation}
r_0 = \zeta r_\mathrm{a} =\zeta  \left(\frac{\mu^4}{2GM\dot{M}^2}\right)^{1/7},
\label{equ:r_a}
\end{equation}
where $\mu$ is the magnetic dipole moment.
Meanwhile, $n$ is usually given as
as a function of ``fastness parameter'' $\omega_\mathrm{s}$,
which is the ratio 
of the pulsar's angular frequency to that of the disk at $r_0$,
and is expressed as
\begin{equation}
\omega_\mathrm{s} = \frac{2\pi\nu_\mathrm{s}}{\sqrt{GMr_0^{-3}}} = \left(\frac{r_0}{r_\mathrm{c}}\right)^{3/2}
\label{equ:ws}
\end{equation}
where $r_\mathrm{c}$ is the corotation radius.
In the slow-rotator condition with $\omega_\mathrm{s}\ll 1$,
$n$ is expected to become almost constant at $\sim 1$.

The value of $\dot{M}$
can be estimated from the observed $L$
of equation (\ref{equ:lumiobs}).
Taking account of the gravitational redshift
on the neutron star surface,
$\dot{M}$ is related to $L$ as
\begin{eqnarray}
\label{equ:LMdot}
L &=& \dot{M}c^2\left(1-\sqrt{1-x^{-1}}\right) \\
&\simeq& \dot{M}c^2\left(\frac{1}{2}x^{-1}-\frac{1}{8}x^{-2}+\frac{1}{32}x^{-3}-...\right), \nonumber
\end{eqnarray}
where $x$ refers to equation (\ref{equ:rsch}).
The first term in the
Taylor expansion of equation (\ref{equ:LMdot}) 
corresponds to the non-relativistic limit.

Substituting equations (\ref{equ:r_a}) and (\ref{equ:LMdot}),
equation (\ref{equ:ntorque}) is reduced to
\begin{equation}
\dot{\nu}_{12} = 2.0\,
  n\zeta^{1/2}
  \mu_{30}^{2/7} \radns^{6/7} M_{1.4}^{-3/7}I_{45}^{-1} 
  L_{37}^{6/7} 
\label{equ:nudot2}
\end{equation}
where 
$\dot{\nu}_{12}$, $\mu_{30}$, $\radns$, $M_{1.4}$, $I_{45}$,
and
$L_{37}$
are 
given in units of
$10^{-12}$ Hz s$^{-1}$, 
$10^{30}$ G cm$^{3}$, 
$10^6$ cm,
1.4 $M_\odot$, 
$10^{45}$ g cm$^2$,
and
$10^{37}$ erg s$^{-1}$, 
respectively.
The factor of the relativistic effect in equation (\ref{equ:LMdot})
is taken into account by assuming $x\simeq 2.4$ from the canonical
values of $M_{1.4}=1$ and $\radns =1$.

\subsubsection{Power-law fit to the observed $\dot{\nu}_\mathrm{s}$-$L$ relation}
\label{sec:powerlawfit}

Although equation (\ref{equ:nudot2}) implies
$\dot{\nu}_\mathrm{s}\propto L^{6/7}$ 
if both $n$ and $\zeta$ are constant against $L$,
observational results obtained so far 
often suggest a larger power-law index 
$\alpha >0.9$ in equation (\ref{equ:nudotLpower})
(e.g. \cite{1997ApJS..113..367B}).
To examine the present data for this possibility,
we fitted the $\dot{\nu}$-$L$ relations in figure \ref{fig:pdotlx}
with a power-law fuction,
\begin{equation}
\dot{\nu}_{12}=k L_{37}^\alpha,
\end{equation}
by floating both $\alpha$ and the coefficient $k$.
We here limited the fit to the data
in luminous phases with $L>1\times 10^{37}$ erg s$^{-1}$
where the correlation between $\dot{\nu}_\mathrm{s}$ and $L$ 
is significant against the measurement errors.
As for GRO J1008$-$57, the data were further limited to
$L>2\times 10^{37}$ erg s$^{-1}$,
because its $\dot{\nu}_\mathrm{s}$ data show
a larger scatter for the errors
possibly because of
the insufficient orbital corrections.

In figure \ref{fig:pdotlx}, the best-fit power-law models
are shown in blue, and the best-fit parameters and $\chi^2_\nu$
are listed in table \ref{tab:glfitpar}.
Thus, the model can approximately reproduce the observed data
in all the sources,
but the $\chi^2_\nu$ values are
often too large to make the fit acceptable at 
90\% limits.
This is presumably attributed to additional systematic errors,
associated with individual measurements
of $L$ and $\dot{\nu}_\mathrm{s}$.
Because the GSC data sparsely sample each target
(i.e. for 30--50 s of the scan transit every 92 min of the ISS rotation), 
time variations on a time scale from $\sim$ 30 s to 92 min
are not properly reflected in $L$.
The fluctuation of the GSC background rate would also contribute to
the error, because the observed
GSC data are mostly dominated by charged-particle backgrounds
and their contributions are estimated by assuming that the rate
is constant during individual scan transit of a source.
In short-period XBPs, the $\dot{\nu}_\mathrm{s}$ measurements
could also be subject to 
any residual errors in the orbital Doppler corrections.

Considering the above situation, 
we assumed that the $L$ and $\dot{\nu}_\mathrm{s}$ measurements 
both have additional systematic errors, 
$\Delta L^\prime$ and $\Delta\dot{\nu}_\mathrm{s}^\prime$,
respectively,
which are proportional to
their nominal fitting errors
($\Delta L$ and $\Delta\dot{\nu}_\mathrm{s}$), as
\begin{equation}
\Delta L^\prime =  \xi\Delta L, ~~~~~ \Delta\dot{\nu}_\mathrm{s}^\prime =  \xi\Delta\dot{\nu}_\mathrm{s}, 
\label{equ:syserr}
\end{equation}
where the factor $\xi>1$ is specific to each object.
We repeated the model fits with these revised errors,
increasing $\xi$ until the fits became acceptable
within the 90 \% confidence limit.
This has allowed us to properly estimate
uncertainty of the model parameters.

Table \ref{tab:glfitpar} includes
the obtained $\xi$ when the fits became accepted,
and 1-$\sigma$ errors on $k$ and $\alpha$, thus estimated.
Except for X 0331$+$53, 
the fits became acceptable with $\xi\lesssim 2.5$,
meaning that the systematic errors are not much larger than the
statistical errors.
We revisit the result of X 0331$+$53 in section \ref{sec:glcorr}. 

In 10 out of the 12 sources,
$\alpha$ was estimated as $\gtsim 1.0$, 
which is higher than $6/7=0.86$ in equation (\ref{equ:nudot2}).
The other two sources,  EXO 2030$+$375 and GX 304$-$1, 
exhibit relatively small values of best-fit $\alpha$.
However,  their $\alpha$ values have large uncertainties
(table \ref{tab:glfitpar}),
because they varied over very limited ranges
in $L$, namely,  $ \lesssim 4\times 10^{37}$ erg s$^{-1}$.
The apparently poor $\dot{\nu}_\mathrm{s}$ vs. $L$ correlations of these sources 
are also due to their narrow $L$ swing rather than intrinsic,
because their error renormalization factor $\xi$
does not take particularly large values.
Thus, including these two cases,  
the error-weighted average of $\alpha$ among the 12 sources is
$\langle \alpha \rangle =1.03$.
The results are consistent with those previously reported.

\subsubsection{Comparison with the Ghosh \& Lamb model}
\label{sec:glcorr}

We next compared the observed relations
with the disk-magnetosphere interaction
model proposed by GL79.
Although its predction of $\alpha=6/7$ at $\omega_\mathrm{s}\ll 1$
is somewhat smaller than $\alpha\simeq 1.0$ derived in
section \ref{sec:powerlawfit},
and the employed physical assumptions are
often debated (e.g. \cite{1987A&A...183..257W}; LRB95),
we select the model as a
representative working tool,
because it has been often used in the previous works
(e.g. \cite{1997ApJS..113..367B}),
and also successfully applied to the spin-up/down transitions 
in 4U 1626$-$67 \citep{2016PASJ...68S..13T}.
We discuss other models in section \ref{sec:xbpmodel}.

In the GL79 model, $r_0$ 
is assumed to be 
\begin{equation}
r_{0}^\mathrm{GL} \simeq 0.52\, r_\mathrm{a} ~~~(\textrm{i.e.} ~\zeta = 0.52),
\label{equ:glr0}
\end{equation}
and $n(\omega_\mathrm{s})$ is approximately expressed by
\begin{equation}
n^\mathrm{GL}(\omega_\mathrm{s}) 
\simeq 1.39\,\frac{1-\omega_\mathrm{s}\left[4.03(1-\omega_\mathrm{s})^{0.173}-0.878\right]}{1-\omega_\mathrm{s}}.
\label{equ:glnws}
\end{equation}
Here and hereafter, the parameters specific to the GL79 model are 
given a superscript of $\mathrm{GL}$.
Substituting equation (\ref{equ:glr0}) 
into equations (\ref{equ:ntorque}) and (\ref{equ:ws}),
$\dot{\nu}_\mathrm{s}$ and $\omega_\mathrm{s}$ 
are reduced respectively to
\begin{equation}
  \dot{\nu}_{12}^\mathrm{GL} = 1.4 \, \mu_{30}^{2/7}
  n^\mathrm{GL}(\omega_\mathrm{s})
  \radns^{6/7} M_{1.4}^{-3/7}I_{45}^{-1} 
L_{37}^{6/7} 
\label{equ:glnudot}
\end{equation}
\begin{equation}
\omega_\mathrm{s}^\mathrm{GL} = 1.3 \, \mu_{30}^{6/7} M_{1.4}^{-2/7} \radns^{-3/7} P_\mathrm{s}^{-1} \, L_{37}^{-3/7} 
\label{equ:glws}
\end{equation}

In the equations above, 
$\mu_{30}$ can be estimated from
the surface magnetic field $B_\mathrm{12}$ 
measured by the CRSF.
Because the pulsar magnetosphere extends far from the neutron star surface,
the gravitational redshift between $\mu_{30}$ and $B_\mathrm{12}$ 
needs to be taken into account.
In a simple configuration that the magnetic dipole
axis is aligned to the rotation axis,
$\mu_{30}$ at the magnetosphere is expressed with 
$B_{12}$ and $\radns$ as
\begin{equation}
\mu_{30}~ = \frac{1}{2} B_{12} \radns^3 \Phi(x) 
\label{equ:mu30}
\end{equation}
where $\Phi(x)$ is a correction factor given as 
\begin{eqnarray}
\Phi(x) &=& \left[-3x^3\ln\left(1-x^{-1}\right)-3x^2\left(1+\frac{1}{2}x^{-1}\right)\right]^{-1}\nonumber \\
&\simeq&  \left[1+\frac{3}{4}x^{-1}+\frac{3}{5}x^{-2}+...\right]^{-1}\nonumber
\end{eqnarray}
\citep{1983ApJ...265.1036W}. 
For a typical neutron star with $x\simeq 2.4$, 
we find $\Phi(x)\simeq 0.68$.

In figure \ref{fig:pdotlx}, the dashed black curves show 
the GL79 model relations
calculated from equations
(\ref{equ:lumiobs}), (\ref{equ:LMdot}), (\ref{equ:glnudot}), and (\ref{equ:mu30}), 
employing the magnetic field $B_\mathrm{12}$
in table \ref{tab:bexbpobs}, 
and the canonical neutron-star parameters, 
$\radns=1$, $M_{1.4}=1$, and $I_{45}=1$.
As for GS 0834$-$430, 2S 1417$+$624, and EXO 2030$+$375 from which
CRSFs have not been detected, we assume 
$B_\mathrm{s} = 2.6\times 10^{12}$ G
from the average of the measured ones.
The value of $L$ at $\omega_\mathrm{s}^\mathrm{GL}=0.1$,
calculated with equation (\ref{equ:glws}),
is also indicated in each panel.
Thus, the model of equation (\ref{equ:glnudot}) generally explain the
slope of the $\dot{\nu}_\mathrm{s}$ versus $L$ distribution, but not
necessarily the absolute values of the $\dot{\nu}_\mathrm{s}$
measurements.
This means that the $\dot{\nu}_\mathrm{s}$-to-$L$ coefficient of the
GL79 model is not always consistent with the data.

\subsubsection{The Ghosh \& Lamb model with a correction factor $\eta$}
\label{sec:eta}

The discrepancy in
the $\dot{\nu}_\mathrm{s}$-to-$L$ coefficient
between the data and the GL79 model is
primarily attributable to errors on the assumed parameters,
$\mu_{30}$, $M$, $R$, $I$, $D$, $f_\mathrm{bol}$, and $f_\mathrm{b}$,
included in the model equations (\ref{equ:lumiobs}), (\ref{equ:LMdot}),
(\ref{equ:glnudot}), and (\ref{equ:mu30}).  Another origin may reside
in the assumption of the GL79 model, that the gravity working on the
accreting matter becomes counter-balanced by the pulsar magnetosphere
at the radius of $r_0 = 0.52 r_\mathrm{a}$, 
and there is a broad transition zone where the pulsar's magnetic field
lines penetrate the disk.

\begin{figure*}

\includegraphics[width=56mm]{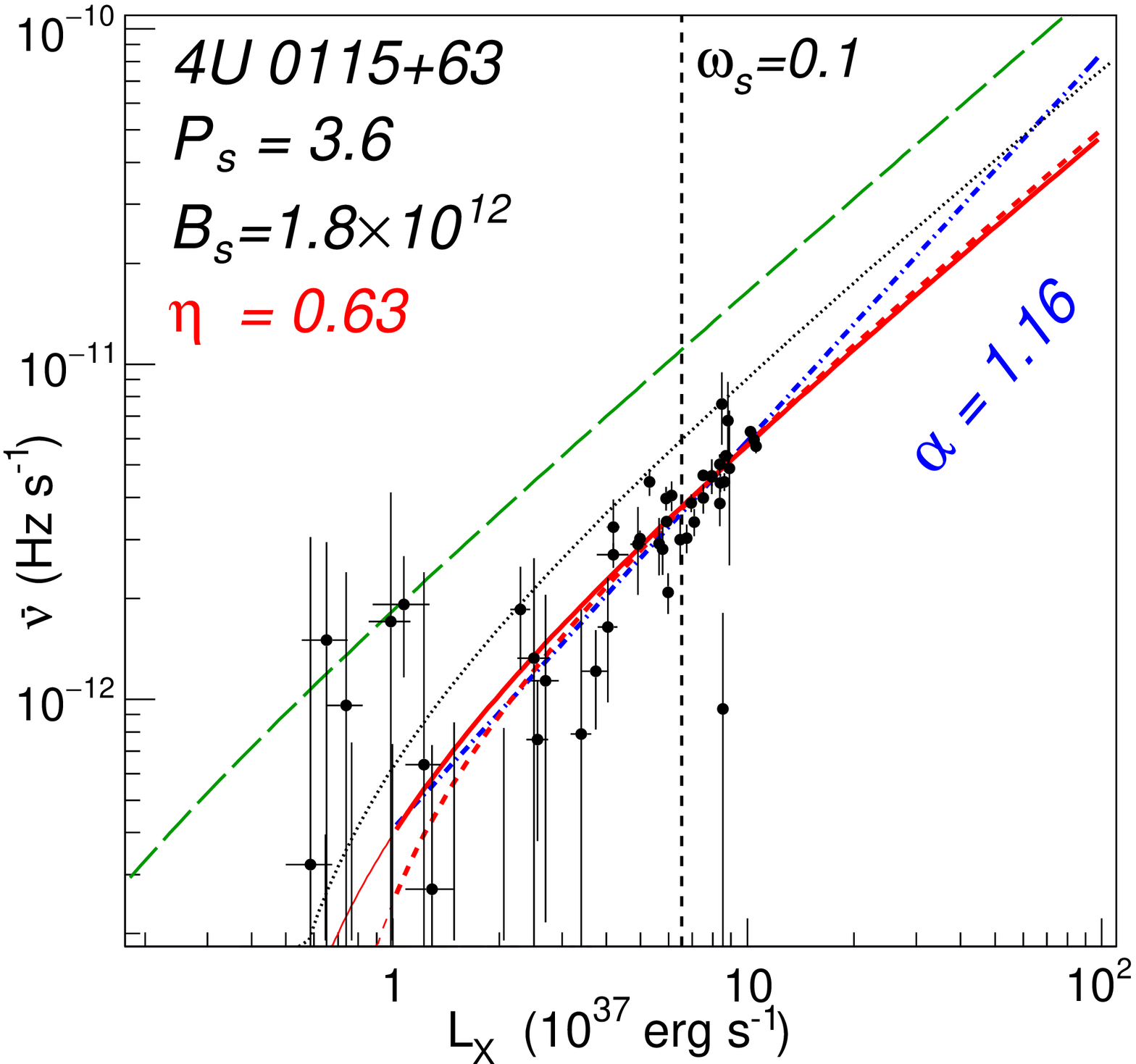}
\includegraphics[width=56mm]{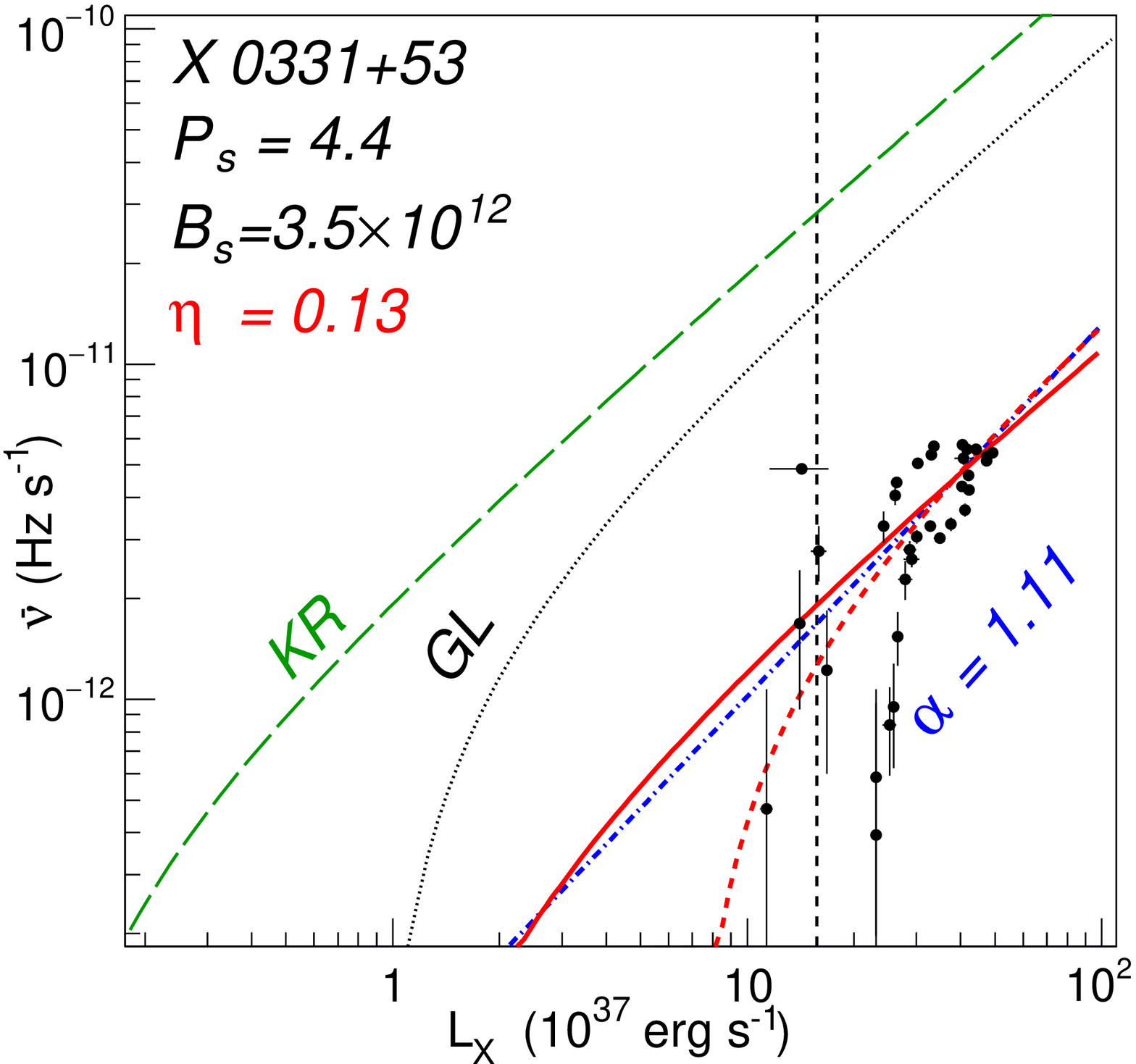}
\includegraphics[width=56mm]{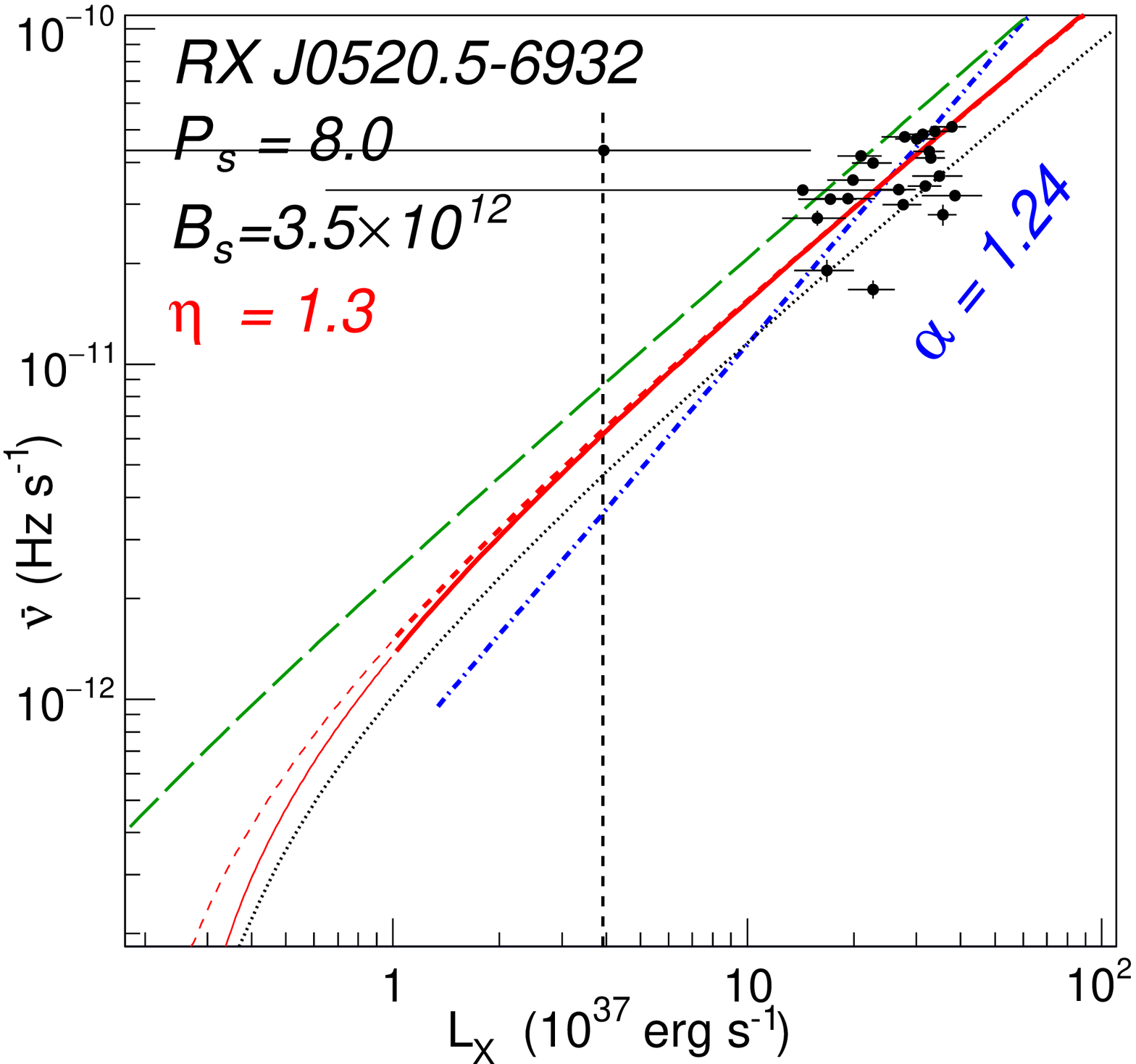}
                                                  
\vspace{3mm}
\includegraphics[width=56mm]{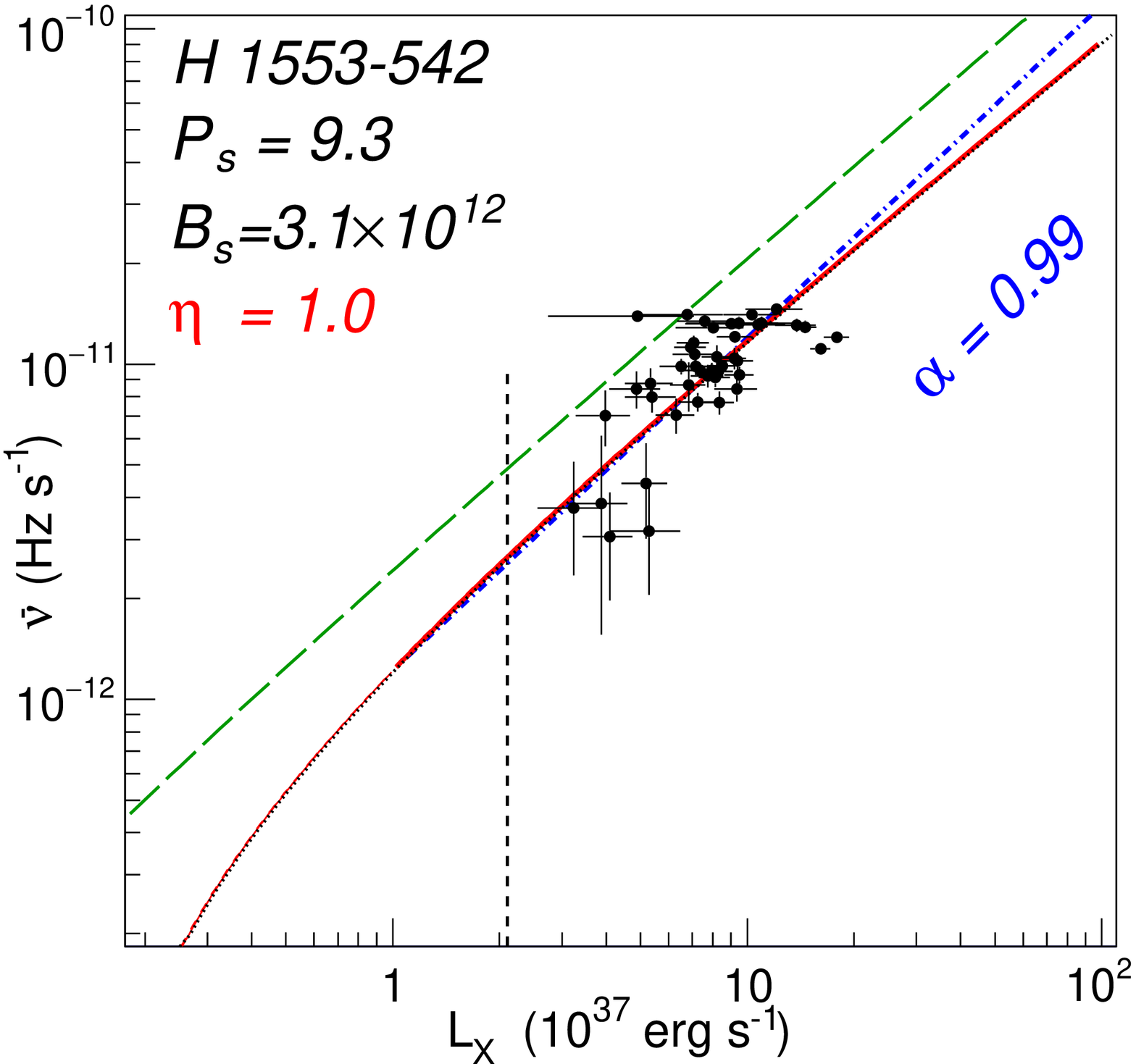}
\includegraphics[width=56mm]{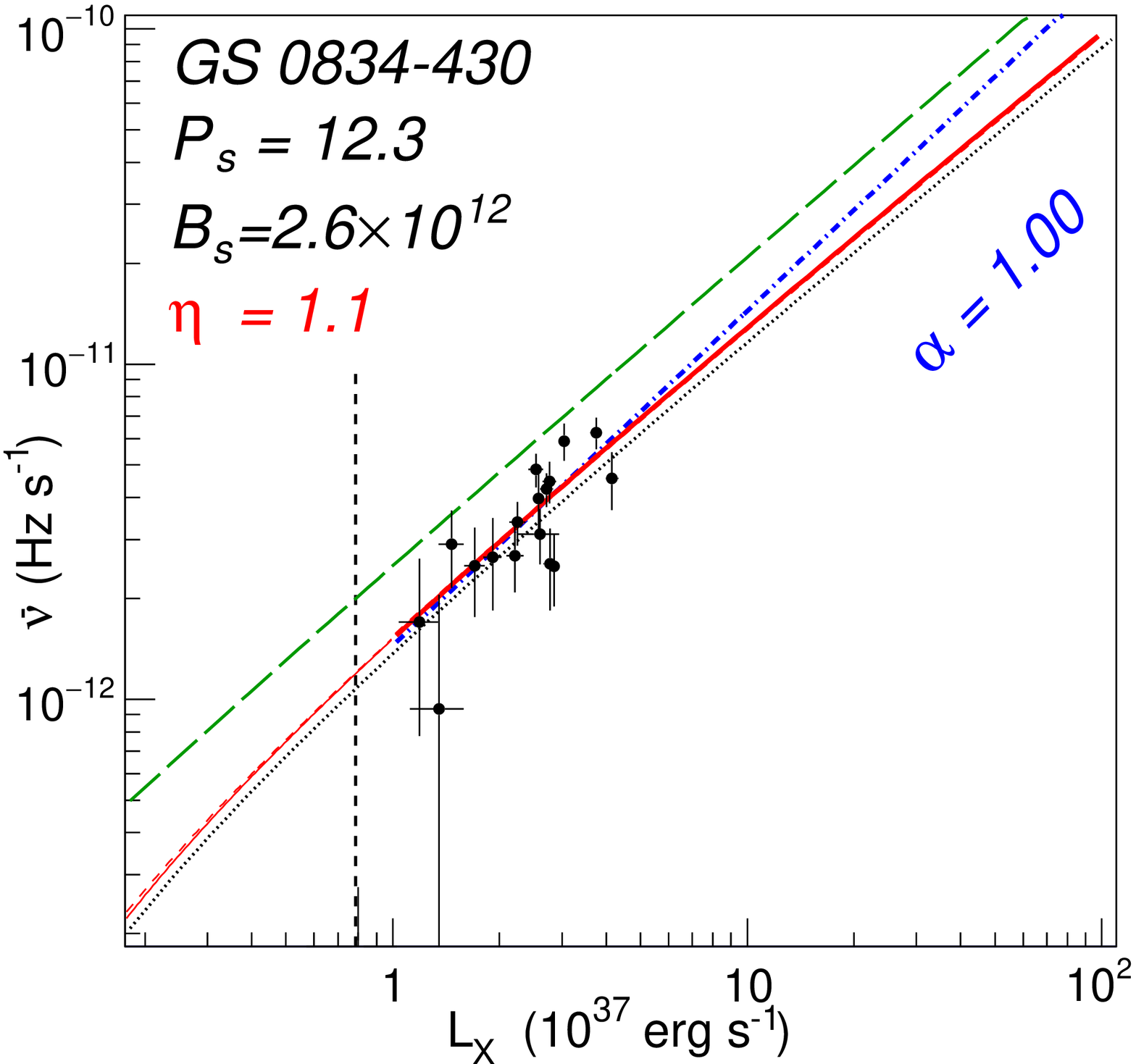}
\includegraphics[width=56mm]{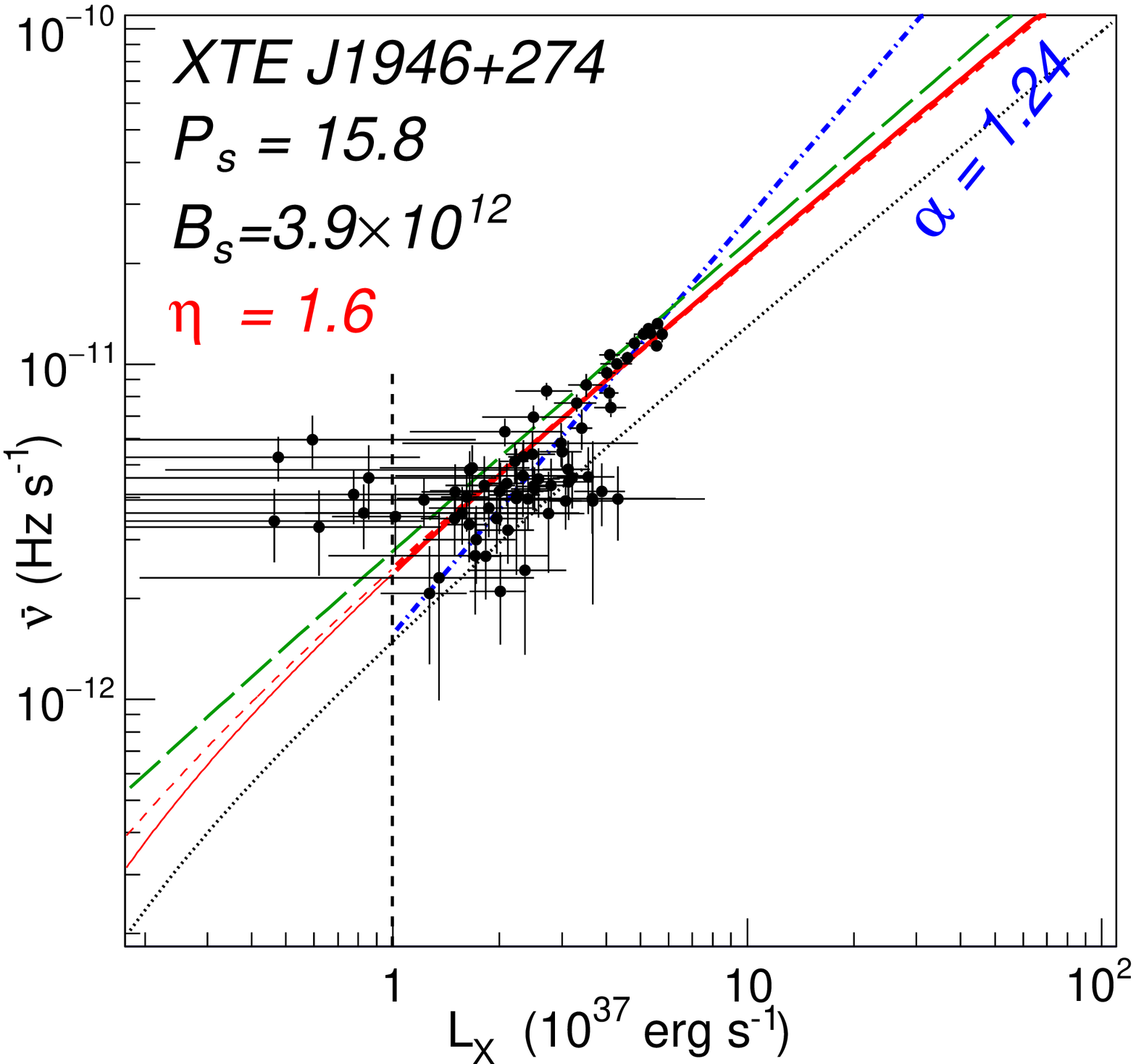}
                                                  
\vspace{3mm}
\includegraphics[width=56mm]{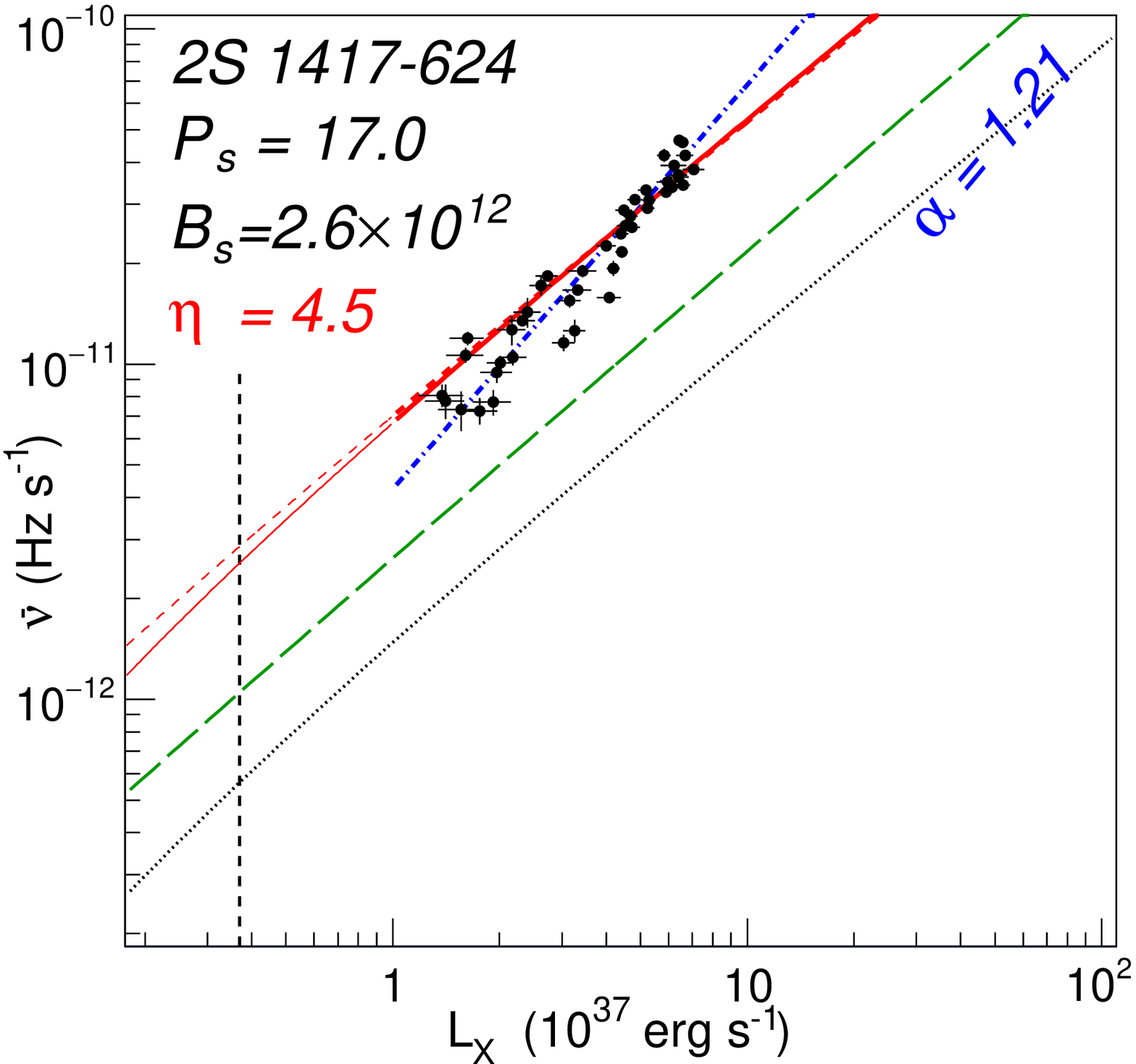}
\includegraphics[width=56mm]{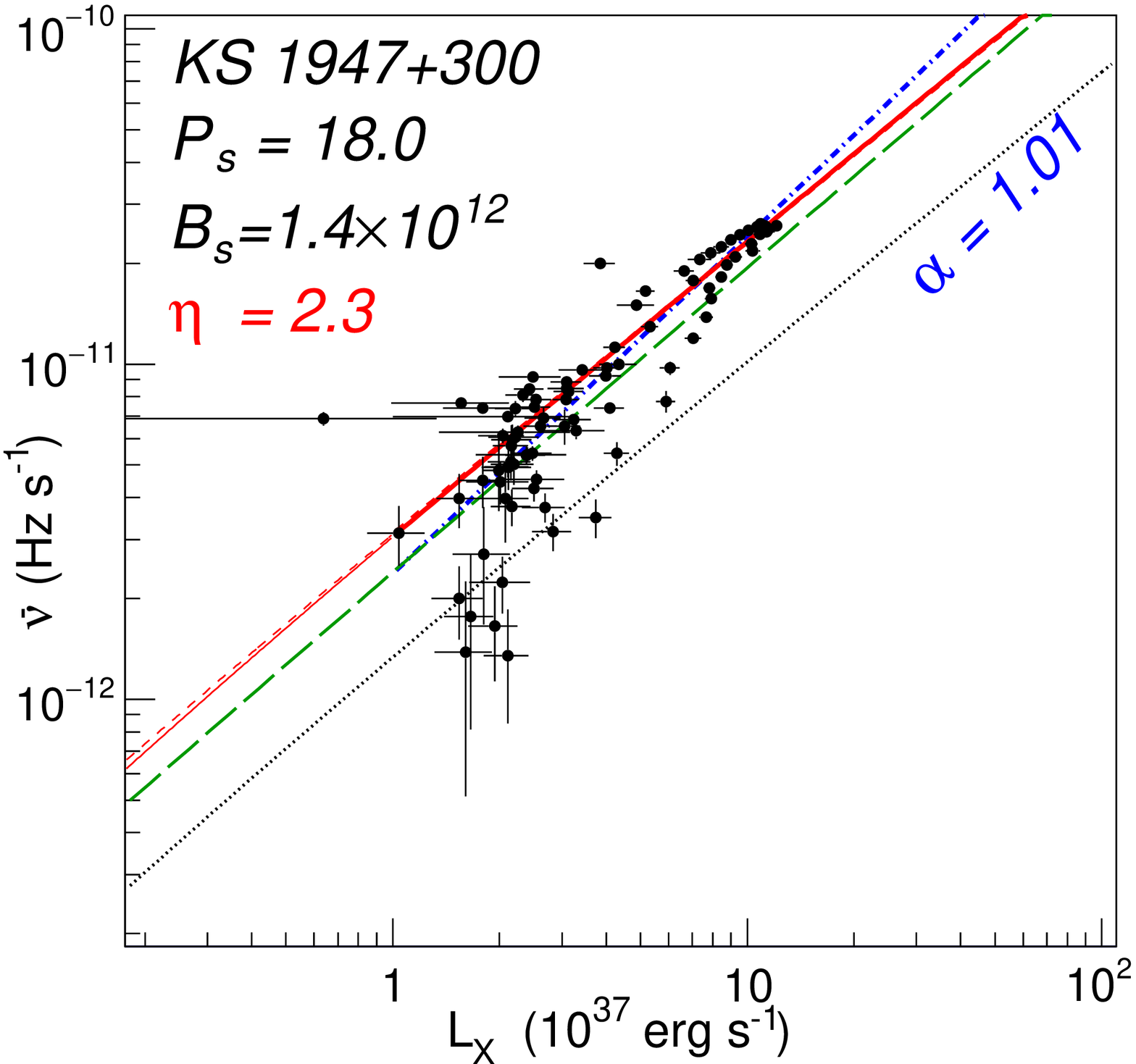}
\includegraphics[width=56mm]{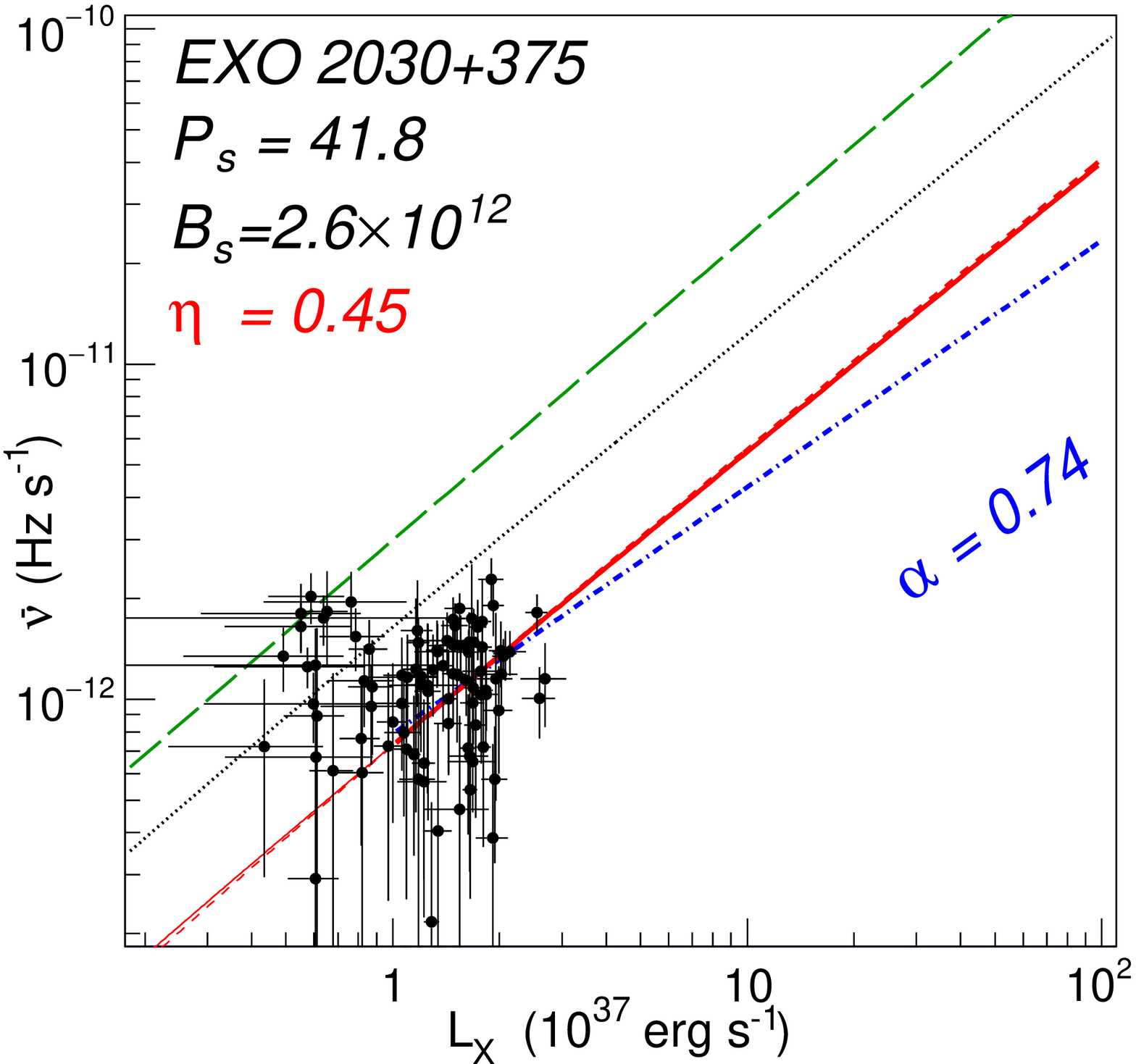}
                                                  
\vspace{3mm}
\includegraphics[width=56mm]{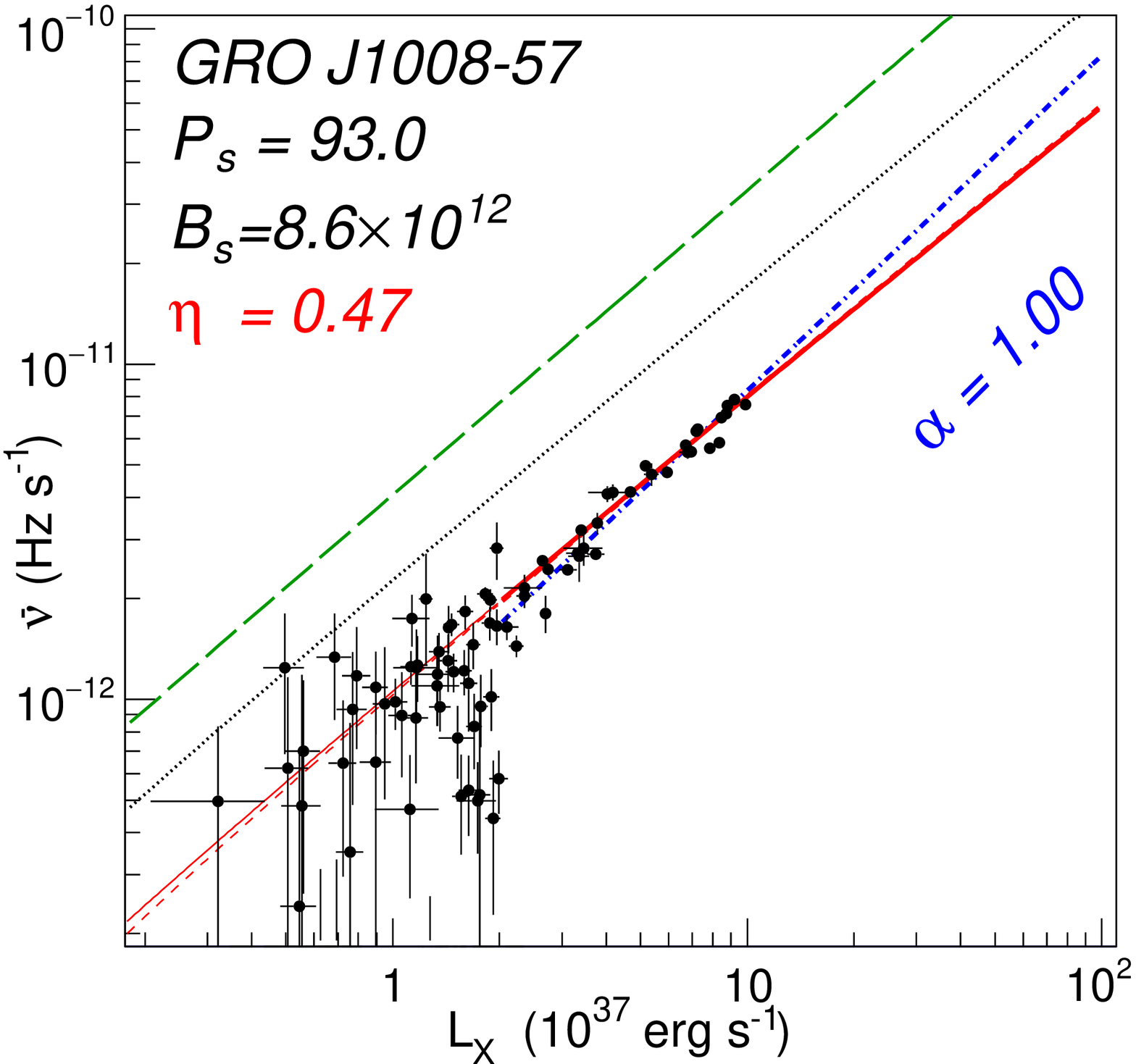}
\includegraphics[width=56mm]{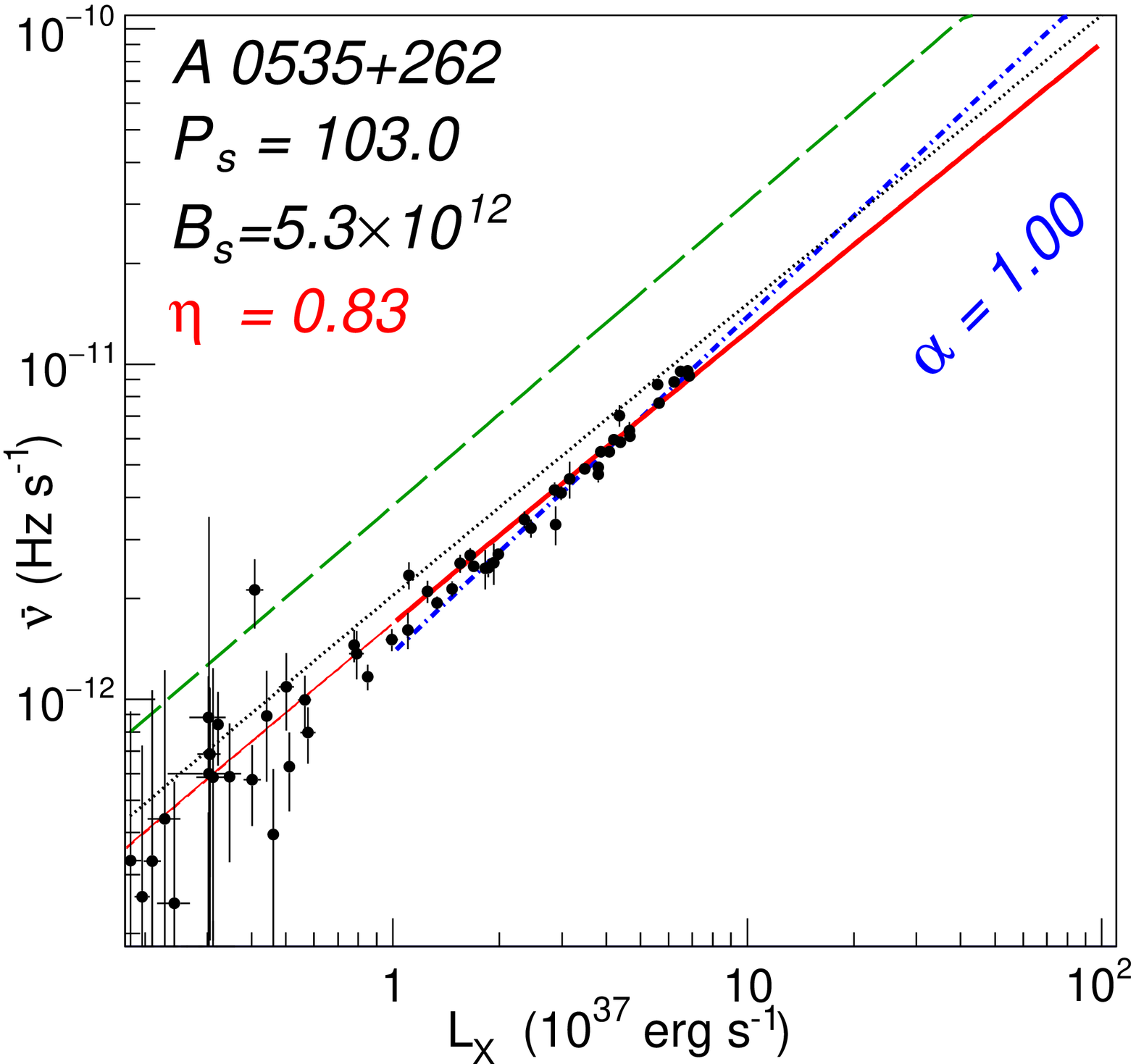}
\includegraphics[width=56mm]{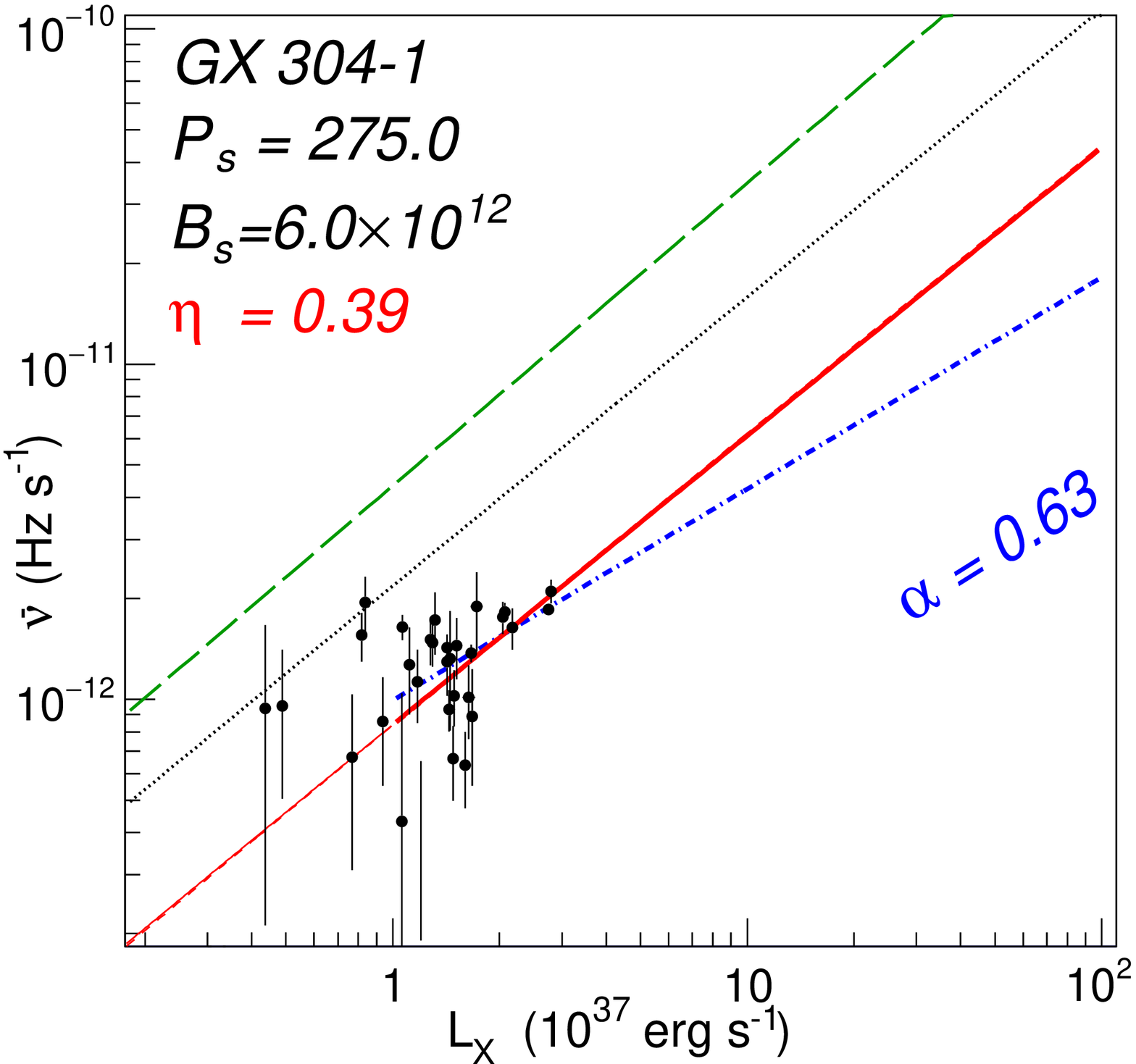}

\vspace{3mm}
\caption{
The $\dot{\nu}$-$L$ diagrams of the 12 Be XBPs.
The best-fit power-law model is shown in a blue dash-dotted line,
together with the value of $\alpha$. 
Black dotted and green long-dashed lines
show predictions by the GL79 and the KR07 models, respectively.
Thick solid and dashed lines in red
are modified GL79 models in which
the correction factor $\eta$ and the beaming fraction $f_{\rm b}$
are allowed to vary, respectively.
The value of $L$ at $\omega_\mathrm{s}^\mathrm{GL}=0.1$ 
is shown by a vertical dashed line, if it is in the plot range.
The spin period $P_{\rm s}$, the surface magnetic field $B_{\rm s}$
utilized in each fit, and the best-fit $\eta$ 
are also given in each panel.
}
\label{fig:pdotlx}

\end{figure*}

To better compare the data and the model, 
we introduce a correction factor $\eta$
to the original GL79 model as
\begin{equation}
  \dot{\nu}_{12} = \eta \dot{\nu}_\mathrm{12}^\mathrm{GL},
\label{equ:etafitfunc}
\end{equation}
and fitted it to the data of each source in figure \ref{fig:pdotlx},
leaving $\eta$ free.
The best-fit values of $\eta$ and $\chi^2_\nu$ 
are summarized in table \ref{tab:glfitpar}.
The $\eta$-corrected models are overlaid in red
on the data in figure \ref{fig:pdotlx}.

In XTE J1946$+$274, 2S 1417$-$624, KS 1947$+$300,
GRO J1008$-$57, and A 0535$+$262,
the fit before renormalizing the error by $\eta$
became somewhat worse than that with the power-law,
because $\alpha$ of these objects is significantly larger than $6/7=0.86$
implied by the GL79 model at $\omega_\mathrm{s}\ll 1$. 
However, in 4U 0115$+$63 and X 0331$+$53,
$\chi^2_\nu$ does not change or gets even better
even though the data indicate $\alpha>1$.
This is because the slow-rotator approximation
of $\omega_\mathrm{s}\ll 1$
is not applicable to these objects,
in which the $\dot{\nu}_\mathrm{s}$-$L$ relaiton
begins bending towards the lower $L$.
This behavior of the data in figure \ref{fig:pdotlx}
is well reproduced by the GL79 model.

The obtained values of $\eta$ distribute from 0.39 to 4.4,
by about an order of magnitude,
except X 0331$+$53 which required an exceptionally
small value of $\eta=0.12$.
As noticed above, 
the $\dot{\nu}_\mathrm{s}$-$L$ diagram of X 0331$+$53
also shows a steepening in 
$L\lesssim 2\times 10^{38}$ erg s$^{-1}$
due to the decrease of $n(\omega_\mathrm{s})$
as $\omega_\mathrm{s}$ approaches unity.
However, the GL79 model, drawn in a dotted line in figure
\ref{fig:pdotlx}, predicts that the steepening of this source would
become significant in $L\lesssim 2\times 10^{37}$ erg s$^{-1}$, which is lower by
one order of magnitude than that in the data.
This suggests that the
values of $L$ calculated from $f_\mathrm{b}=1$ and $D=6$ kpc
in equation (\ref{equ:lumiobs})
are overestimated.
We hence repeated the GL79 model fits to all the sources
by fixing $\eta=1$ but
allowing $f_\mathrm{b}$ to float.
This can express distance uncertainties.

Table \ref{tab:glfitpar} includes the best-fit $f_\mathrm{b}$ values and
their $\chi^2_\nu$.
The two best-fit models with free-$\eta$ and free-$f_\mathrm{b}$ 
are compared in figure \ref{fig:pdotlx}.
Thus, the free-$f_\mathrm{b}$ model better reproduces the data
in X 0331$+$53.
In other words, the data of X 0331$+$53 is better reproduced
by shifting the original GL79 prediction horizontally,
rather than vertically,
because of the steeping distribution of the data points.
In the other sources, the free-$\eta$ and the free-$f_\mathrm{b}$ 
approaches gave nearly the same $\chi^2_\nu$, because 
the data distributions are approximately linear (in the log-log plots).

\begin{figure}
\begin{center}
\includegraphics[width=81mm]{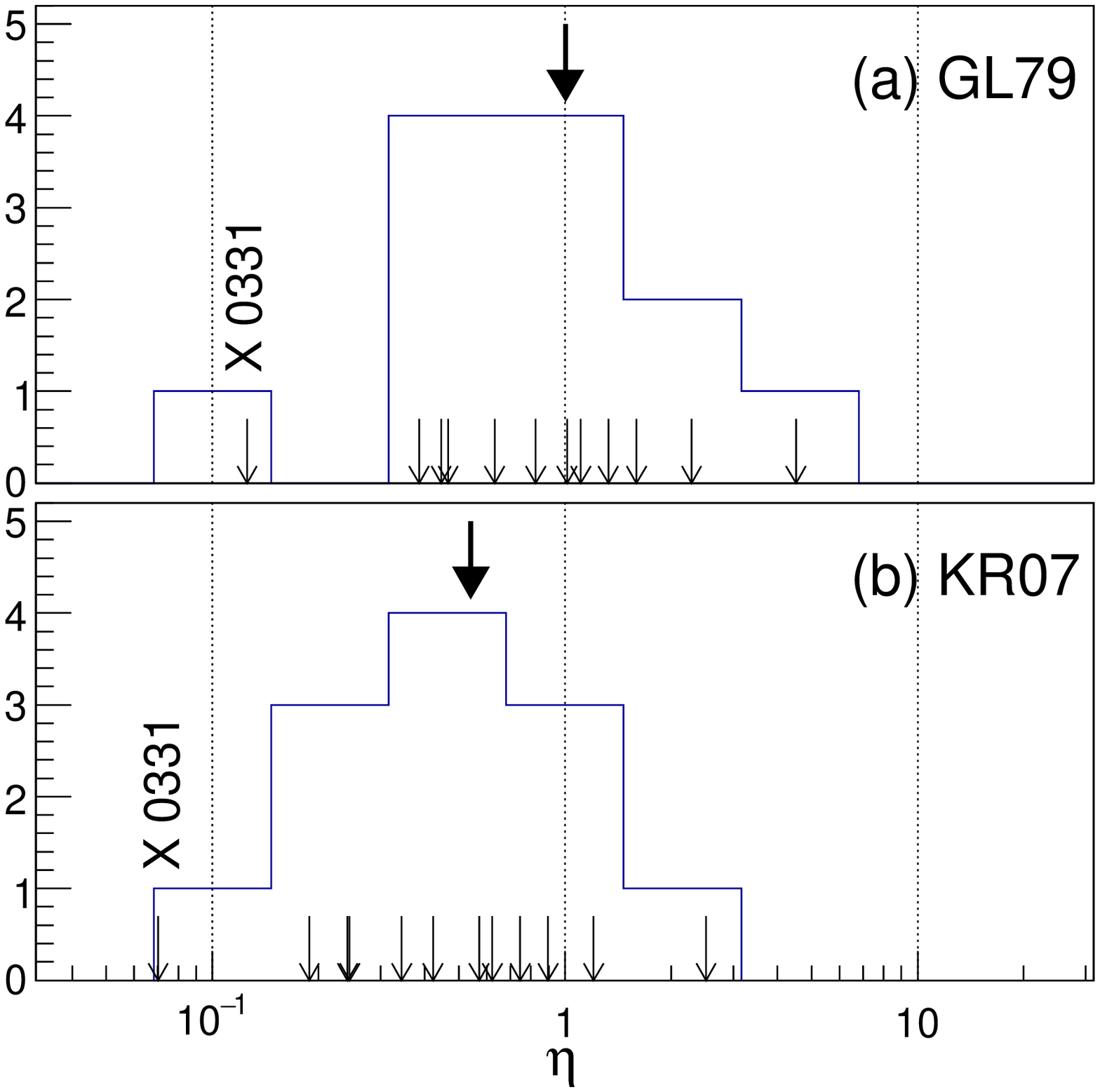}
\end{center}
\caption{
Histograms per logarithmic intervals
of the correction factor $\eta$ to the GL79 model
(panel a)
and to the KR07 model (panel b).
In each panel,
12 arrows at the bottom indicate values for the 12 sources, 
and a thick arrow at the top represent the logarithmic average
excluding the X 0331$+$53 data.
}
\label{fig:nuhist}
\end{figure}

Figure \ref{fig:nuhist}(a) show
a histogram of the 12 best-fit values of $\eta$,
where we employed logarithmic bins because the errors on $\eta$
are mostly proportional to $\eta$ themselves. 
The average and the standard deviation of $\log \eta$ among the 11
sources, with X 0331$+$53 excluded, are $\langle \log \eta
\rangle=0.001$ and $\sigma(\log \eta)=\pm 0.32$, respectively.  Therefore,
the log-average of $\eta$ is estimated to be
$10^{0.001\pm 0.32/\sqrt{11}}\simeq 1.0\pm 0.25$, and the 1-$\sigma$
range is given by a factor of $10^{0.32}= 2.1$.

%

\begin{table*}

\caption{
Summary of model fits to $\dot{\nu}_\mathrm{s}$-$L$ relation.
}
\label{tab:glfitpar}

\footnotesize

\begin{center}
\begin{tabular}{lllcc@{~}clcc@{~}clcc}
\hline
\hline
Source ID &  \multicolumn{12}{c}{Fitting model}\\
& \multicolumn{4}{c}{Power-law: $\dot{\nu}_\mathrm{12}= k \cdot L_{37}^\alpha$}
&& \multicolumn{3}{c}{GL79$^\ddagger$: $\dot{\nu}_\mathrm{12} = \eta \dot{\nu}^\mathrm{GL}(f_\mathrm{b} L_{37})$ } 
&& \multicolumn{3}{c}{KR07: $\dot{\nu}_\mathrm{12} = \eta \dot{\nu}^\mathrm{KR}$ } 
\\
\cline{2-5} \cline{7-9} \cline{11-13} 
& \multicolumn{1}{c}{$k^*$} & \multicolumn{1}{c}{$\alpha^*$} & $\xi^\dagger$ & $\chi^2_\nu$ ($\nu$)
&& \multicolumn{1}{c}{$\eta^*$ or $f_\mathrm{b}^*$} & $\xi^\dagger$ & $\chi^2_\nu$ ($\nu$)
&& \multicolumn{1}{c}{$\eta^*$} & $\xi^\dagger$ & $\chi^2_\nu$ ($\nu$) \\
\hline
   4U 0115 & $0.41\pm 0.07$ & $1.16\pm 0.08$ & 1.6 &  3.0 (46) && $0.632\pm 0.012$ & 1.6 &  3.1 (47) && $0.344\pm 0.007$ & 1.7 &  3.4 (47) \\
           &                &                &     &           && $0.633\pm 0.012$ & 1.6 &  3.0 (47) \\
\hline
    X 0331 & $0.08\pm 0.04$ & $1.11\pm 0.14$ & 5.2 & 35.3 (35) && $0.125\pm 0.004$ & 5.2 & 34.6 (36) && $0.070\pm 0.002$ & 5.2 & 35.0 (36) \\
           &                &                &     &           && $0.134\pm 0.004$ & 5.1 & 32.9 (36) \\
\hline
RX J0520.5 & $0.66\pm 0.32$ & $1.24\pm 0.15$ & 1.6 &  3.4 (21) && $ 1.33\pm  0.06$ & 1.7 &  3.8 (22) && $ 0.75\pm  0.03$ & 1.7 &  3.7 (22) \\
           &                &                &     &           && $ 1.37\pm  0.07$ & 1.7 &  3.8 (22) \\
\hline
    H 1553 & $1.22\pm 0.45$ & $0.99\pm 0.17$ & 1.8 &  3.8 (42) && $ 1.01\pm  0.04$ & 1.8 &  3.8 (43) && $ 0.57\pm  0.02$ & 1.8 &  3.8 (43) \\
           &                &                &     &           && $ 1.01\pm  0.03$ & 1.8 &  3.8 (43) \\
\hline
   GS 0834 & $1.45\pm 0.36$ & $1.00\pm 0.24$ & 1.2 &  1.8 (15) && $ 1.11\pm  0.06$ & 1.1 &  1.7 (16) && $ 0.62\pm  0.03$ & 1.1 &  1.7 (16) \\
           &                &                &     &           && $ 1.12\pm  0.06$ & 1.1 &  1.7 (16) \\
\hline
 XTE J1946 & $1.56\pm 0.13$ & $1.24\pm 0.06$ & 1.0 &  1.0 (63) && $ 1.60\pm  0.02$ & 1.2 &  1.7 (64) && $0.896\pm 0.016$ & 1.2 &  1.7 (64) \\
           &                &                &     &           && $ 1.66\pm  0.02$ & 1.2 &  1.7 (64) \\
\hline
   2S 1417 & $4.24\pm 0.34$ & $1.21\pm 0.05$ & 1.6 &  3.2 (43) && $ 4.52\pm  0.04$ & 2.2 &  5.9 (44) && $ 2.51\pm  0.05$ & 2.1 &  5.6 (44) \\
           &                &                &     &           && $ 5.52\pm  0.13$ & 2.3 &  6.4 (44) \\
\hline
   KS 1947 & $2.37\pm 0.18$ & $1.01\pm 0.04$ & 2.4 &  6.6 (90) && $2.287\pm 0.016$ & 2.5 &  7.4 (91) && $ 1.21\pm  0.02$ & 2.5 &  7.0 (91) \\
           &                &                &     &           && $ 2.59\pm  0.05$ & 2.6 &  7.5 (91) \\
\hline
  EXO 2030 & $0.79\pm 0.12$ & $0.74\pm 0.28$ & 1.3 &  1.9 (68) && $ 0.45\pm  0.02$ & 1.3 &  1.8 (69) && $0.242\pm 0.011$ & 1.3 &  1.9 (69) \\
           &                &                &     &           && $0.401\pm 0.014$ & 1.3 &  1.8 (69) \\
\hline
 GRO J1008 & $0.84\pm 0.07$ & $1.00\pm 0.04$ & 2.4 &  7.0 (32) && $0.466\pm 0.007$ & 2.6 &  8.6 (33) && $0.245\pm 0.003$ & 2.5 &  7.8 (33) \\
           &                &                &     &           && $0.418\pm 0.007$ & 2.6 &  8.3 (33) \\
\hline
    A 0535 & $1.38\pm 0.05$ & $1.00\pm 0.02$ & 1.8 &  4.0 (32) && $0.826\pm 0.009$ & 2.5 &  7.9 (33) && $0.422\pm 0.004$ & 2.2 &  5.9 (33) \\
           &                &                &     &           && $0.802\pm 0.004$ & 2.5 &  7.9 (33) \\
\hline
    GX 304 & $0.99\pm 0.10$ & $0.63\pm 0.13$ & 1.6 &  3.5 (23) && $0.386\pm 0.015$ & 1.7 &  3.7 (24) && $0.188\pm 0.008$ & 1.7 &  3.8 (24) \\
           &                &                &     &           && $0.332\pm 0.007$ & 1.7 &  3.7 (24) \\
\hline
\end{tabular}
\end{center}
$^*$ Errors represent 1-$\sigma$ confidence limits of the fitting parameters.\\
$^\dagger$ Artificial factor to inflate the measurement errors $\Delta\dot{\nu}_\mathrm{s}$ and $\Delta L$ as equation (\ref{equ:syserr})
that can bring the model fit to the 90\% confidence limit.\\ 
$^\ddagger$ Top and bottom lines in each column present the results of the GL79 model fits 
with free-$\eta$, $f_\mathrm{b}=1$ and with $\eta=1$, free-$f_\mathrm{b}$, respectively. 
\end{table*}


\section{Discussion}
\label{sec:discussion}

Using the data taken by the MAXI GSC all-sky survey 
and the Fermi GBM pulsar project
for over the 6 years since 2009 August,
we analyzed the long-term X-ray intensity and 
pulse-period changes of the well-defined 12 Be XBPs.
In all the 12 sources, the $\dot{\nu}_\mathrm{s}$-$L$ diagrams,
obtained from large outbursts with $L \gtsim 10^{37}$ erg s$^{-1}$,
show the expected positive correlations close to the direct proportionality.
We performed model fits to the $\dot{\nu}_\mathrm{s}$-$L$ data
with a power-law function
and also a representative theoretical model given by GL79,
leaving $\eta$ or $f_\mathrm{b}$ free.
Below, we discuss validity of some representative theoretical models
including GL79,
and then consider
possible origins of the scatter of $\eta$ among the sample.

\subsection{
Comparison among different theoretical models 
}
\label{sec:xbpmodel}

\subsubsection{The Ghosh \& Lamb model}
\label{sec:discuss_GL79}

The $\dot{\nu}_\mathrm{s}$-$L$ relation of Be XBPs
has been known to largely agree with the GL79 model prediction
within an order of magnitude
(e.g. \cite{1996A&A...312..872R,  1997ApJS..113..367B}).
Through a uniform analysis of a large data sample,
we improved the knowledge,
in particular, the distribution of the correction factor 
$\eta\simeq$ 0.1--4 to the GL79 model
among the 12 sources.
The parameter $\eta$ 
is needed to bring the observed relation of each object
in agreement with the GL79 model that incorporates the canonical
neutron-star parameters,
together with the observationally estimated $\mu$,
$D$, and $f_\mathrm{bol}$ in equations (\ref{equ:lumiobs}),
(\ref{equ:LMdot}), (\ref{equ:glnudot}), and (\ref{equ:mu30}).
As obtained in section \ref{sec:eta}, the log-average
of $\eta$ and its 1-$\sigma$ error among
the 11 sources, excluding X 0331$+$53, are $1.0 \pm 0.25$.
Therefore, the GL79 model very well explains
the average behavior of our sample.
Because the factor $\eta$ mostly depends only on $r_0$,
the results indicate that $r_0\simeq 0.5\,r_\mathrm{a}$
is a reasonable approximation in average.
The 1-$\sigma$ range of $\eta$ given by a factor 2.1 is
discussed in section \ref{sec:MRrange}.

When $L$ approaches the torque equilibrium
($\omega_\mathrm{s}^\mathrm{GL}\approx 0.35$),
the $\dot{\nu}_\mathrm{s}$-$L$ relations
of 4U 0115$+$63 and X 0331$+$53
start deviating from the direct proportionality.
The GL79 model has successfully explained this
important feature, behavior of 4U 1626$-$27 across
the spin-up/down threshold \citep{2016PASJ...68S..13T}.
In contrast,
the other models to be considered later do not
provide as successful account as GL79 of
this observations.

\subsubsection{The Klu{\'z}niak \& Rappaport model}
\label{sec:discuss_KR07}

The success of the GL79 model
in explaining the observed $\dot{\nu}_\mathrm{s}$-$L$ relation
does not necessarily mean that
the assumed physical conditions as a whole
are correct.
In fact,
\citet{1987A&A...183..257W} and LRB95 pointed out that
GL79 assume unrealistically large slip
between the disk and magnetic field lines
in the region between $r_0$ and $r_\mathrm{a}$. 
Following Wang (1987, 1995),
KR07 developed alternative models 
in which toroidal magnetic fields are dissipated by
either (A) turbulent diffusion in the disk, 
or (B) recombination outside the disk.
Because
the two KR07 assumptions lead to similar predictions,
we here examine the representative one, the turbulent-disk model (A).

To visualize differences between the GL79 and KR07 models, 
in figure 5 we show their $\dot{\nu}_\mathrm{s}$-$L$ predictions,
for typical values of  $P_\mathrm{s}$ and $B_\mathrm{s}$,
and canonical neutron-star parameters. 
Thus, we notice three differences between the two models.
\begin{enumerate}
\item[(i)]
In the slow-rotator regime ($\omega_\mathrm{s}\ll 1$),
both models predict straight $\dot{\nu}_\mathrm{s}$-$L$ relations,
but the slope is slightly different;
$\alpha=0.86$ by GL97 and  $\alpha= 0.9$ by KR07.
\item[(ii)]
In the same regime, 
the KR07 model predicts about a factor 2 higher 
$\dot{\nu}_\mathrm{s}$  than the GL79 model. 
\item[(iii)]
As $L$ decreases towards the torque equilibrium,
the predictions by both models start steepening.
However, this bending in KR07 takes place
at a much lower luminosity ($\omega_\mathrm{s}\gtsim 0.9$)
than in the GL79 model ($\omega_\mathrm{s}\sim$ 0.1).
\end{enumerate}

With the above three differences in mind,
we performed the KR07 model  fits to the data,
first without using the correction factor $\eta$.
The results, presented in figure 3 in green,
confirms the above property (ii).
Therefore, we next incorporated $\eta$ 
in the same as in section \ref{sec:eta},
and obtained the fit results as summarized in table \ref{tab:glfitpar}.
(The best-fit models are not shown in figure \ref{fig:pdotlx}
to avoid making the plots too confusing).
Thus, the KR07 model with floating $\eta$ generally gave
somewhat better fits to the data than the GL79 model,
because of the property (i).
However, the low-luminosity bending in
4U 0115$+$63 and X 0331$+$53
is better reproduced by the GL79 model,
reflecting the property (iii).
Furthermore, as presented in figure \ref{fig:nuhist}(b),
the values of $\eta$ with the KR07 model became
on average $\sim 0.5$ due to the property (ii),
making a contrast to the GL79 result of
$\langle\eta\rangle\sim 1.0$.

These comparisons,
together with the success of \citet{2016PASJ...68S..13T},
are thought to provide an {\it a posteriori} 
justification to our choice of GL79
as the representative accretion torque formalism.

\begin{figure*}
\begin{center}
\includegraphics[width=56mm]{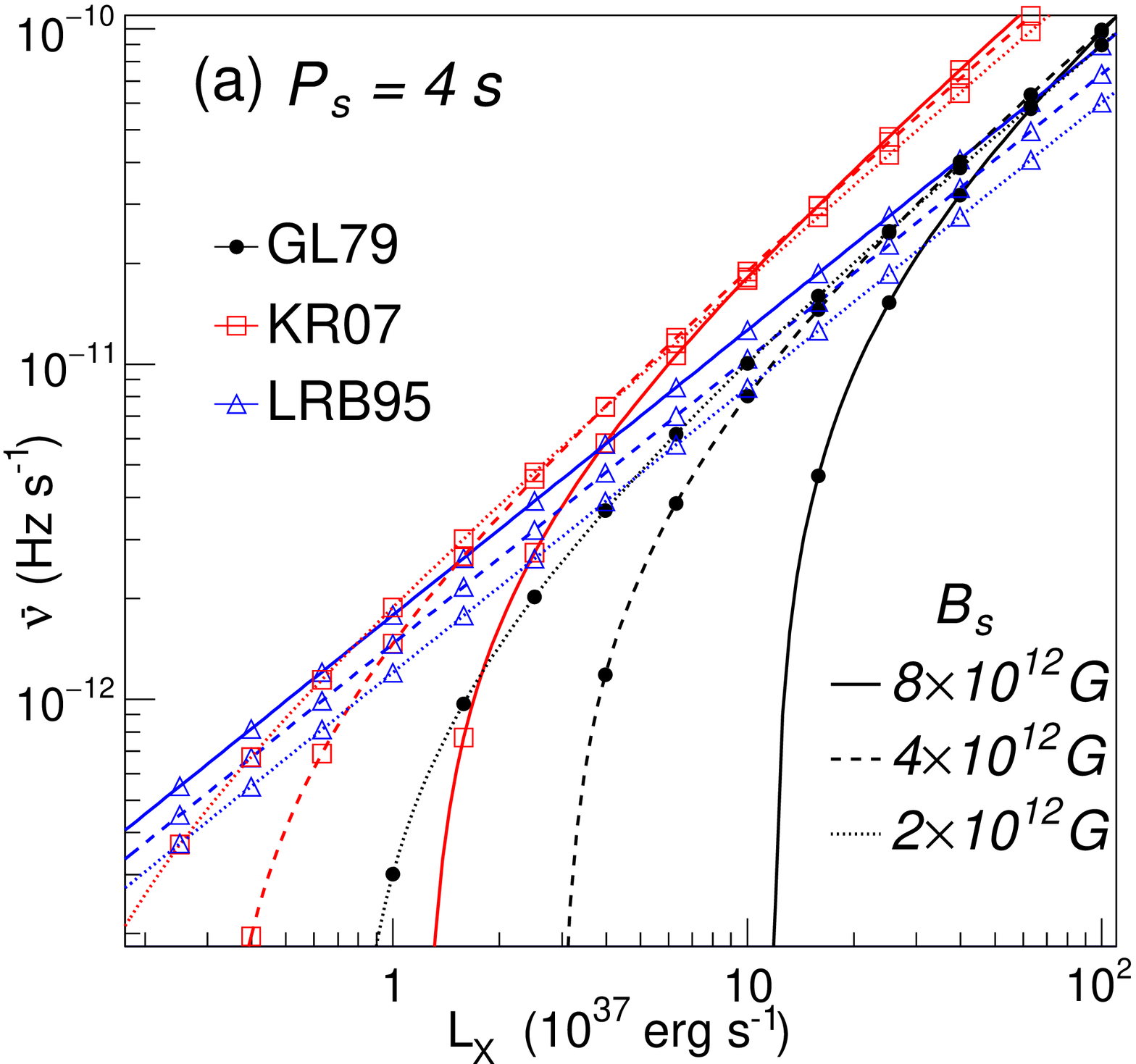}
\includegraphics[width=56mm]{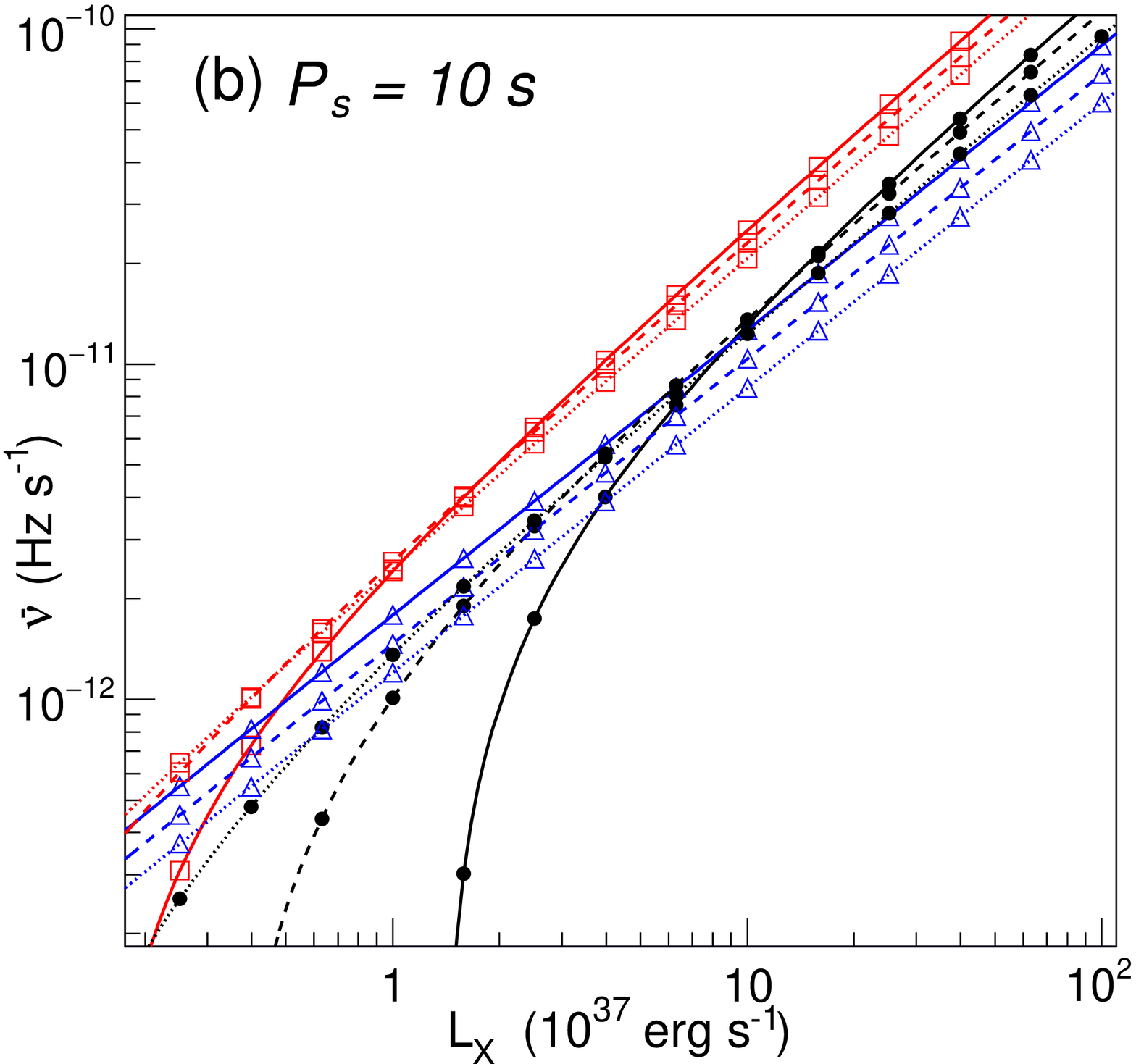}
\includegraphics[width=56mm]{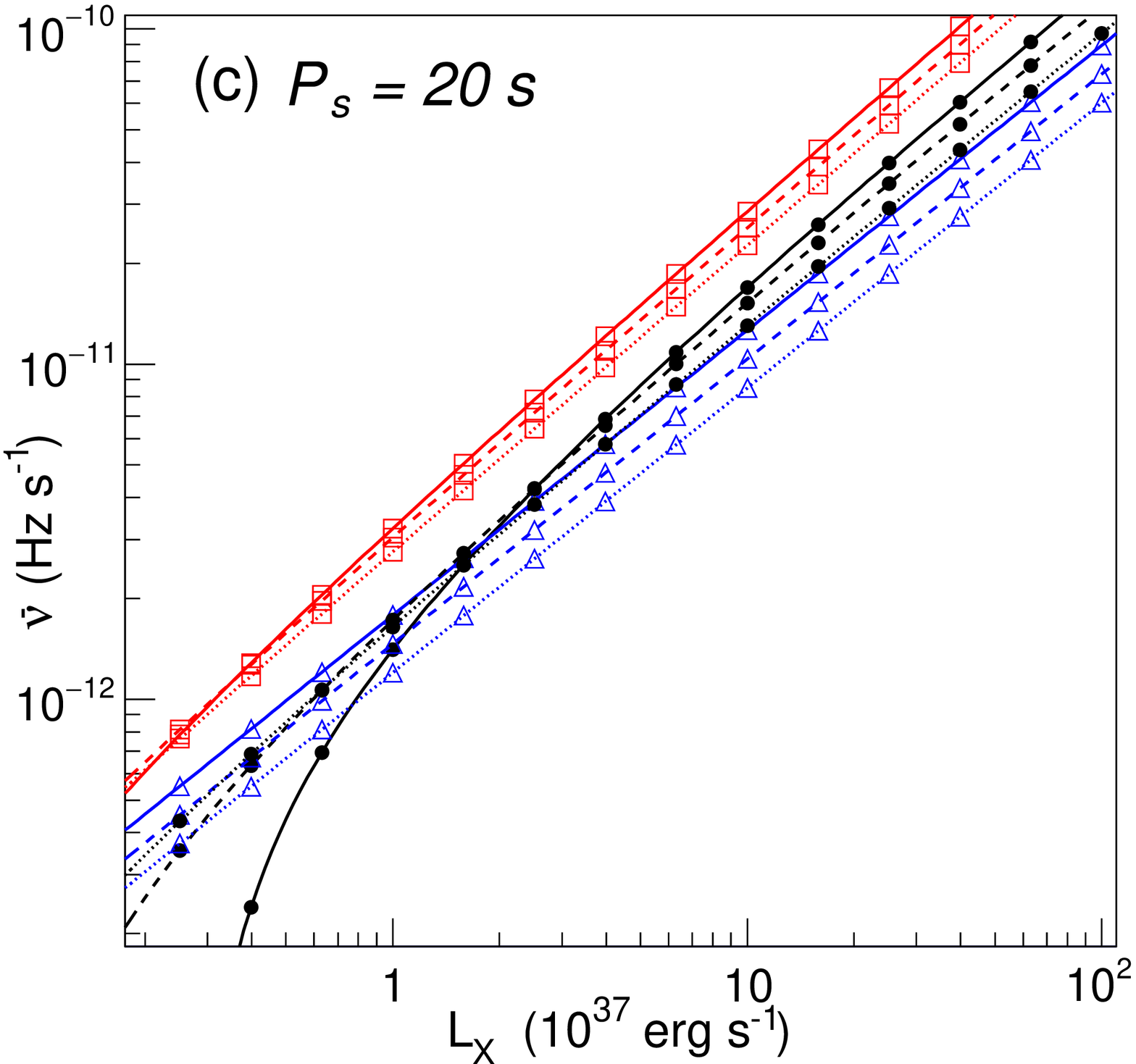}
\end{center}
\caption{
  Theoretical $\dot{\nu}_\mathrm{s}$-$L$ relations
  in XBPs with $P_\mathrm{s}=$ 4 s (panel a), 10 s (panel b) , and 20 s (panel c)
  calculated 
  from the GL79 (black line with dot), 
  the KR07 (red line with triangle), 
  and the LRB95 (blue line with box) models.
  Solid, dashed, and dotted lines are relations
  calculated for $B_\mathrm{s}=$ 2, 4, and 8 $\times 10^{12}$ G. 
}
\label{fig:nudotmodel}
\end{figure*}

\subsubsection{The Lovelace model}

LRB95 developed a turbulent-disk model
considering the open field lines that lead to 
magnetically driven outflows.
It is characterized by a parameter representing
the magnetic diffusivity in the disk,
$\alpha_\mathrm{m} D_\mathrm{m}\simeq$ 0.01--0.1.
In figure \ref{fig:nudotmodel}, the $\dot{\nu}$-$L$ relations of the
LRB95 model assuming  $\alpha_\mathrm{m} D_\mathrm{m}=0.1$
are plotted together with those of GL79 and KR07.
In the slow-rotator regime, 
the LRB95 model predicts
constantly $\simeq 0.7$ times smaller $\nu_\mathrm{s}$ than GL79.
Considering that the GL79 fits gave $\langle\eta\rangle\simeq 1.0$,
we need to increase the LRB95 prediction by a factor of $1/0.7$.
This could be done by choosing $\alpha_\mathrm{m} D_\mathrm{m}\simeq 1$,
but this falled outside its nominal range of 0.01--0.1.
The model cannot either reproduce the data bending in 4U 0115$+$63 and
X 0331$+$53.
Yet another disadvantage of LRB95 is its failure to explain
the spin-up/down transition observed
from 4U 1626$-$67 \citep{2016PASJ...68S..13T}.
Hence, we do not employ this model.

\subsubsection{Other models}

\citet{2012MNRAS.420.1034C} propose another disk-magnetosphere
interaction model considering the angular-momentum feedback
from the accreting matter to the disk. 
The model suggests that the accretion torque 
is reduced by a factor $\omega_\mathrm{s}\propto L^{-3/7}$
from that in equation (\ref{equ:ntorque}),
and thus the $\dot{\nu}_\mathrm{s}$-$L$ relation becomes
$\dot{\nu}_\mathrm{s}\propto L^{3/7}$.
Thus, the model is not applicable to the present data,
which demand $\alpha\simeq 1$.

Motivated by an apparent double-valued $\dot{\nu}_\mathrm{s}$-$L$ relation
observed from the slow rotator GX 304$-$1,
\citet{2015MNRAS.446.1013P} proposed a quasi-spherical accretion picture, 
which predicts $\alpha=7/11$ \citep{2012MNRAS.420..216S}.
In table \ref{tab:glfitpar}, GX 304$-$1 indeed exhibits 
$\alpha\simeq 0.6$ (though with the large error)
in an agreement with that prediction.
However, as presented in figure \ref{fig:pdotlx},
the double-valued behavior has been explained away
when using the refined orbital elements \citep{2015PASJ...67...73S}. 
Therefore, it remains inconclusive
whether the object prefers the model by \citet{2012MNRAS.420..216S}.

\subsection{Reconsideration of X 0331$+$53 analysis results}
\label{sec:x0331dist}

Among the 12 values of $\eta$ for our Be XBP sample,
$\eta=0.12$ of X 0331$+$53 is unusually deviated from unity.
The source also looks strange in 
that the estimated outburst-peak luminosity, $\sim 5\times
10^{38}$ erg s$^{-1}$, is significantly higher than those of the
others ($\lesssim 10^{38}$ erg s$^{-1}$), and also exceeds the
Eddington luminosity, $\simeq 2\times 10^{38}$ erg s$^{-1}$,
for a $1.4M_\odot$ neutron star
(figures \ref{fig:lcper} and \ref{fig:pdotlx}).
These facts suggest that the employed source distance, 6 kpc,
from the optical photometry of the companion, BQ Cam
\citep{2015A&A...574A..33R},
is overestimated.
For example,
\citet{1985PASJ...37...97K} optically estimated it
as 3.5 kpc, or even smaller, just after the source was re-discovered
in X-rays \citep{1990PASJ...42..295M}. 
However, all these measurements 
had a problem 
of contaminations of infra-red emission from the
Be disk \citep{1999MNRAS.307..695N, 2015A&A...574A..33R}.

In section \ref{sec:eta},
we found that the GL79 model better reproduces
the X 0331$+$53 data 
with the bolometric correction factor $f_\mathrm{b}=0.12$ than
with the factor $\eta=0.12$ to the $\dot{\nu}_\mathrm{s}$-to-$L$ coefficient.
This means that
the true $L$ is likely to be $\sim 0.12$ times the nominal one,
and hence the actual $D$ is $6\times \sqrt{0.12}\simeq 2.4$ kpc.
Considering these facts,
we suggest that X 0331$+$53 is located at $D=$2--3 kpc

\subsection{Estimate of physical parameter ranges}
\label{sec:MRrange}

As obtained in section \ref{sec:eta},
the 1-$\sigma$ range of the
correction factor $\eta$
among the 11 sources (excluding X 0331$+$53)
is given by $\sigma(\log \eta)= 0.31$,
which means the range from $10^{-0.31}=0.49$ to $10^{0.31}=2.1$.
Then, a key question is whether this scatter in $\log \eta$ can be explained
by taking into account possible uncertainties in the parameters involved
in the equations 
(\ref{equ:lumiobs}), (\ref{equ:glnudot}), (\ref{equ:mu30}), and (\ref{equ:LMdot}),
or requires some corrections
to the GL79 model itself.
In section \ref{sec:xbpmodel}, we examined several alternative
disk-magnetosphere models, and found 
that the differences among them 
are mostly represented by systematic differences in $\eta$.
Therefore, the scatter of $\log \eta$
obtained from the 11 objects (excluding X 0331$+$53) 
does not depend on these models.
Below, let us examine the equation for $\eta$ in
a somewhat simpler form, neglecting for simplicity
the general relativistic effects.

The values of $I_{45}$ are mostly determined by $M_{1.4}$ and $\radns$.
We employed the approximating equation with them,
\begin{equation}
  I_{45}\simeq 1.0 M_{1.4}\radns^2(1-x^{-1})^{-1}
  \simeq 1.0 M_{1.4}\radns^2,
\label{equ:inertia}
\end{equation}
which is applicable to most of the 
major models describing the neutron-star interior
\citep{1994ApJ...424..846R}.

In the GL79 model, the magnetic dipole 
is assumed to be aligned to the spin axis.
This is however not exactly correct, because the
observed X-ray fluxes are generally pulsating.
Therefore, $\mu_{30}$ in equation
(\ref{equ:glnudot}) needs some corrections.
At distances far from the neutron-star surface, the field strength can
change by a factor of 2 according to the dipole axis orientation
\citep{1997ApJ...475L.135W}.
We thus introduce a factor $f_\mu$, which takes a value from 1 to 2,
and rewrite equation (\ref{equ:mu30}) as 
\begin{equation}
  \mu_{30} = \frac{1}{2} f_\mu B_{12} \radns^3 \Phi(x).
  \simeq \frac{1}{2} b f_\mu E_\mathrm{a} \radns^3,
  \label{equ:mu30b}
\end{equation}
where $E_\mathrm{a}$ refers to equation (\ref{equ:bcyc}),
and $b$ is a conversion constant in the equation.

Substituting 
equations (\ref{equ:lumiobs}), (\ref{equ:inertia}) and (\ref{equ:mu30b})
into equation (\ref{equ:glnudot}), we obtain
\begin{eqnarray}
\label{equ:glnudot_v2}
\dot{\nu}_{12}^\mathrm{GL}  
&=&  1.9 \, n(\omega_\mathrm{s}) \left( \frac{1}{2} b f_\mu E_\mathrm{a} \right)^{2/7} 
\radns^{-2/7} M_{1.4}^{-10/7} \nonumber \\
&&  
\cdot \left( 4\pi D^2 f_\mathrm{bol} f_\mathrm{b} \right)^{6/7} C_\mathrm{2-20}^{6/7}
\end{eqnarray}
If the GL79 model equation (\ref{equ:glnudot}) is accurate enough, 
$\eta$ will be accounted for by uncertainties or biases
in the various parameters involved in equation (\ref{equ:glnudot_v2}).
Assuming that the values of
$f_\mu$, $E_\mathrm{a}$, $D$, $f_\mathrm{bol}$, and $f_\mathrm{b}$
employed above
are different from their true values by 
factors of
$10^{\pm\delta\!f_\mu}$,
$10^{\pm\delta\!E_\mathrm{a}}$,
$10^{\pm\delta\!D}$,
$10^{\pm\delta\!f_\mathrm{bol}}$,
and $10^{\pm\delta\!f_\mathrm{b}}$, 
respectively,
we can express $\eta$ as
\begin{eqnarray}
\label{equ:eta_v2}
\eta  &=&
\radns^{-2/7} M_{1.4}^{-10/7} \nonumber \\
&&
\cdot
({10}^{\delta\!f_\mu} {10}^{\delta\!E_\mathrm{a}})^{2/7} 
(10^{\delta\!f_\mathrm{b}} 10^{\delta\!f_\mathrm{bol}})^{6/7} 
(10^{\delta\!D})^{12/7} 
\end{eqnarray}
The dispersion of $\log \eta$ is then approximately reduced to
\begin{eqnarray}
  \sigma^2\left(\log \eta\right)
  &&\simeq \left(\frac{10}{7}\right)^2  \sigma^2\left( \log ( M_{1.4}\radns^{1/5} ) \right) \nonumber\\
  && +\left(\frac{2}{7}\right)^2
  \left[
    \sigma^2\left( \delta\!f_\mu \right) + \sigma^2\left(\delta\!E_\mathrm{a} \right)
    \right] \nonumber\\
  && +\left(\frac{6}{7}\right)^2 \left[
    \sigma^2\left(\delta\!f_\mathrm{b} \right) + \sigma^2\left( \delta\!f_\mathrm{bol} \right)
    \right] \nonumber\\
  && + \left(\frac{12}{7}\right)^2 \sigma^2\left(\delta\!D \right),
\label{equ:log_eta}
\end{eqnarray}
where the function $\sigma^2()$ means the variance of a given parameter
among the sample of 11 sources, and 
the parameter $M_{1.4}\radns^{1/5}$ is left as a single variable 
because $M$ and $R$ cannot vary independently.

In equation (\ref{equ:log_eta}),
the left-hand side shows a scatter of $\sigma(\log \eta)=0.31$
(section \ref{sec:eta}).
Then, how about the right hand side ?
Let us consider the involved parameters one by one.
\begin{enumerate}

\item As discussed above,
  the correction factor $f_\mu$ for $\mu_{30}$ is considered to 
  take a value from 1 to 2.
  We hence assume $\sigma(\delta\!f_\mu )\simeq \log 1.5 =0.18$.  
  
\item The observed CRSF energy, $E_\mathrm{a}$,
  depends on the source luminosity to some extent
  (section \ref{sec:target_select}).
  Among the 9 sources whose CRSF has been detected in our sample,
  4U 0115$+$63 exhibits the largest $E_\mathrm{a}$ change by 40\%
  (e.g. \cite{2006ApJ...646.1125N}).  The values of $E_\mathrm{a}$
  determined by model fits to X-ray spectra also depend on the employed
  model functions for the continuum and the absorption feature.
  However, differences among the model functions are estimated at most 10\%
  (e.g. \cite{2004ApJ...610..390M}),
  which is smaller than the change by the luminosity.
  We here employ the 1-$\sigma$ error range of 30\%, and thus
  $\sigma(\delta\!E_\mathrm{a} ) \simeq \log 1.3 =0.11$.

\item According to table \ref{tab:specparam},
  the 1-$\sigma$ error on $f_\mathrm{bol}$
  is at most 10 \%,
  which means $\sigma(\delta\!f_\mathrm{bol})\simeq \log 1.1 = 0.04$.

\item Although we have assumed $f_\mathrm{b}=1$ in equation (\ref{equ:lumiobs})
  for simplicity, the assumption is not necessarily
  warranted because
  the source are clearly pulsating.
  \citet{1975A&A....42..311B} suggested 
  that it can change by a factor $\sim 2$,
  based on their theoretical model. 
  Assuming that it has an 1-$\sigma$ range given by a factor 2,
  we obtain $\sigma(\delta\!f_\mathrm{b} )\simeq \log 2 = 0.30$.

\item The errors on the source distances $D$ estimated from the optical
  observations are listed in table \ref{tab:bexbpobs}.
  They are typically $\sim 20$\% although their confidence
  levels are not clearly given in some cases.
  We here assume that the 1-$\sigma$ error is $\sim 20$\%, and thus
  $\sigma(\delta\!D) \simeq \log 1.2 = 0.079$
  
\end{enumerate}

Accumulating the variances of logarithmic uncertainties in 
$f_\mathrm{b}$, $f_\mathrm{bol}$, $f_\mu$, $E_\mathrm{a}$, and $D$,
as estimated above,
and assuming that their errors are all independent from one another,
the right side of equation (\ref{equ:log_eta}) 
becomes
\begin{eqnarray}
  \label{equ:etavarval}
  \sigma^2
  &=&  \left(\frac{6}{7}\right)^2 (0.30^2 + 0.04^2)\nonumber\\
  && +\left(\frac{2}{7}\right)^2 ( 0.18^2 + 0.11^2 )\nonumber\\
  && + \left(\frac{12}{7}\right)^2 (0.079^2)\nonumber\\
  &\simeq& 0.0673 + 0.0036 + 0.0183 \nonumber\\
  &=& 0.0892 \simeq 0.30^2. 
\end{eqnarray}
The value is very close to the  observed one, $\sigma^2(\log \eta)=0.31^2$.
Therefore,
the present high-quality data are still consistent with
the GL79 model within the uncertainties considered above,
and we do not need to involve a significant variance in $M_{1.4}\radns^{1/5}$,
which has been neglected.

In equation (\ref{equ:etavarval}), the total variance
mostly owes to the two parameters, the beaming fraction
$f_\mathrm{b}$ and the distance $D$.
Further studies of these parameters will allow us to perform 
more accurate calibration of the GL79 formalism.

\subsection{Correlations between $\eta$ and other parameters}

\begin{figure*}
\begin{center}
\includegraphics[width=83mm]{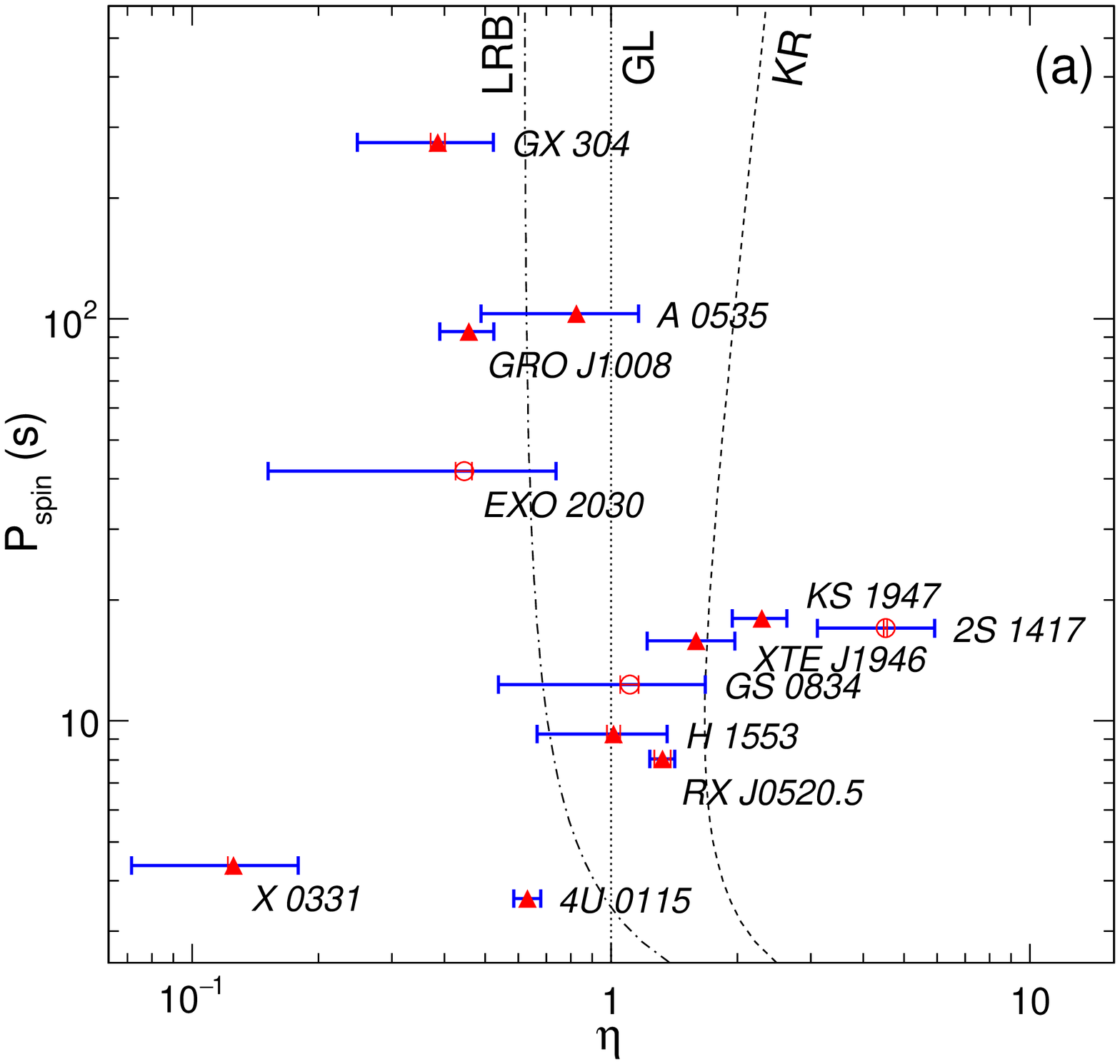}
\hspace{2mm}
\includegraphics[width=83mm]{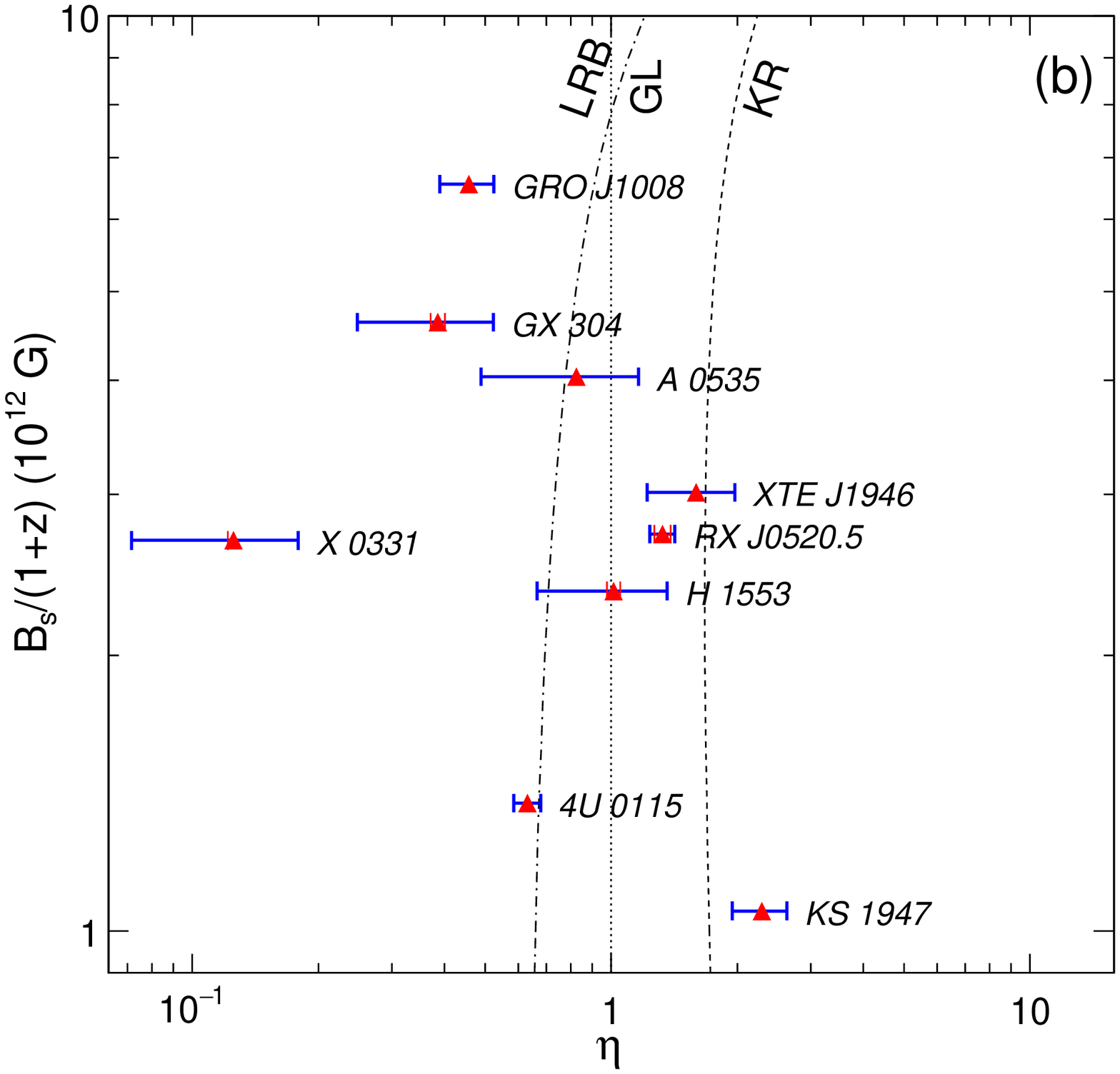}
\end{center}
\caption{ 
Distributions of $\eta$ over the 12 Be XBPs, presented versus
the pulse period $P_\mathrm{s}$
(panel a) and the magnetic field $B_\mathrm{s}$ (panel b).  
Sources whose $B_\mathrm{s}$ are estimated by the
CRSFs, are marked with solid triangles, and the others with open
circles.  Horizontal error bars in red represent the 
1-$\sigma$ errors in the model fit to the $\dot{\nu}_\mathrm{s}$-$L$ data,
and those with blue represent uncertainties
expected from the errors in the parameters
involved in the GL79 model formalism.  
Dash and dot-dash lines represent
the expected behavior of $\eta$ by KR07 and LBR95, respectively,
both relative to the GR79 predictions.
}
\label{fig:nudist}

\end{figure*}

Although we have shown that the scatter in $\eta$ can be
explained by uncertainties in the involved parameters,
it is still worth examining 
whether high-$\eta$ and low-$\eta$ XBPs have any
systematic differences in their properties.
For this purpose, we plot in figure \ref{fig:nudist}
the two basic parameters, 
$P_\mathrm{s}$ and $B_\mathrm{s}$, as a function of $\eta$.
In the $\eta$-$B_\mathrm{s}$ plot,
9 sources with secure $B_\mathrm{s}$ measurements were used.
Figure \ref{fig:nudist} 
also shows the behavior of $\eta$ from the KR07 and the LRB95 models
relative to the GL79 model
against $P_\mathrm{s}$ and $B_\mathrm{s}$.
It clearly reveals
$\dot{\nu}_\mathrm{s}^\mathrm{KR}\simeq 2 \dot{\nu}_\mathrm{s}^\mathrm{GL}$
and $\dot{\nu}_\mathrm{s}^\mathrm{LRB}\simeq 0.7 \dot{\nu}_\mathrm{s}^\mathrm{GL}$,
as discussed in section \ref{sec:xbpmodel}.

We observe weak negative correlations both in
the $\eta$-$P_\mathrm{s}$  and $\eta$-$B_\mathrm{s}$ diagrams,
in such a way that higher-field and longer-period XBPs tend
to show lower $\eta$ (i.e., more difficult to be spun up).
Since we already know that $P_\mathrm{s}$ and $B_\mathrm{s}$ of XBPs
positively correlate with each other (e.g.  \cite{Makishima1999}),
the two correlations may not be independent.

One possible interpretation of figure \ref{fig:nudist} 
is to consider that higher-field objects with longer pulse periods
may have lower values of $f_\mathrm{b}$,
because the emission is more tightly beamed
under the stronger magnetic fields,
and the beam axis  sweeps away from us.
Yet another, more speculative possibility is
to assume that higher-field objects
somehow have slightly higher mass,
and hence smaller values of $\eta$ via equation (\ref{equ:eta_v2}).

Even putting aside such specific causes,
the negative $\eta$-$P_\mathrm{s}$ correlation 
may be explained in the following way.
Some XBPs, for unspecified reasons, 
may intrinsically have somewhat higher values of $\eta$.
Such XBPs would be more efficiently spun up by accretion,
to achieve faster rotation.
In contrast, those with intrinsically lower $\eta$
may end up with having long pulse periods.

\section{Conclusions}

To examine the validity of the pulsar spin-up models
due to the interaction between the pulsar magnetosphere
and the accretion disk in XBPs,
we analyzed the X-ray lightcurves and pulse-period variations of the
12 Be XBPs whose distance and orbital elements are well determined.
The X-ray intensity was derived from the MAXI GSC data, and the timing
information was derived from the Fermi GBM, both for more than 6 years
since 2009.
In all these objects, 
closely proportional relations between $\dot{\nu}_\mathrm{s}$ and $L$,
which are expected theoretically,
were confirmed.  Except in X 0331$+$53, the
coefficient $\eta$ of proportionality 
between $\dot{\nu}_\mathrm{s}$ and $L$
agrees, within a factor of 3, with that
predicted by the GL79 model.
When averaged over the 11 sources,
$\eta$ becomes close to the GL79 prediction,
and its scatter 
can be explained
by uncertainties in the involved parameters, including
in particular, $D$ and $f_\mathrm{b}$.
The large discrepancy found with X 0331$+$53 is likely to arise
from its distance overestimation.

\begin{ack}
%

The authors thank all the MAXI team members for their dedicated work
on the ISS MAXI operation. 
Their thanks are also due to the Fermi/GBM pulsar
project for providing the useful results to the public. 
The present work is partially supported by the Ministry
of Education, Culture, Sports, Science and Technology (MEXT),
Grant-in-Aid No. 25400239.

\end{ack}


\clearpage
\appendix 

\section*{Improvements of the orbital elements}
\label{sec:refine_oelem}

Observed pulse-period variations in XBPs include two distinct effects,
the intrinsic spin-period change and the orbital Doppler
shifts.
In Be XBPs, both of them often correlate with the orbital
phase.  Therefore, it is not easy to separate the two effects
from the observed period data.
Actually, some of the Be XBPs analyzed here
were found to show period variations coupled with the orbital modulation,
even though the orbital Doppler effects had been already removed
in the Fermi GBM pulsar data.

We hence 
construct a numerical pulse-period model,
taking into account both the effects, 
and then fit it to the data, in an attempt to simultaneously determine
the spin-period changes and improve the orbital elements.
The method has been utilized in
\citet{2015PASJ...67...73S} and \citet{2015ApJ...815...44M}.
The analysis procedure using the MAXI GSC and Fermi GBM data,
is described below, together with
the results obtained from
4U 0115$+$63, GS 0834$+$430, KS 1947$+$300, and GRO J1008$+$57.

\subsection*{Analysis procedure}
We employed
the empirical power-law
model of equation (\ref{equ:nudotLpower})
to express the intrinsic spin-period change.
As discussed there,
the frequency change during large outbursts
with $L\gtsim 10^{37}$ erg s$^{-1}$
can be approximated as
$\dot{\nu}_{\rm s} = k L^\alpha$,
where $\alpha$ is $\simeq$0.85--1
and $k$ is constant.
The spin frequency $\nu_{\rm s}(t)$ at a given time $t$
is then expressed by
\begin{equation}
\nu_{\rm s}(t) = \nu_i - \int_{\tau_i}^{t} k \left\{L(t^\prime)\right\}^\alpha dt^\prime,
\label{equ:numodel}
\end{equation} 
where $\nu_i = \nu_{\rm s}(\tau_i)$ are
values at reference epochs $\tau_i$, $i=1,2..$.
We defined $\tau_i$ for each outburst separately,
because Be XBPs usually spin down gradually 
between the adjacent outbursts
by the propeller effects.

The period modulation due to the orbital motion is calculated with
the orbital elements, namely, $P_{\rm B}$, $e$, $a_{\rm x}\sin i$,
$\tau_0$ and $\omega_0$ (see table \ref{tab:beparam}).
The velocity of the pulsar orbital motion along
the line of sight, $v_{\rm l}(t)$, is represented by
\begin{equation}
v_{\rm l}(t)
= \frac{2\pi a_{\rm x} \sin i}{ P_{\rm B} \sqrt{1-e^2}} \left\{\cos\left(\theta(t)+\omega_0\right)+e\cos\omega_0\right\}
\label{equ:vl}
\end{equation}
where $\theta(t)$ is a parameter called ``true anomaly'' 
associated with an elliptical orbit, and calculated from the Kepler's
equation.
The observed barycentric pulse frequency, $\nu_{\rm obs}(t)$, is then expressed by
\begin{equation}
\nu_{\rm obs}(t) \simeq \nu_{\rm s}(t) \left(1+\frac{v_{\rm l}(t)}{c}\right)^{-1}.
\label{equ:pobsmod}
\end{equation}

The model represented by equations (\ref{equ:numodel}), (\ref{equ:vl}), and (\ref{equ:pobsmod}),
includes at least 7 parameters,
$\nu_0$, $\alpha$, $k$ in (\ref{equ:numodel}), 
and 5 orbital elements in (\ref{equ:vl}).
We estimated the source luminosity, $L(t)$, in (\ref{equ:vl}) from 
the GSC 2--20 keV light-curve data in 1-d time bin assuming
that the emission averaged over the time bin is approximately
constant and isotropic,
and then fit the model to the barycentric periods from the Fermi GBM data.

In some sources, all the orbital parameters cannot be determined
only from the present data.
If some of the parameters are considered to be better determined in the past, 
we treated them as fixed ones.
The details on each source are presented, in the below.

\subsection*{4U 0115$+$63}

\begin{figure*}
\includegraphics[width=83mm]{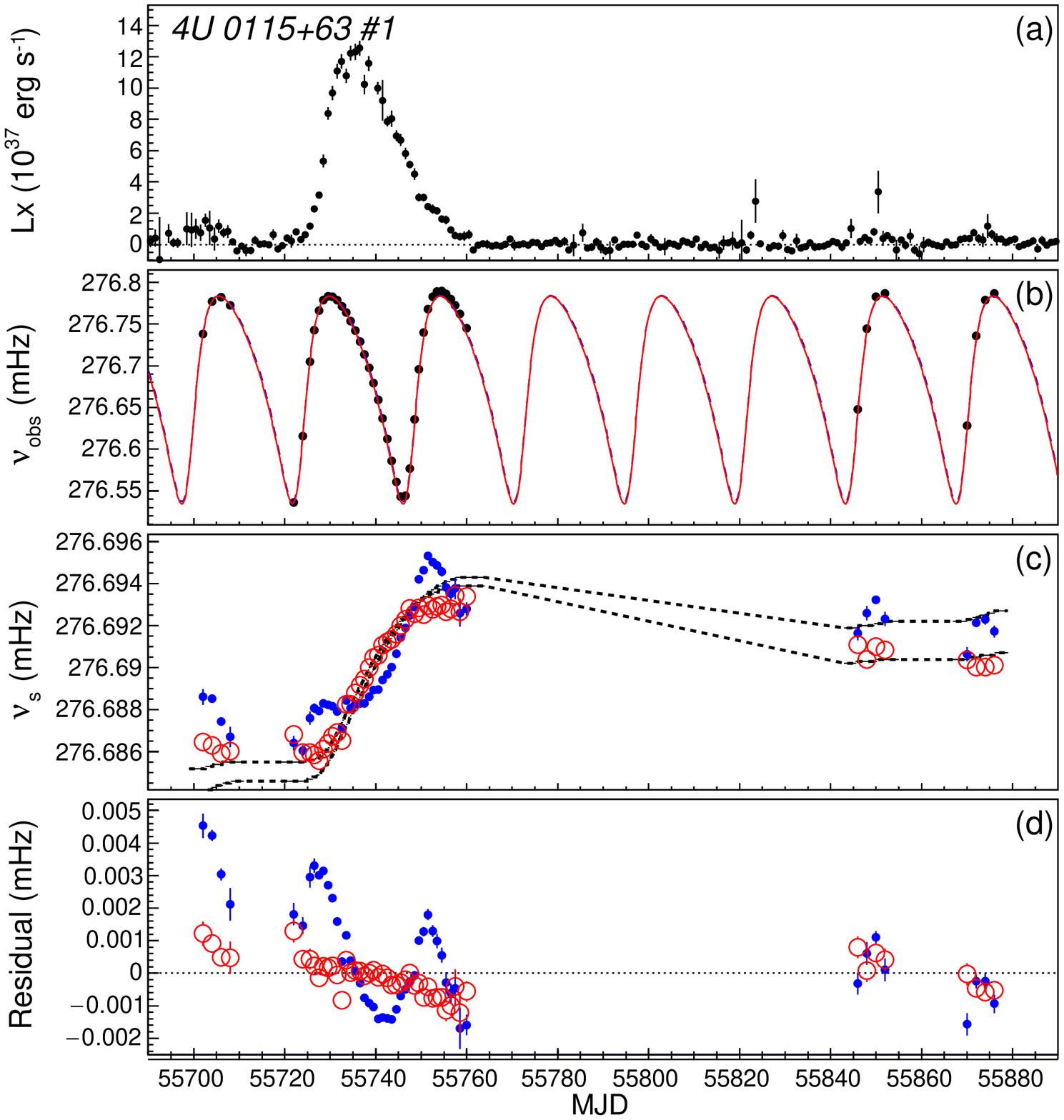}
\hspace{2mm}
\includegraphics[width=83mm]{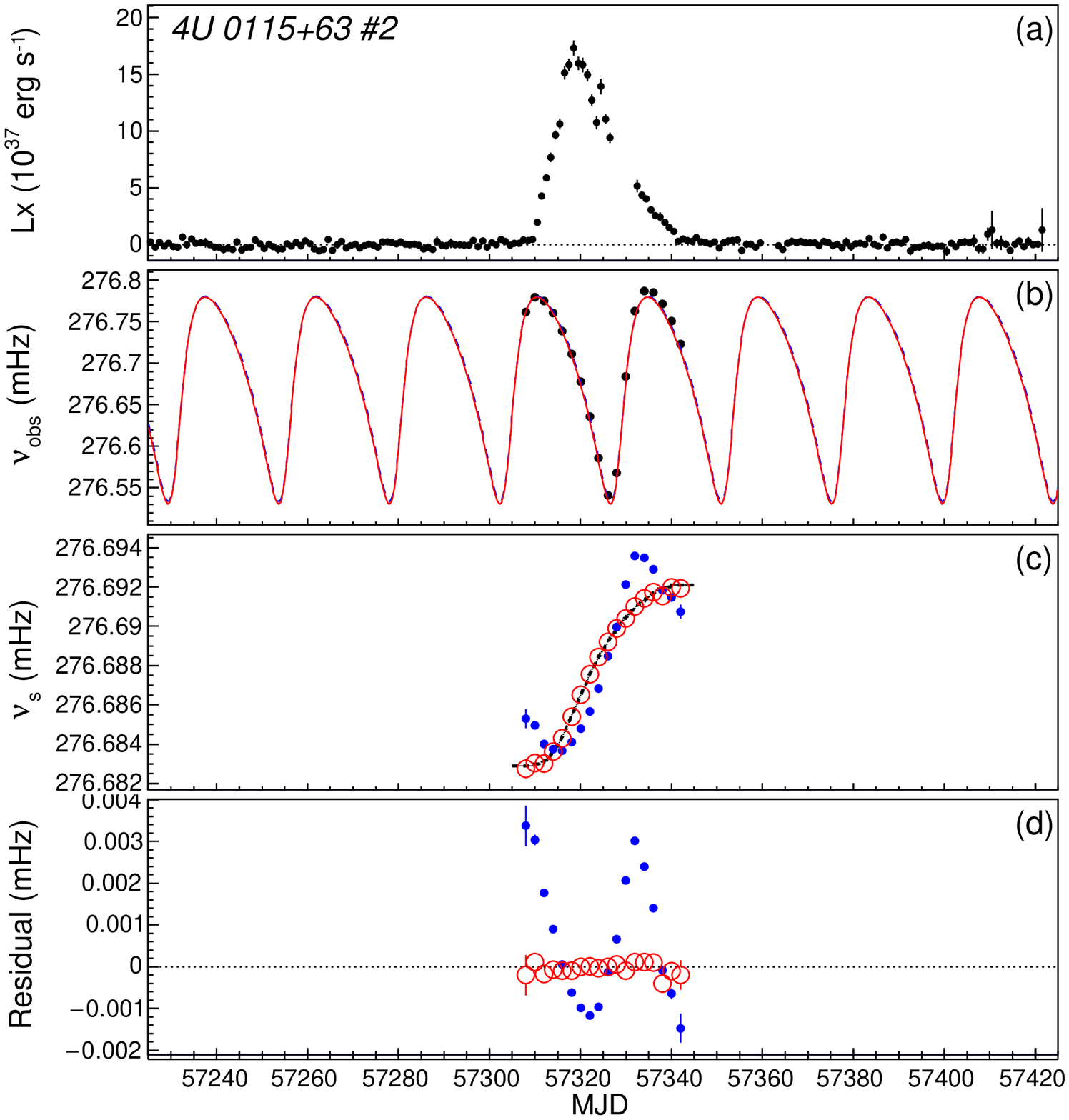}
\caption{
Model fits to the observed pulse-period evolution of 4U 0115$+$63,
during the outburst in 2011 June--July
(left panels)
and that in 2015 October--November
(right panels).
(a) MAXI GSC 2-20 keV light curves. The data are the same as those in figure \ref{fig:lcper}.
(b) Barycentric pulse frequencies $\nu_\mathrm{obs}$ observed by the Fermi GBM 
and the expected modulation due to the orbital motion.
Two modulation curves calculated from the previous orbital elements (dashed blue) in the
literature and those refined in the present analysis (solid red), 
both of which are shown in table \ref{tab:beparam}, cannot be resolved in this
scale.
(c) Pulsar spin frequencies $\nu_\mathrm{s}$ corrected for the orbital effects,
employing the previous orbital elements (blue dots), and those
obtained in the present analysis (red circles).
A dashed line represents the best-fit spin-frequency
model with equation (\ref{equ:numodel}).
(d) 
Residuals of the best-fit spin frequency models,
where symbols mean the same as in panel (c).
}
\label{fig:pmodfit_4u0115}
\end{figure*}

Figure \ref{fig:pmodfit_4u0115} shows 
the fit with equations (\ref{equ:numodel}), (\ref{equ:vl}), and (\ref{equ:pobsmod})
to the data of 4U 0115$+$63. 
The data are the same as in figure \ref{fig:lcper}, 
but focused on periods of two giant outbursts.
Both outbursts lasted longer than 30 d, and thus
covered the entire orbital cycle of 24.3 d.  
We performed the period model fit by allowing all the parameters free.  
Thus, the fit has indeed been improved
(figure \ref{fig:pmodfit_4u0115} d) by adjusting the orbital
parameters.
The refined orbital
parameters are listed in table \ref{tab:beparam},
in comparison with the previous ones.

\subsection*{GS 0834$+$430}

Figure \ref{fig:pmodfit_ks1947_gs0834} left panels present the model fit 
to the data of GS 0834$+$430.  A significant outburst has
been detected once by the two instruments, in 2012 July.  
The outburst lasted $\sim$ 30 d, which did not cover the entire
orbital phase of 105.8 d.  We thus performed the model fit with 
the orbital period fixed at 105.8 d, which had been obtained previously
by \citet{1997ApJ...479..388W}, and $\alpha=6/7$.
Again, the fit has been improved significantly by refining the orbital
parameters.  
The refined parameters are listed in table \ref{tab:beparam}.

\subsection*{KS 1947$+$300}

\begin{figure*}
\includegraphics[width=83mm]{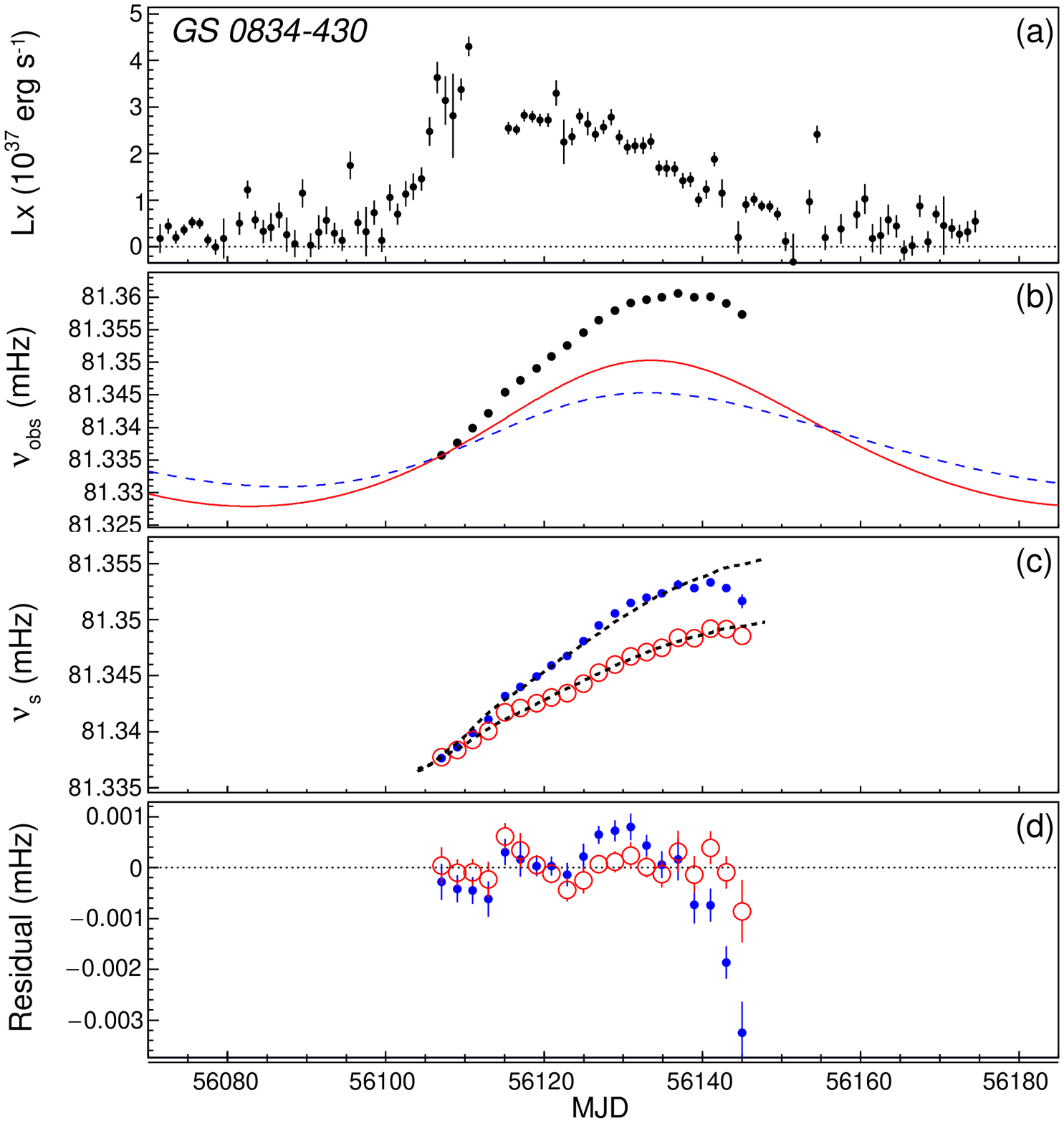}
\hspace{2mm}
\includegraphics[width=83mm]{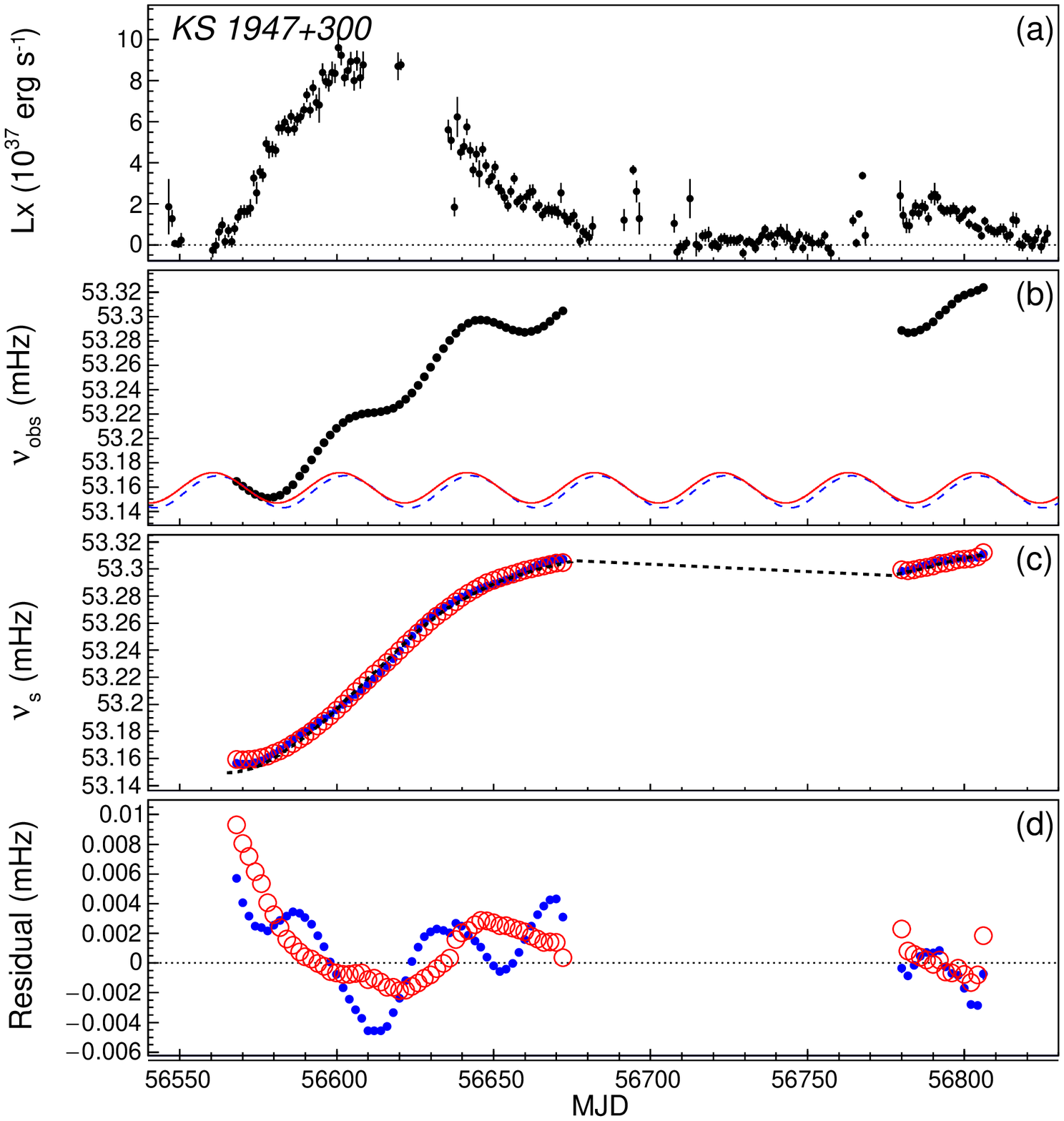}
\caption{
Model fits to observed pulse-period evolution
in KS 1947$+$300 (left panels) and GS 0834$-$430 (right panels).
The meanings of all symbol are same as
in figure \ref{fig:pmodfit_4u0115}.
In panels (b), red and blue lines
represent orbital modulation 
calculated from the previously orbital elements
and those refined in this analysis, respectively.
}
\label{fig:pmodfit_ks1947_gs0834}
\end{figure*}

Figure \ref{fig:pmodfit_ks1947_gs0834} right panels show the period
fit for KS 1947$+$300.  The source exhibited on outburst activity
since 2013 September to 2015 March, where the first major outburst was
followed by three minor ones.  The first outburst lasted for about
100 d, which covers about two cycles of the 40.5 d orbital period.  
The model fit was performed by allowing all the parameters free.  
%
The data-to-model residuals at the bottom of figure \ref{fig:pmodfit_ks1947_gs0834}
reveal that the artificial modulation coupled with the orbital Doppler effect
has been successfully reduced.

\subsection*{GRO J1008$+$57}

\begin{figure*}
\includegraphics[width=83mm]{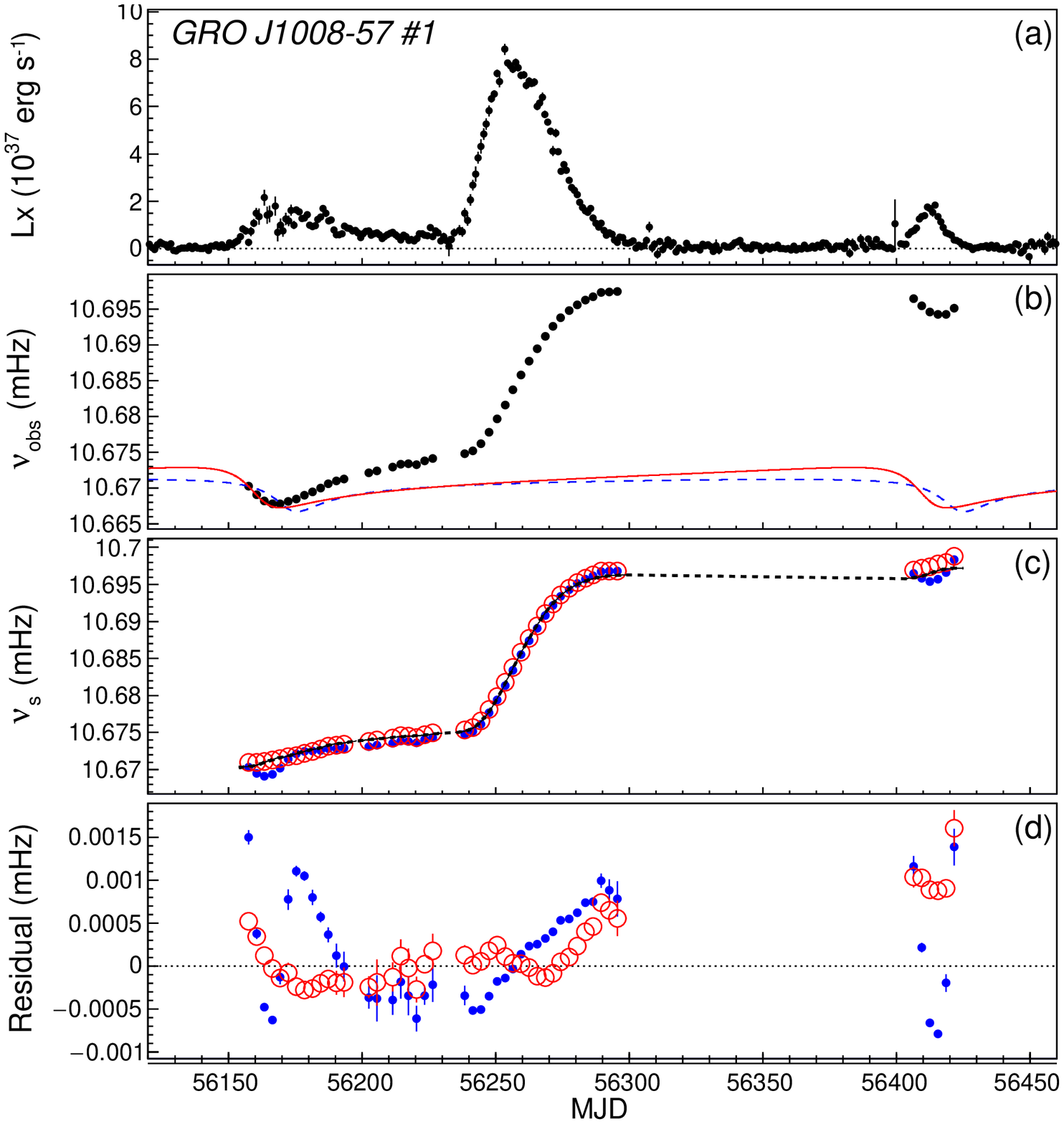}
\hspace{2mm}
\includegraphics[width=83mm]{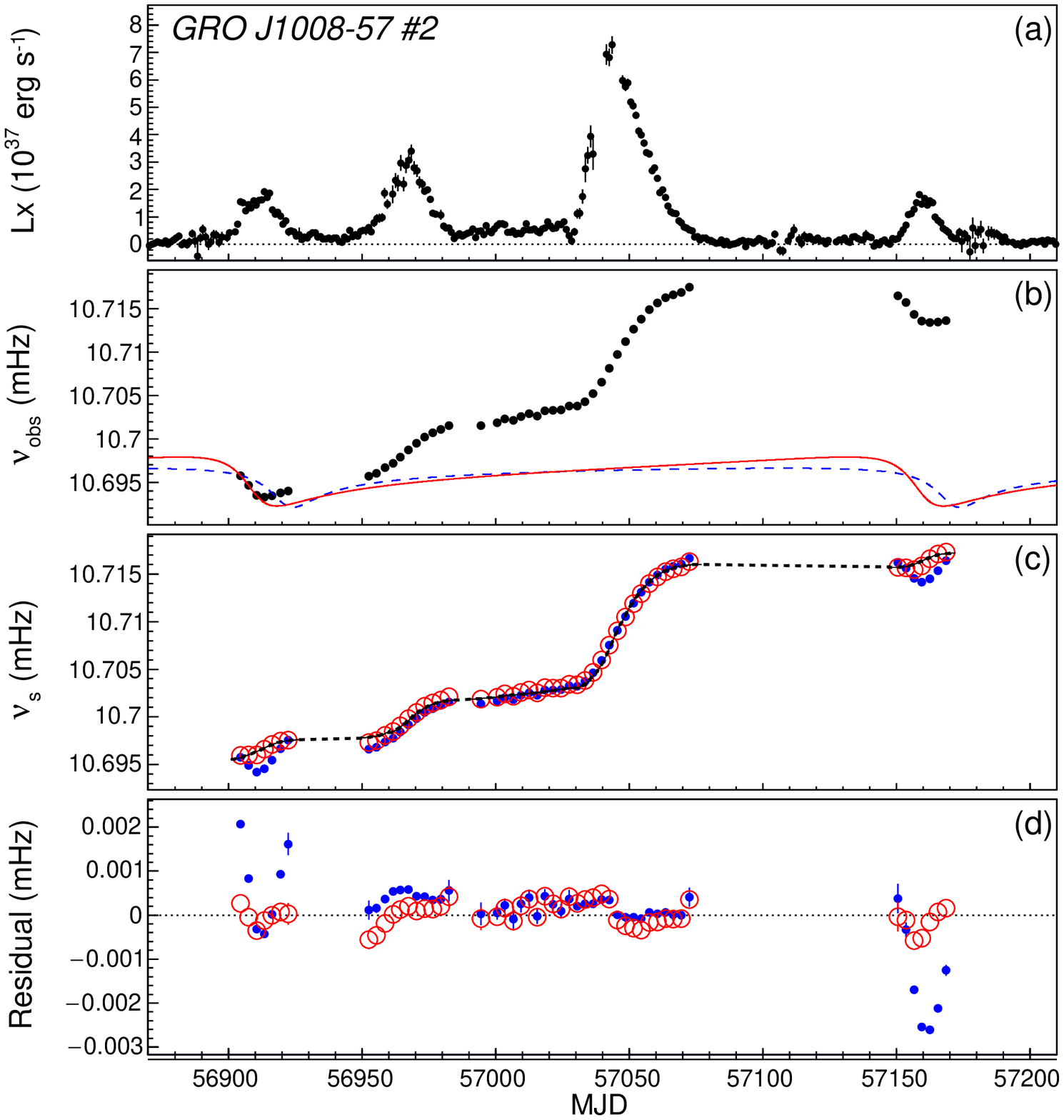}
\caption{
The same as figure \ref{fig:pmodfit_4u0115}, but for 
GRO J1008$-$57 in the outburst from
2012 August to 2013 January (left panels)
and that from 2014 August to 2015 February (right panels).
}
\label{fig:pmodfit_groj1008}
\end{figure*}

The results on GRO J1008$+$57 are presented in figure \ref{fig:pmodfit_groj1008},
covering two extended active periods.
As seen in figure \ref{fig:lcper}, the source normally
repeated outbursts every periastron passage by the 249.48 d orbital cycle.
However, during these extended active periods,
the source exhibited multiple flares 
almost throughout the entire orbital cycle.
The orbital period is precisely determined by the pulse arrival time
analysis \citep{2013A&A...555A..95K}.
We thus performed the model fit with the orbital period at 
this value, 249.48 d.
%
The model-fit residuals clarify that 
the refined orbital parameters better reproduce the 
observed pulse-period modulation, 
in particular, near the periastron phases.


\begin{thebibliography}{}



\bibitem[Arnaud(1996)]{1996ASPC..101...17A} Arnaud, K.~A.\ 1996,
  Astronomical Data Analysis Software and Systems V, 101, 17



\bibitem[Basko \& Sunyaev(1975)]{1975A&A....42..311B} Basko, M.~M., \&
  Sunyaev, R.~A.\ 1975, \aap, 42, 311

  

\bibitem[Baykal et al.(2000)]{2000ApJ...544L.129B} Baykal, A., Stark,
  M.~J., \& Swank, J.\ 2000, \apjl, 544, L129

  
\bibitem[Bildsten et al.(1997)]{1997ApJS..113..367B} 
Bildsten, L., Chakrabarty, D., Chiu, J., et al.\ 1997, \apjs, 113, 367 


\bibitem[Bonnet-Bidaud \& Mouchet(1998)]{1998A&A...332L...9B}
  Bonnet-Bidaud, J.~M., \& Mouchet, M.\ 1998, \aap, 332, L9

\bibitem[Bozzo et al.(2009)]{2009A&A...493..809B}
  Bozzo, E., Stella, L., Vietri, M., \& Ghosh, P.\ 2009, \aap, 493, 809 


  

\bibitem[Caballero et al.(2007)]{2007A&A...465L..21C} 
Caballero, I., Kretschmar, P., Santangelo, A., et al.\ 2007, \aap, 465, L21 


\bibitem[Camero-Arranz et al.(2010)]{2010ApJ...708.1500C} 
Camero-Arranz, A., Finger, M.~H., Ikhsanov, N.~R., Wilson-Hodge, C.~A., 
\& Beklen, E.\ 2010, \apj, 708, 1500 


\bibitem[Campbell(2012)]{2012MNRAS.420.1034C} Campbell, C.~G.\ 2012,
  \mnras, 420, 1034



\bibitem[Coburn et al.(2002)]{2002ApJ...580..394C} Coburn, W., Heindl, 
W.~A., Rothschild, R.~E., et al.\ 2002, \apj, 580, 394 


\bibitem[Coe et al.(1988)]{1988MNRAS.232..865C}
Coe, M.~J., Payne, B.~J., Longmore, A., \& Hanson, C.~G.\ 1988, \mnras, 232, 865

\bibitem[Coe et al.(1994)]{1994MNRAS.270L..57C}
Coe, M.~J., Roche, P., 
Everall, C., et al.\ 1994, \mnras, 270, L57 

  
\bibitem[Coe et al.(2001)]{2001MNRAS.324..623C}
Coe, M.~J., Negueruela, I., 
Buckley, D.~A.~H., Haigh, N.~J., 
\& Laycock, S.~G.~T.\ 2001, \mnras, 324, 623 


\bibitem[Coe et al.(2007)]{2007MNRAS.378.1427C}
Coe, M.~J., Bird, A.~J., Hill, A.~B., et al.\ 2007, \mnras, 378, 1427 




\bibitem[Doroshenko et al.(2016)]{2016A&A...589A..72D}
Doroshenko, V., Tsygankov, S., \& Santangelo, A.\ 2016, \aap, 589, A72





\bibitem[Finger et al.(1996)]{1996A&AS..120C.209F} 
Finger, M.~H., Wilson, R.~B., \& Chakrabarty, D.\ 1996, \aaps, 120, 209

\bibitem[Finger et al.(1996)]{1996ApJ...459..288F} 
Finger, M.~H., Wilson, R.~B., \& Harmon, B.~A.\ 1996, \apj, 459, 288 


\bibitem[Finger et al.(2009)]{2009arXiv0912.3847F} 
Finger, M.~H., Beklen, E., Narayana Bhat, P., et al.\ 2009, arXiv:0912.3847


\bibitem[F{\"u}rst et al.(2014)]{2014ApJ...784L..40F} 
  F{\"u}rst, F., et al.\ 2014, \apjl, 784, LL40 


\bibitem[F{\"u}rst et al.(2015)]{2015ApJ...806L..24F} F{\"u}rst, F.,
  Pottschmidt, K., Miyasaka, H., et al.\ 2015, \apjl, 806, L24


\bibitem[Galloway et al.(2004)]{2004ApJ...613.1164G} 
  Galloway, D.~K., at al.\ 2004, \apj, 613, 1164 


\bibitem[Ghosh \& Lamb(1979)]{1979ApJ...232..259G}
  Ghosh, P., \& Lamb, F.~K.\ 1979a, \apj, 232, 259 

\bibitem[Ghosh \& Lamb(1979)]{Ghosh_Lamb1979}
  Ghosh, P., \& Lamb, F.~K.\ 1979b, \apj, 234, 296 




\bibitem[Grindlay et al.(1984)]{1984ApJ...276..621G} 
Grindlay, J.~E., Petro, L.~D., \& McClintock, J.~E.\ 1984, \apj, 276, 621 
  


\bibitem[Heindl et al.(1999)]{1999ApJ...521L..49H} 
Heindl, W.~A., Coburn, W., Gruber, D.~E., et al.\ 1999, \apjl, 521, L49 

\bibitem[Heindl et al.(2001)]{2001ApJ...563L..35H} 
Heindl, W.~A., Coburn, W., Gruber, D.~E., et al.\ 2001, \apjl, 563, L35 



\bibitem[{\.I}nam et al.(2004)]{2004MNRAS.349..173I} 
{\.I}nam, S.~{\c C}., Baykal, A., Matthew Scott, D., Finger, M., 
\& Swank, J.\ 2004, \mnras, 349, 173 



\bibitem[Israel et al.(2000)]{2000MNRAS.314...87I} Israel, G.~L., Covino, 
S., Campana, S., et al.\ 2000, \mnras, 314, 87 





\bibitem[Klochkov et al.(2012)]{2012A&A...542L..28K} 
Klochkov, D., Doroshenko, V., Santangelo, A., et al.\ 2012, \aap, 542, L28 

\bibitem[Klu{\'z}niak \& Rappaport(2007)]{2007ApJ...671.1990K}
  Klu{\'z}niak, W., \& Rappaport, S.\ 2007, \apj, 671, 1990


\bibitem[Kodaira et al.(1985)]{1985PASJ...37...97K} 
Kodaira, K., Nishimura, S., Kondo, M., et al.\ 1985, \pasj, 37, 97


\bibitem[Kuehnel et al.(2014)]{2014ATel.5856....1K} 
Kuehnel, M., Finger, 
M.~H., Fuerst, F., et al.\ 2014, The Astronomer's Telegram, 5856,  

\bibitem[K{\"u}hnel et al.(2013)]{2013A&A...555A..95K} 
K{\"u}hnel, M.,  M{\"u}ller, S., Kreykenbohm, I., et al.\ 2013, \aap, 555, A95


\bibitem[Lamb et al.(1973)]{1973ApJ...184..271L}
  Lamb, F.~K., Pethick, C.~J., \& Pines, D.\ 1973, \apj, 184, 271 



\bibitem[Liu et al.(2006)]{2006A&A...455.1165L}
Liu, Q.~Z., van Paradijs, J., \& van den Heuvel, E.~P.~J.\ 2006, \aap, 455, 1165


\bibitem[Lovelace et al.(1995)]{1995MNRAS.275..244L} Lovelace,
  R.~V.~E., Romanova, M.~M., \& Bisnovatyi-Kogan, G.~S.\ 1995, \mnras,
  275, 244


\bibitem[Lutovinov et al.(2016)]{2016MNRAS.462.3823L}
Lutovinov, A.~A., Buckley, D.~A.~H., Townsend, L.~J., Tsygankov, S.~S., \&
Kennea, J.\ 2016, \mnras, 462, 3823

  
\bibitem[Makishima et al.(1990)]{1990PASJ...42..295M} 
Makishima, K., Ohashi, T., Kawai, N., et al.\ 1990, \pasj, 42, 295 

\bibitem[Makishima et al.(1999)]{Makishima1999}
Makishima, K., Mihara, T., Nagase, F. \& Tanaka, Y., 1999, \apj, 525, 978


\bibitem[Marcu-Cheatham et al.(2015)]{2015ApJ...815...44M}
Marcu-Cheatham, D.~M., Pottschmidt, K., K{\"u}hnel, M., et al.\ 2015, \apj, 815, 44 


\bibitem[Mason et al.(1978)]{1978MNRAS.184P..45M} Mason, K.~O., Murdin, 
P.~G., Parkes, G.~E., \& Visvanathan, N.\ 1978, \mnras, 184, 45P 


\bibitem[Matsuoka et al.(2009)]{Matsuoka_pasj2009} 
Matsuoka, M., et al. \ 2009, \pasj, 61, 999


\bibitem[McBride et al.(2006)]{2006A&A...451..267M} McBride, V.~A.,
  Wilms, J., Coe, M.~J., et al.\ 2006, \aap, 451, 267


\bibitem[Meegan et al.(2009)]{Meegan2009} 
  Meegan, C., et al.\ 2009, \apj, 702, 791 


\bibitem[Mihara et al.(1991)]{1991ApJ...379L..61M}
Mihara, T., Makishima, K., Kamijo, S., et al.\ 1991, \apjl, 379, L61

\bibitem[Mihara et al.(1998)]{1998AdSpR..22..987M}
Mihara, T., Makishima, K., \& Nagase, F.\ 1998, Advances in Space Research, 22, 987 


\bibitem[Mihara et al.(2004)]{2004ApJ...610..390M} 
Mihara, T., Makishima, K., \& Nagase, F.\ 2004, \apj, 610, 390 

  
\bibitem[Mihara et al.(2011)]{Mihara_pasj2011} 
Mihara, T., et al.\  2011, \pasj, 63, 623




\bibitem[Motch \& Janot-Pacheco(1987)]{1987A&A...182L..55M} 
Motch, C., \& Janot-Pacheco, E.\ 1987, \aap, 182, L55 



\bibitem[Naik et al.(2013)]{2013ApJ...764..158N} Naik, S., Maitra, C., 
Jaisawal, G.~K., \& Paul, B.\ 2013, \apj, 764, 158 



\bibitem[Nakahira et al.(2012)]{2012PASJ...64...13N}
Nakahira, S., Koyama, S., Ueda, Y., et al.\ 2012, \pasj, 64,


\bibitem[Nakajima et al.(2006)]{2006ApJ...646.1125N} 
Nakajima, M., Mihara, T., Makishima, K., \& Niko, H.\ 2006, \apj, 646, 1125 


\bibitem[Nakajima et al.(2010)]{2010ApJ...710.1755N}
Nakajima, M., Mihara, T., \& Makishima, K.\ 2010, \apj, 710, 1755 


\bibitem[Negueruela et al.(1999)]{1999MNRAS.307..695N} 
Negueruela, I., Roche, P., Fabregat, J., \& Coe, M.~J.\ 1999, \mnras, 307, 695 



\bibitem[Negueruela \& Okazaki(2001)]{2001A&A...369..108N} 
Negueruela, I., \& Okazaki, A.~T.\ 2001, \aap, 369, 108 


\bibitem[Negueruela et al.(2003)]{2003A&A...397..739N} Negueruela, I.,
  Israel, G.~L., Marco, A., Norton, A.~J., \& Speziali, R.\ 2003,
  \aap, 397, 739



\bibitem[Pakull et al.(2003)]{2003ATel..202....1P} Pakull, M.~W.,
  Motch, C., \& Negueruela, I.\ 2003, The Astronomer's Telegram, 202,

\bibitem[Parkes et al.(1980)]{1980MNRAS.190..537P} Parkes, G.~E., Murdin, 
P.~G., \& Mason, K.~O.\ 1980, \mnras, 190, 537 

\bibitem[Postnov et al.(2015)]{2015MNRAS.446.1013P} Postnov, K.~A.,
  Mironov, A.~I., Lutovinov, A.~A., et al.\ 2015, \mnras, 446, 1013


\bibitem[Raichur \& Paul(2010)]{2010MNRAS.406.2663R} 
Raichur, H., \& Paul, B.\ 2010, \mnras, 406, 2663 


\bibitem[Rappaport \& Joss(1977)]{1977Natur.266..683R} Rappaport, S.,
  \& Joss, P.~C.\ 1977, \nat, 266, 683


\bibitem[Ravenhall \& Pethick(1994)]{1994ApJ...424..846R} Ravenhall,
  D.~G., \& Pethick, C.~J.\ 1994, \apj, 424, 846


  
\bibitem[Reig et al.(2004)]{2004A&A...421..673R} Reig, P., Negueruela,
  I., Fabregat, J., et al.\ 2004, \aap, 421, 673

\bibitem[Reig et al.(2005)]{2005A&A...440.1079R} Reig, P., Negueruela,
  I., Fabregat, J., Chato, R., \& Coe, M.~J.\ 2005, \aap, 440, 1079


\bibitem[Reig (2011)]{Reig2011} 
  Reig, P.\ 2011, \apss, 332, 1

\bibitem[Reig et al.(2011)]{2011A&A...533A..23R} Reig, P., Nespoli,
  E., Fabregat, J., \& Mennickent, R.~E.\ 2011, \aap, 533, A23

  
\bibitem[Reig \& Fabregat(2015)]{2015A&A...574A..33R} Reig, P., \&
  Fabregat, J.\ 2015, \aap, 574, A33



\bibitem[Revnivtsev \& Mereghetti(2015)]{2015SSRv..191..293R}
Revnivtsev, M., \& Mereghetti, S.\ 2015, \ssr, 191, 293 



\bibitem[Reynolds et al.(1996)]{1996A&A...312..872R} 
Reynolds, A.~P., et al.\ 1996, \aap, 312, 872 


\bibitem[Riquelme et al.(2012)]{2012A&A...539A.114R} 
Riquelme, M.~S., Torrej{\'o}n, J.~M., \& Negueruela, I.\ 2012, \aap, 539, A114 


\bibitem[Santangelo et al.(1999)]{1999ApJ...523L..85S} 
Santangelo, A., Segreto, A., Giarrusso, S., et al.\ 1999, \apjl, 523, L85 

\bibitem[Shakura et al.(2012)]{2012MNRAS.420..216S} Shakura, N.,
  Postnov, K., Kochetkova, A., \& Hjalmarsdotter, L.\ 2012, \mnras,
  420, 216


\bibitem[Shi et al.(2015)]{2015ApJ...813...91S} Shi, C.-S., Zhang,
  S.-N., \& Li, X.-D.\ 2015, \apj, 813, 91


\bibitem[Staubert et al.(2007)]{2007A&A...465L..25S}
Staubert, R., Shakura, N.~I., Postnov, K., et al.\ 2007, \aap, 465, L25 


\bibitem[Steele et al.(1998)]{1998MNRAS.297L...5S}
Steele, I.~A., 
Negueruela, I., Coe, M.~J., \& Roche, P.\ 1998, \mnras, 297, L5 



\bibitem[Sugizaki et al.(2011)]{sugizaki_pasj2011} 
  Sugizaki, M. et al.\ 2011, \pasj, 63, 635

\bibitem[Sugizaki et al.(2015)]{2015PASJ...67...73S} 
Sugizaki, M., Yamamoto, T., Mihara, T., Nakajima, M., 
\& Makishima, K.\ 2015, \pasj, 67, 73 



\bibitem[Takagi et al.(2016)]{2016PASJ...68S..13T} Takagi, T., Mihara,
  T., Sugizaki, M., Makishima, K., \& Morii, M.\ 2016, \pasj, 68, S13


\bibitem[Tendulkar et al.(2014)]{2014ApJ...795..154T} 
Tendulkar, S.~P., 
F{\"u}rst, F., Pottschmidt, K., et al.\ 2014, \apj, 795, 154 


\bibitem[Terada et al.(2006)]{2006ApJ...648L.139T} 
Terada, Y., Mihara, T., Nakajima, M., et al.\ 2006, \apjl, 648, L139 



\bibitem[Tsygankov et al.(2012)]{2012MNRAS.421.2407T} Tsygankov,
  S.~S., Krivonos, R.~A., \& Lutovinov, A.~A.\ 2012, \mnras, 421, 2407


\bibitem[Tsygankov et al.(2016)]{2016MNRAS.457..258T} 
Tsygankov, S.~S., 
Lutovinov, A.~A., Krivonos, R.~A., et al.\ 2016, \mnras, 457, 258 



\bibitem[Verrecchia et al.(2002)]{2002A&A...393..983V} 
Verrecchia, F., Israel, G.~L., Negueruela, I., et al.\ 2002, \aap, 393, 983 


\bibitem[Walter et al.(2015)]{2015A&ARv..23....2W} Walter, R.,
  Lutovinov, A.~A., Bozzo, E., \& Tsygankov, S.~S.\ 2015, \aapr, 23, 2


\bibitem[Wang(1987)]{1987A&A...183..257W} Wang, Y.-M.\ 1987, \aap, 183, 257 
\bibitem[Wang(1995)]{1995ApJ...449L.153W} Wang, Y.-M.\ 1995, \apjl, 449, L153 

\bibitem[Wang(1997)]{1997ApJ...475L.135W} Wang, Y.-M.\ 1997, \apjl, 475, L135 
  

\bibitem[Wasserman \& Shapiro(1983)]{1983ApJ...265.1036W} Wasserman,
  I., \& Shapiro, S.~L.\ 1983, \apj, 265, 1036


\bibitem[Wilson et al.(1997)]{1997ApJ...479..388W} Wilson, C.~A., Finger, 
M.~H., Harmon, B.~A., et al.\ 1997, \apj, 479, 388 

\bibitem[Wilson et al.(2002)]{2002ApJ...570..287W} 
  Wilson, C.~A., et al.\ 2002, \apj, 570, 287


\bibitem[Wilson et al.(2008)]{2008ApJ...678.1263W} 
Wilson, C.~A., Finger, 
M.~H., \& Camero-Arranz, A.\ 2008, \apj, 678, 1263-1272 




\bibitem[Yamamoto et al.(2011)]{2011PASJ...63S.751Y} 
Yamamoto, T., et al.\ 2011, \pasj, 63, 751 

\bibitem[Yamamoto et al.(2014)]{2014PASJ...66...59Y} 
Yamamoto, T., et al.\ 2014, \pasj, 66, 59 





  





















































































\end{thebibliography}
\end{document}